\documentclass[10pt]{article}
\usepackage[T1]{fontenc}

\usepackage{authblk}

\usepackage{geometry}
\usepackage{alphabeta} 
\usepackage{physics}
\usepackage{amsmath}  
\usepackage{amsthm}
\usepackage{graphicx}
\usepackage{amssymb}
\usepackage{slashed}
\usepackage{float}
\usepackage{hyperref}
\usepackage{breqn}   
\usepackage{upgreek}
\usepackage[dvipsnames]{xcolor}
\usepackage{listings}
\usepackage{braket}
\usepackage{mathtools}
\usepackage{pgfplots}
\usepackage{comment}
\usepackage[normalem]{ulem}
\usepackage[compat=1.1.0]{tikz-feynman}
\usepackage{silence} 
\newcommand{\be}{\begin{equation}}
\newcommand{\ee}{\end{equation}}
\newcommand{\bea}{\begin{eqnarray}} 
\newcommand{\eea}{\end{eqnarray}}
\newcommand{\bal}{\begin{aligned}}
\newcommand{\eal}{\end{aligned}}

\def\a{\alpha}
\def\b{\beta}
\def\g{\gamma}
\def\G{\Gamma}
\def\d{\delta}
\def\D{\Delta}
\def\ve{\varepsilon}
\def\m{\mu}
\def\n{\nu}
\def\l{\lambda}

\def\r{\rho}
\def\s{\sigma}
\def\calT {\mathcal{T}}
\def\rmd {\mathrm{d}}
\title{Energy momentum tensor correlators in $\phi^4$ theory II:\\
 The spin-two sector}

 \author[]{Nikos Irges\thanks{irges@mail.ntua.gr} }
\author[]{Leonidas Karageorgos\thanks{leokarageorgos@mail.ntua.gr}}

\affil[]{Department of Physics, National Technical University of Athens\\

\normalsize Zografou Campus, GR-15780, Greece
}
\date{}
\begin{document}

\maketitle

\begin{abstract}
We extend the computation of the $C_T$ charge of the 2-point function of the Energy-Momentum Tensor to 4-loops.
We show that $C_T$ decomposes into two sectors, the conformal sector, which encodes the value of the central charge at fixed points 
and an RG-sector that contains logarithmic and constant corrections proportional to the $\beta$-function.
This latter constitutes the main new result of this work and is inaccessible via CFT methods alone.
Furthermore, we demonstrate that $C_T$ satisfies an eigenvalue-like equation analogous to that of the spin-0 charge, as discussed in  part I, though with a different 
in general eigenvalue.
Finally we present three possible applications.
\end{abstract}
\newpage
\tableofcontents

\newpage
\section{Introduction}

In part I of this work \cite{spin0} we (re)examined several correlation functions of operators in the spin-0 sector 
of the four-dimensional $\l \phi^4$ theory. Our main interest lied in the correlators of the trace $\Theta$ of the Energy-Momentum Tensor (EMT) $T_{\m\n}$,
defined as $\Theta\equiv \eta_{\m\n} T^{\m\n}$\footnote{We use the mostly minus metric $\eta_{\m \n} = (+,-,-,-)$ }. 
The computations were performed in Dimensional Regularization (DR), in the $\ve$-expansion around four dimensions, with $\ve = 4 - d$.
This is a computational scheme that is valid near the UV, or Gaussian fixed point where the system can be described by a free 
Conformal Field Theory (CFT) but it is thought that it has some validity also near the IR, or interacting, or Wilson-Fisher (WF) fixed point where the system can also be interpreted as a CFT.
This latter fixed point is denoted in the present work by the symbol ${\Large \star}$. 
Any fixed point in a quantum theory obeys the equation $\b_i = 0$, where $\b_i$ are the beta functions of the couplings.
In $\lambda\phi^4$ theory, the value of the quartic coupling on the Gaussian fixed point is 
$\l=0$ while on the WF it is $\l^* = \frac{16\pi^2}{3}\ve + \cdots$. 
The fixed point value of $\l^*$ for $\ve=1$ reveals the subtle nature of the IR fixed point in the $\ve$-expansion, a three-dimensional, interacting CFT.
For general operators, the abrupt change of dimensionality as the system touches the WF fixed point is perhaps related
to the absence of a smooth conformal limit in perturbation theory and it was one of the reasons
behind the step function-like $\frac{p^2}{\m^2} \to p^2$ (here $p$ is an external momentum and $\m$ is a regulating scale)
substitution rule that had to be applied to correlators in \cite{spin0} so that they transit from their QFT expression to their corresponding CFT expression.
The technique used in \cite{spin0} is the operator insertion method where one constructs the operator as the coincident point limit of the fields and
derivatives that it contains, inserted in a correlator together with another such operator or a number of fundamental fields $\phi$.
The renormalization process resulted in correlators of renormalized fields and operators that were shown to obey appropriate Callan-Symanzik (CS) equations.
The above substitution rule together with its derivative version $ \m \frac{\partial }{\partial\m}\to - p \frac{\partial}{\partial p}+d$
guarantee the correct transition from the correlators obeying Callan-Symanzik equations to obeying Dilatation Ward identities.
Perhaps the most striking consequence of this is that all correlators that contain $\Theta$ must vanish at the fixed point.
This can be guaranteed only if the renormalized trace operator itself is proportional to the couplings's beta-function: $\Theta \sim \b_\l$.
In the next section we summarize the main steps of \cite{spin0} that lead to the spin-0 charge $C_\Theta$ 
via the correlator $\braket{\Theta \Theta}\sim p^d C_\Theta$.

In part II we extend the analysis to the spin-2 sector. 
An immediate issue is how to define a renormalized EMT.
The classic paper \cite{Improvement} constructed an improved EMT with finite matrix elements to all orders of $\l$.
In addition, through an explicit 4-loop calculation it has been confirmed
that the traceless part of ${T}_{\m \n}$ has vanishing anomalous dimension \cite{ManashovTraceless}. 
However, these authors took into consideration only correlation functions of the form $\braket{T_{\m \n} \phi \phi \cdots}$.
Our target is the correlator of the EMT with itself (called $TT$-correlator for short) and in particular the charge $C_T$,
hereafter referred to as the "spin-2 charge", that it contains. 
This means that it is sufficient to look at the traces $\braket{T_{\m\n} T^{\m\n} }$ and $\eta^{\m \n} \eta^{\r \s} \braket{T_{\m\n} T_{\r \s} } $, 
and construct an appropriate linear combination of them in order to isolate the spin-2 charge. 
This approach simplifies the evaluation process, as it avoids the need to manipulate loop integrals with free Lorentz indices. 

In the present analysis we use the vertex corresponding to the EMT operator insertion to evaluate the relevant loop diagrams. 
This technique is equivalent to the coincident point limit method and offers a more tractable computational strategy.
In this work, by $T_{\m\n}$ we denote the reducible representation. For its traceless part we 
will use the notation ${\cal T}_{\m\n}$. The $TT$-correlator can be decomposed as 
$\braket{T_{\m\n} T_{\rho\sigma}} = [C_T \Pi^T_{\m\n\rho\sigma} 
+ C_\Theta \Pi^\Theta_{\m\n\rho\sigma}]p^d$ with the tensor structures
compatible with the conservation of $T_{\m\n}$ and some other constraints that we will make precise in the next section.

The leading order computation of $C_T$ in QFT is the 3-loop, $O(\l^2)$ calculation of \cite{spectralTT}. 
Therefore, to go beyond it, we must compute $C_T$ to 4-loops.\footnote{There are several works on the evaluation of the spin-2 charge, but mainly in the context of a CFT 
\cite{Giombi1, Giombi2, Tassos,4ptemtSerinoa, ClaudioReview, SkenderisCFT} or in the context of a QFT 
with generalised conformal structure \cite{Claudio}, though the latter is constrained to two-loop order.}
The qualitative change that we expect to see at this order is the appearance, for the first time, of logarithmic corrections depending explicitly on the ratio of 
the external momentum over the regulating scale. 
The logarithmic corrections, along with associated constant terms originating from the finite part of the 4-loop integrals and the renormalization 
of the coupling constant, constitute what we refer to as the "RG-sector" of the spin-2 charge.
These corrections represent the main result of this study.
As we detail in Section~\ref{tmntrs}, the RG-sector is proportional to the $\beta$-function, identifying it with the source of the breaking of scale invariance.
Furthermore, the fact that the RG-sector is proportional to the $\beta$-function guarantees a smooth CFT limit: 
$C_T$ becomes automatically $\mu$-independent when the system reaches a fixed point.

We would also like to give some physical meaning to our calculations. Out of the many possible applications, we will present three,
in an increasing order of non-conventionality. The first is a technical operation which allows one to normalize $C_\Theta$
to $C_T^{\rm free}$, the conformal charge of the free CFT at the Gaussian fixed point, chosen here to equal to 1.
The second application is to the running of $C_T$ between the UV and IR fixed points, which is affected by the logarithms.
The third is a possible application to holographic cosmology, where the charges $C_\Theta$ and $C_T$ play a central role, 
governing the scalar and tensor indices $n_S$, $n_T$ and the ratio $r$ of tensor to scalar CMB fluctuations. 

\section{Review of the spin-0 sector and related comments}

The massless $\l \phi^4$ theory is determined by the action
	\be
	S= \int \rmd^d x \mathcal{L}=\int \rmd^d x \left[\frac{1}{2}\eta^{\m \n}\partial_\m \phi \partial_\n \phi 
	-\frac{\l}{4!}\phi^4
	 \right]\, .
	\ee
The Lagrangean in the square brackets
is classicaly scale invariant in $d = 4$, which implies that the trace of the EMT, initially defined as the canonical tensor
$T^{\rm can}_{\m\n}=\frac{\partial {\cal L}}{\partial (\partial_\m \phi)}\partial_\n \phi - \eta_{\m\n}{\cal L}$ is supposed to vanish in four dimensions.
However, from the above Lagrangean, one obtains the EMT 
	\begin{align}
		T_{\m \n}^{\rm can} &= 
		\partial_\m \phi \partial_\n \phi 
		+ \eta_{\m \n} \frac{\l}{4!} \phi^4 
		- \eta_{\m \n} \frac{1}{2}\partial_\a \phi \partial^\a \phi \, 
	\end{align}
that is conserved, but with non-vanishing trace: $\Theta^{\rm can}=(1-\frac{d}{2}) (\partial \phi)^2 + d \frac{\l}{4!} \phi^4$.
The cure for this is to add an improvement term that respects its conservation and at the same time 
makes it traceless.
There are two equivalent ways to do this. The first is to introduce the Belinfante tensor \cite{Improvement}.
 The second is to consider a conformally coupled scalar in a generic curved background spacetime
 	\be\label{Lgravphixi}
	S_{\rm curved}= \int \rmd^d x \sqrt{|g|} \left[\frac{1}{2}g^{\m \n}\partial_\m \phi \partial_\n \phi 
	-\frac{\l}{4!}\phi^4
	+ \frac{1}{2}\xi R \phi^2
	 \right]
	\ee
along with the definition
	\be
	T_{\m \n} = 
	\left. 2\frac{\d S_{\rm curved}}{\sqrt{|g|} \d g^{\m \n}} \right|_{g_{\m \n} = \eta_{\m \n}}\, .
	\ee
 Both of these approaches lead to the same improved EMT
 	\be\label{Tmnxi}
	T_{\m \n}= T_{\m \n}^{\rm can} 
		- \xi \left( \partial_\m \partial_\n - \eta_{\m \n} \Box\right) \phi^2\, ,
	\ee
with $\xi$ the so-called conformal coupling, $\xi = \frac{(d-2)}{4(d-1)}$.
Indeed, for $d = 4$, one can check that the above EMT is conserved and traceless: $\eta^{\m\n} T_{\m\n} \equiv \Theta = 0$.
This is a welcome improvement but it comes at a price.
Because of its vanishing in the classical theory, there is no obvious definition of a bare, quantum operator $\Theta_0$. 
That is to say, there is no classical object to be promoted to a bare operator and then follow the standard renormalization algorithm: $\text{classical}\to\text{bare} \to \text{renormalized}$.
At the same time, the scale symmetry is broken at the quantum level.
In the quantum theory the Langrangean must be renormalized, introducing a renormalization scale $\m$, where a non-vanishing $\b$-function triggers the RG-flow.
From this point of view it is rather obvious that the renormalized operator\footnote{Typically, the renormalization process proceeds 
by promoting the classical couplings and fields to bare quantum quantities, denoted by
adding a label $0$ to the quantity. Then after renormalization, the label is dropped, bringing the notation back to its original form.
Since the notation for classical and renormalized quantities is the same, the distinction between the two should be made from the context.
This lightens quite a bit the notation and it is the convention we will use here. } 
representing the trace of the EMT (the charge associated with the scale transformations) 
should be proportional to the $\b$-function:
\be \label{Theta beta}
\Theta \sim \b_\l \, .
\ee
By dimensional analysis, the trace operator has engineering dimension equal to $d$.
Therefore, one way to define a quantum trace operator is to consider a linear combination of all the 
(independent) operators of dimension $d$. In the case of the massless $\l\phi^4$-theory, 
exploiting the equations of motion, the number of independent operators can be reduced to two. The initial number of operators with
dimension $d$ is four: $(K_1=\partial_\m \phi \partial^\m \phi, K_2=\Box \phi^2, K_3=\phi\Box \phi, K_4=\l\m^\ve \phi^4)$. Utilizing the equations of motion, we get that 
$\phi \Box \phi = -\frac{\l\m^\ve}{6}\phi^4$, so there are three remaining operators. These three operators are connected by the derivative identity 
$\Box \phi^2 = 2\phi \Box \phi + 2 \partial_{\m} \phi \partial^\m \phi$. 
We call the former the $E$-identity and the latter the $F$-identity.\footnote{These identities were shown to hold at the quantum operator level in \cite{spin0}.}
As a result, the basis of independent operators with mass 
dimension $d$ consists of two operators. Any two of the $K_1,\cdots, K_4$ can be chosen as the independent set.
The trace can be therefore represented for example as\footnote{$\phi\Box \phi$ stands for the 
renormalized version of $\phi\Box \phi$ operator and not for the product of the renormalized field $\phi$ 
with the operator $\Box \phi$. The same applies to the  $\Box \phi^2$ operator.}: 
\be
\Theta = \b_\l \left[c_1 \phi\Box \phi+ c_2 \Box \phi^2 \right]\, ,
\ee
where $c_1$ and $c_2$ are coefficients that may depend on $\l$.
The Callan-Symanzik equation satisfied by correlation functions that contain the trace operator determines these coefficients.
Because of the conserved current nature of $T_{\m\n}$, the trace operator $\Theta$, as part of the EMT,  has an exactly vanishing anomalous dimension, i.e. $\Gamma_\Theta=0$.
For example, the Callan-Symanzik equation for the relatively simple correlator $\braket{\Theta\phi\phi}$ is 
\be\label{CS theta 3pt}
\left[ \m \frac{\partial}{\partial \m} + \b_\l \frac{\partial }{\partial \l} + 2 \g_\phi\right]
\braket{\Theta \phi \phi} =0.
\ee
When applying this equation, one should not neglect the mixing effects between the $\phi \Box \phi$ and $\Box \phi^2$ operators.
In this case, the mixing contribution comes from the following loop topology:

\begin{figure}[H]\label{3loopmixing}
\centering
\begin{tikzpicture}[scale= 2.5]
\begin{feynman}
\vertex (x) at (0,-0.15) ;
\vertex (y) at (1,-0.15) ;
\vertex (z) at (0.5,0) ;
\diagram{(x)--(z)--(y)};
\end{feynman}
\draw[color=black](0.5,0.25) circle (0.25);
\draw[color=black](0.5,0.5) circle (0.125);
\node[draw, fill=black, scale=0.7] at (0.38,0.48) {};
\end{tikzpicture}
\caption{The 3-loop diagram contributing to the mixing between the 
$\phi\Box \phi$ and $\Box \phi^2$ operators. The black square denotes a $\phi^4$  insertion.}
\end{figure}
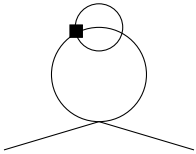

In the presence of mixing the anomalous dimension is a matrix and 
we can write the Callan-Symanzik equation satisfied by the $\Box \phi^2$ and $\phi\Box \phi$ operators as
	\be \label{CS of K-operators}
	\left[ \m \frac{\partial}{\partial \m}+ \b_\l \frac{\partial}{\partial \l}+ 2\g_\phi \right]
	\braket{Q_{I}\phi \phi}
	 + \G_{IJ} 
	\braket{Q_{J}\phi \phi}=0\, , \hskip .5cm I,J=1,2
	\ee
with
	\begin{align} \label{mixing matrix}
	Q = \begin{bmatrix}
		\Box\phi^2 \\
		\phi \Box \phi
		\end{bmatrix} \, \, , \,  \,
		\G = \begin{bmatrix}
		\G_{\phi^2} & 0 \\
		\frac{\l^3}{4(4\pi)^6} & \G_{\phi^4} - \frac{\b_\l}{\l}
		\end{bmatrix}\, .
	\end{align} 
	The off-diagonal mixing term was first determined by \cite{spectralTT} and confirmed in \cite{spin0}.
Combining \eqref{CS theta 3pt}  and \eqref{CS of K-operators} we get a $2 \times 2$ system of differential equations for the coefficients $c_1$ and $c_2$ that can be solved
using the perturbative results for the $\b$-function and the anomalous dimensions of the operators. We list these in Appendix \eqref{quantumaction}. The solution is 
	\begin{align}
	c_1 &=c \frac{\b_\l}{\l}\\
	c_2 &=c\b_\l \left( \frac{\l}{32(4\pi)^4} + O(\l^2) \right)\, ,
	\end{align}
with $c$ an overall normalization constant. So the leading order expression for the trace operator becomes \cite{spin0}:
	\be\label{Thetarenormalized}
		\Theta = c \b_\l \left[-\frac{1}{6} \phi^4
		+ \left(\frac{\l}{32 (4\pi)^4}+O(\l^2) \right) \Box \phi^2 \right]\, .
	\ee
The coefficient of the $\Box \phi^2$ operator above
receives contributions from its anomalous dimension, the renormalization of $\lambda$ and the 
off-diagonal mixing term in \eqref{mixing matrix}. 
Of course, knowing that there are only two independent operators of dimension 4,
we could have expressed $\Theta$ as a combination of $(\partial\phi)^2$ and $\phi^4$, using the $E$ and $F$ identities.
This would be enough reason to prohibit the addition of a $\xi \Box \phi^2$ effective term in the Lagrangean, but
let us elaborate a bit more on this, as it will clarify the connection of our result \eqref{Thetarenormalized} with some related important work.

One could think that if there is a classical operator of the form \eqref{Tmnxi} that in the quantum theory 
gets renormalized term by term, the Zamolodchikov statement
\be
\Theta = \sum_i \b_{g_i} \frac{\partial {\cal L}}{\partial g_i} = \sum_i \b_{g_i} O_{g_i}\, 
\ee
would require a $\xi$-term in the Lagrangean, say in the form $\xi \Box \phi^2$. 
In fact, L\"uscher showed \cite{Luscher} that the conformal coupling $\xi$ in \eqref{Lgravphixi} receives corrections
in the curved space action. The puzzling fact is that even though the $\xi$-term disappears from the Lagrangean in the flat limit,
it survives in $\Theta$ as a $\xi \Box \phi^2$ term even in the flat limit, as shown in \cite{spectralTT}. 
In \cite{Luscher} the beta function of $\xi$ was extracted. In $d=4$ curved space the result was
	\be\label{Luscherbeta}
	\b_\xi = \left(\xi -\frac{1}{6} \right)\G_{\phi^2} + \frac{\l^2}{18(4\pi)^4} + \cdots
	\ee
with the perturbative trajectory of $\xi$ in the vicinity of the Gaussian fixed point given by
	\be\label{Luscherxi}
	\xi = \frac{1}{6} +\frac{\l}{36(4\pi)^2}+\cdots\, ,
	\ee
the solution of $\m \frac{\partial \xi}{\partial \m} = \b_\xi$ with boundary condition $\xi=\xi_0=\frac{1}{6}$ when $\l=0$.
This last result should be another alarm since $\xi$ does not depend on a regulating scale, its running being determined by the running
of $\l$. This is a consequence of the fact that it is a parameter (rather than a coupling) 
associated with an exactly marginal operator, like $\Box \phi^2$. Consistently with that, $\Box \phi^2$ is not an independent operator
in the Lagrangean and its anomalous dimension is determined by the anomalous dimension of $\phi^2$ \cite{spin0}.
Ignoring these warnings, one could argue that the correct form of the renormalized trace operator is of the form
$\Theta \sim \b_\l \phi^4 + \b_\xi \Box\phi^2$, which would clearly contradict the direct computation that lead to \eqref{Thetarenormalized}, where $\b_\l$ is an overall factor.
The contradiction can also be understood by noticing that simultaneous solution to the Wilson-Fisher equations
$\b_\l=0$ and $\b_\xi=0$ does not exist. The non-trivial fixed point would be therefore lost, introducing a quantum Weyl anomaly 
of the form $\Theta^* = \b_\xi^* \Box \phi^2$. 
For all these reasons we will be using the classical $\xi_0$ just as a useful "improvement parameter"
which is however eventually replaced in the renormalized $\Theta$ by the specific $\l$-dependent corrections that the
flat space quantum effects impose. As a consequence, the curved space order parameter of spontaneous scale symmetry breaking $\b_\xi$,
will not be relevant in our flat space approach. The short excursion in curved space that yielded \eqref{Tmnxi} 
is merely a useful trick to obtain an improved EMT.

The leading order  2-point function of $\Theta$ is entirely determined by the self 2-point function of the $K_3=\phi \Box \phi$ operator 
because the other operators (only $K_2$ in this case) contribute higher order corrections. Its value is \cite{spin0}:
	\be\label{theta theta}
		\braket{\Theta \Theta} =c^2 p^4\b_\l^2 
		\left[1 + \frac{6\l}{(4\pi)^2} \ln \left( \frac{-p^2}{\m^2}\right) \right] 
		+ O(\l^6) \, .\, 
	\ee
The minus sign in the logarithm comes from the convention on the renormalization condition at $p^2 =-\m^2$, 
which corresponds to spacelike momenta.
We choose the renormalization condition at this value of the momentum in order to avoid on-shell poles in the correlation functions.
The charge $C_\Theta$ of the leading order correlator 
\be
\braket{\Theta \Theta} \equiv C_\Theta (d-1)^2p^d
\ee
obeys the following system of eigenvalue-like  equations:
	\bea
	 \label{Theta eigenvalue eq}
		\m\frac{\partial}{ \partial \m}C_\Theta &=& -  e_\Theta C_\Theta\nonumber \\
		 \beta_\l\frac{\partial}{ \partial \l}C_\Theta &=& + e_\Theta C_\Theta
	\eea
The two equations sum to the Callan-Symanzik equation. 
The solution for \eqref{theta theta} is
	\be \label{Theta eigenvalue}
		e_\Theta = 2\G_{\phi^4} + O(\l^2)\,.
	\ee
In general, the eigenvalue-like quantity (actually a function of $\l$) $e_\Theta$ 
determines a specific RG flow, as the system \eqref{Theta eigenvalue eq} of differential equations can be solved, 
at least in principle.
	
$C_\Theta$ is fixed up to an overall normalization constant $c$.
As mentioned in the Introduction, this normalization can not be fixed in perturbation theory but it can be related to the normalization 
of the spin-2 charge, as we will see later on.

\section{The spin-2 sector}\label{tmntrs}

This section is divided into two parts. In the first part, we revisit the calculation of the 2-point function of the EMT to order $\mathcal{O}(\lambda^2)$,
originally performed in~\cite{spectralTT}, in order to set the stage for the second part, which presents the 4-loop calculation.

The decomposition of the bare 2-point function in momentum space, for a QFT with vanishing 1-point functions,
\footnote{The one point function $\braket{ T_{\m \n} }$ is given by a scaleless integral, which vanishes in DR.
A clear explanation of the vanishing of scaleless integrals within dimensional regularization is provided in \cite{Gorishnii}.}
 is fixed to the following form:
	\be \label{Tmn Trs decomposition}
		\braket{T^{(0)}_{\m \n} T^{(0)}_{\r \s}}=
		\left[C_{T}^{(0)}(p) \Pi_{\m \n \r \s} 
		+ C_{\Theta}^{(0)}(p) \pi_{\m \n} \pi_{\r \s} \right] p^d\, ,
	\ee
where $C_T$ and $C_\Theta$ are the spin-2 and spin-0 charges respectively and 
	\bea
		\label{pimn} \pi_{\m \n} &=& \eta_{\m \n} - \frac{p_\m p_\n}{p^2}\\
		\label{pimnrs} \Pi^T_{\m \n \r \s }&=& \frac{1}{2} \left( \pi_{\m\r}\pi_{\n\s} +\pi_{\m\s}\pi_{\n\r} 
		- \frac{2}{d-1}\pi_{\m \n}\pi_{\r \s} \right)\, .
	\eea
The spin-2 charge is associated with the traceless part of the EMT's 2-point function,
while the spin-0 charge is associated with the trace.
To evaluate the  charges it is sufficient to consider the traces of the two-point function \eqref{Tmn Trs decomposition}, 
according to
	\begin{align}
\label{ctheta}	C_\Theta^{(0)} &=  
	\frac{ \eta^{\m \n} \eta^{\r \s} \braket{T^{(0)}_{\m \n} T^{(0)}_{\r \s} }}
	{(d-1)^2} p^{-d} \\
\label{ct}	C_T^{(0)} &= \frac{2}{(d+1)(d-2)} 
	\left[ \braket{ T^{(0)\m \n} T^{(0)}_{\m \n} } 
	- \frac{\eta^{\m \n} \eta^{\r \s} \braket{T^{(0)}_{\m \n} T^{(0)}_{\r \s} } }
	{d-1} \right] p^{-d}\, .
	\end{align}
In Feynman diagrams the insertion of the operator $T^{(0)}_{\m\n}$ is realized by the vertex \eqref{Tmn vertex}.
\footnote{Notice that due to the form of \eqref{ct} the trace necessarily contaminates the computation.
A calculation without involving the trace would be possible to do only in the spirit of $\cite{spin0}$, projecting the traceless irrep 
${\cal T}_{\m\n}$ on the basis of dimension-$d$ traceless operators and solving the mixing problem. 
This has been essentially done in \cite{ManashovTraceless} for the 4-loop calculation of the anomalous dimension of symmetric, 
traceless operators in the context of the $\braket{{\cal T}_{\m\n}\cdots\phi \phi}$ correlator.}

In \cite{spin0} it was shown that a maximal loop computation could be avoided in the case of $\braket{\Theta\Theta}$ 
because four powers of $\l$, 
equivalent to 2 orders of perturbation theory could be bypassed by the boundary condition that required the result \eqref{theta theta} to be proportional to $\b_\l^2$,
with the boundary condition imposed by the CS equation.
This is no longer possible in the spin-2 sector where ${\cal T}_{\m\n}$ has no reason to vanish on fixed points.
Thus, the order of the required computation is more straightforward to deduce.
A diagrammatic analysis shows that, as far as $\braket{T_{\m\n}T_{\rho\sigma}}$ is concerned, for a given power of $\l$ in perturbation theory 
the number of required loops follows the rule
\be\label{TTorderrule}
C_T\,\,\,\, {\rm from}\,\,\,\,\braket{T^{(0)}_{\m\n}T^{(0)}_{\r\s}}\,\, {\rm in}\,\, \eqref{ct}: \hskip 1cm O(\l^n) \to (n+1)\text{-loops}\, .
\ee
In \eqref{Theta eigenvalue eq}, we presented the eigenvalue-like equation satisfied by $C_{\Theta}$.
It is natural to ask what is the analogous implication for $C_T$.
 Due to its vanishing anomalous dimension, EMT satisfies the following CS equation
	\be
	\left[ \m \frac{\partial}{\partial \m} + \b_\l \frac{\partial}{\partial \l} \right]\braket{T_{\mu \nu} T_{\rho \sigma}}=0\, .
	\ee
From the decomposition \eqref{Tmn Trs decomposition}, we observe that all the dependence of the correlator on the renormalization scale $\mu$ and the coupling constant $\lambda$ is encoded in the renormalized charges $C_\Theta$ and $C_T$.
The projectors $\pi_{\mu \nu}$ and $\Pi^T_{\mu \nu \rho \sigma}$ merely fix the Lorentz structure of the correlator and are independent of the dynamics.
Hitting the correlator with the Callan-Symanzik operator we obtain
	\be \bal
		\left[\m \frac{\partial}{\partial \m} + \b_{\l} \frac{\partial}{ \partial \l} \right]
		\left[  C_T\Pi^T_{\m \n \r \s } + C_\Theta \pi_{\m \n}\pi_{\r \s} \right] =0
		\Rightarrow 
			\begin{cases}
				\left(\m \frac{\partial}{\partial \m}
				 + \b_{\l} \frac{\partial}{ \partial \l} \right)C_T=0 \\
				\left(\m \frac{\partial}{\partial \m} 
				+ \b_{\l} \frac{\partial}{ \partial \l} \right)C_\Theta =0
			\end{cases}
	\eal\ee
and it is reasonable to ask whether $C_T$ satisfies an analogous to \eqref{Theta eigenvalue eq} eigenvalue-like system of equations
	\bea
	 \label{ct eigenvalue eq}
		\m \frac{\partial}{\partial \m}C_T &=& - e_T C_T\nonumber\\
		\beta_\l \frac{\partial}{\partial \l}C_T &=& e_T C_T
	\eea
with a different than $e_\Theta$, in general, eigenvalue $e_T$. 
As in the spin-0 sector, the above system of differential equations can be highly non-trivial for general $e_T$.
As stated in \cite{spin0, Ising}, the proper way to express this is that solutions to 
\eqref{Theta eigenvalue eq} and \eqref{ct eigenvalue eq} define RG flows on the phase diagram that approach the fixed point(s).
Generic trajectories connecting the Gaussian and WF fixed points may deviate from the perturbative flow.

\subsection{$C_T$ at 3-loops }

Like $C_\Theta$, away from fixed points, $C_T$ is not constant along the RG flow.
In $\l\phi^4$ theory it receives corrections starting at the 3-loop level, with the result \cite{spectralTT, Tassos}:
\be\label{FriedanCT}
C_T = \left[1 - \frac{5}{36}\frac{\l^2}{(4\pi)^4} +O(\ve)  \right] + O(\l^3) \, .
\ee
The normalization here is that the charge in the free theory is equal to 1.
We are going to repeat the derivation of this result and there are three reasons for this.
The first is of a technical nature. Among the integrals that enter the calculation, one turns out to 
be irreducible, in the sense that it can not be expressed in terms of more elementary integrals.
In \cite{spectralTT} this integral was computed using a representation in terms of Gegenbauer polynomials.
Here we will compute it using the Mellin-Barnes representation. This is not only a more modern, easier to generalize method 
but it also serves as an introduction to the master integrals in next to leading order computations that follow.
The second reason is of a qualitative nature. In \cite{spectralTT} the 3-loop calculation was explicitly done for 
the conformal value $\xi = \frac{d-2}{4(d-1)}$ and it was simply stated that it was repeated for the 
value $\xi=0$, leading to the same result. From this the authors concluded that the $O(\l^2)$ result is $\xi$-independent.
Here we will perform the calculation for generic value of $\xi$, proving the statement.
To this end, we divide $C_T$ into two parts, the $\xi$-independent and the $\xi$-dependent parts:
	\be
	C_T = C_T^{\rm (i)} + C_T^{(\xi)} \, ,
	\ee
with the corresponding vertices vefined in \eqref{calVi} and \eqref{calVxi}. 
We will show, order by order, that the $\xi$-dependent parts cancel.
The third reason concerns the derivation of the $O(\ve)$ contributions to \eqref{FriedanCT}.
These terms are of particular interest, since they ensure the smooth fixed point limit of the correlation function 
at the WF fixed point, at the next order.
Before delving into the details of the 3-loop calculation, we briefly present some preliminaries.

The renormalization of $ C_T$  follows the standard procedure used for generic composite operators. 
Specifically, we introduce a bare charge of the spin-two sector, $C_T^{(0)}$, along with a renormalization factor $Z_T$,
which absorbs the divergences arising from loop diagram evaluations:
	\be
	C_T^{(0)} = Z_T C_T\, .
	\ee
To complete the renormalization, an appropriate renormalization condition must be imposed.
This condition should be specified at the same energy scale used for the renormalization of the coupling constant.
	\be \label{renorm cond}
		C_T(p)= 1 + \sum_{n}\l^n h_{n,0} \,\,\,\, \text{at } \, p^2 =-\tilde{\m}^2 \, ,
	\ee
where the renormalization scale $\tilde\m$ is a modified scale $\m$ defined for the renormalization of $\l$ in the Appendix \ref{quantumaction}.
The $h_{n,0}$ are coefficients that arise from the finite part of the loop diagrams.
Their values at the fixed point can be determined via independent CFT methods, see for example \cite{Tassos, HenrikssonCT1, HenrikssonCT, GopakumarCT}. The QFT result for $C_T$ includes however an RG-flow sector that is proportional to the $\b$-function
and consists of powers of the coupling constant accompanied by logarithmic corrections, which are not possible to see in the CFT.

In the computations that follow we will encounter tensorial loop integrals. These integrals can be systematically reduced to scalar loop integrals, where the tensorial structure is expressed in terms of the external momenta and the metric tensor. Several open-source \texttt{Mathematica} packages are available to facilitate this reduction. Up to three-loop order, the reduction can be performed manually by employing standard techniques such as integration by parts (IBP) identities \cite{ChetyrkinI}, and the resulting scalar integrals are generally tractable. In the case of the three-loop calculation, the reduction was independently cross-verified using both \texttt{FeynCalc} \cite{FeynCalc1,FeynCalc2,FeynCalc3} and \texttt{FIRE6} \cite{FIRE} .

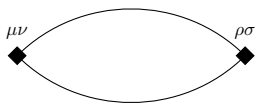
\begin{figure}[H]
     \centering
     \begin{tikzpicture}
         \begin{feynman}
             \vertex (x) at (0, 0);       
             \vertex (y) at (3, 0);       
             \diagram {(x) -- [quarter left] (y) -- [quarter left] (x)};
         \end{feynman}
         \node[draw, fill=black, scale=0.7, rotate=45] at (x) {};
         \node[draw, fill=black,,scale=0.7, rotate=45 ] at (y) {}; 
         \node[scale=0.7]at (0,0.3){$\m \n$};
           \node[scale=0.7]at (3,0.3){$\r \s$};
     \end{tikzpicture}
     \caption{Topology of 1-loop diagrams, contributing at $O(1)$ to $\braket{{T^{(0)}}_{\m \n}{T^{(0)}}_{\r \s}}$.  }
     \label{fig: one loop topology}
\end{figure}

We begin with the $O(1)$ contribution. Fig. \ref{fig: one loop topology} presents
the topology of the 1-loop diagram contributing to
$\braket{{T}_{\mu\nu} {T}_{\rho\sigma}}$.
This topology is proportional to a simple $B$-type integral, presented in Appendix \ref{IntegralsB}. 
The black rhombuses denote insertions of ${T}_{\m \n}$. Then the loop integral
that determines the $O(1)$ 2-point function of the EMT is
	\be
		\braket{{T}^{(0)}_{\m \n}(p){T}^{(0)}_{\r \s}(-p)}_{O(1)}= 
		 2  \int \frac{\rmd^dk}{(2\pi)^d}V_{\m \n}(p,k) V_{\r \s}(p,k)
		 \frac{i^2}{k^2 (p+k)^2}\, .
	\ee
The evaluation of the $\xi$-independent and $\xi$-dependent parts at this order is straightforward, 
with the former being proportional to the 1-loop $B_0$ integral and the latter reducing to a scaleless integral.
Therefore,
	\begin{align}
		C_T^{\rm( i)}|_{O(1)}  &=- \frac{p^{-d+4}}{4(d^2-1)} B_0(p)\\
		C_T^{(\xi)}|_{O(1)} &=0\, .
	\end{align}
Substituting the value of the $B_0$ integral (see \eqref{B0}), we obtain the expression for $C_T^{(0)}(p)$,
	\be
		\left.C_T^{(\rm i)} \right|_{O(1)}=
		- \frac{1}{4(d^2-1)}  b_{1,1}  
		\,  .
	\ee
It is important to note that although  $\left.C_T^{(\rm i)}\right|_{O(1)}$ is divergent for
$d\to 4$, this divergence is momentum-independent. 
As a result, the renormalization of this contribution does not introduce any dependence
on the renormalization scale, thus an anomalous dimension for the operator.
Applying the  renormalization condition  \eqref{renorm cond} and solving  for $Z_{T}$ we obtain:
	\be \label{Z_T}
		Z_T^{(0)}= \left.C_T^{(\rm i)}\right|_{O(1)}  = - \frac{1}{4(d^2-1)}b_{1,1} 
		= \frac{-i}{30 (4\pi)^2} \left[\frac{1}{\ve}
		- \frac{1}{2} \ln \left( \frac{-e^\gamma}{4\pi} \right) + \frac{23}{15}  + O(\ve) \right]\, .
	\ee
Using this result we can find the $O(1)$ renormalized expression:
	\be\label{TmnTrs O1}
		\braket{{T}_{\m \n}(p) {T}_{\r \s}(-p) } =p^d \Pi_{\m \n \r \s}\, .
	\ee
Indeed, the $O(1)$ renormalized $\braket{T_{\m \n}(p) T_{\r \s}(-p)}$ is $\m$-independent. 
As expected, this implies a scale-invariant 2-point function, since the $O(1)$ correlation functions
are equivalent to those of the free theory, which is a CFT.

We should emphasize an important distinction concerning the divergences that arise in the evaluation of correlation functions involving composite operators. These divergences fall into two categories.
 The first category consists of momentum-independent divergences, such as those encountered in the above example of the 
 $O(1)$ EMT correlation function.
 The renormalization of such divergences does not introduce any dependence on the renormalization scale $\mu$ in the correlation functions.
In contrast, the second category includes momentum-dependent divergences, which may generate terms proportional to $\ln \tilde{\mu}^2$
in the renormalized expressions. 
These logarithmic terms reflect the breaking of scale invariance and consequently, of the full conformal symmetry.
 However, not every momentum-dependent divergence necessarily leads to a logarithmic term in the renormalized result. 
 As we will demonstrate in the following sections, there are instances where non-trivial cancellations eliminate the logarithmic contributions,
 resulting in no explicit dependence on the energy scale $\tilde{\mu}$.
Therefore, each case must be carefully examined, as the emergence of logarithmic or finite $\tilde{\mu}$-independent contributions 
from momentum-dependent divergences depends sensitively on the structure of the loop diagrams involved.
In the following, we will show that the renormalization factor given in\eqref{Z_T} is sufficient to renormalize the spin-2 charge up to $O(\lambda^3)$, as non-trivial cancellations occur, rendering the final result finite.

\begin{figure}[h]
		\centering
		\begin{tikzpicture}[scale=1]
         			\begin{feynman}
             				\vertex (x) at (0, 0); 
             				\vertex(z) at(1.5,0);
             				\vertex (y) at (3, 0);       
			            		 \diagram {(x) -- [half left] (z)	
					 	-- [half left] (y)--[half left] (z)--[half left] (x)};
        				 \end{feynman}
       					\node[draw, fill=black, scale=0.7, rotate=45] at (x) {};
         				\node[draw, fill=black, scale=0.7, rotate=45] at (y) {};
          				\node[scale=0.7]at (-0.2,0.3){$\m \n$};
           				\node[scale=0.7]at (3.2,0.3){$\r \s$};
    			 \end{tikzpicture}
			 \caption{\small The Caterpillar topology}
			 \label{caterpillar}
	\end{figure}
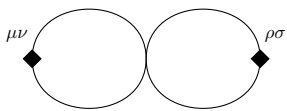

Proceeding in the same manner for the $O(\lambda)$ contribution, which is represented by the Caterpillar topology shown in Figure~\ref{caterpillar}, we find:
	\begin{align}
	C_T^{\rm( i)}|_{O(\l)}  &=\frac{8-4d}{16(1-d)}p^4[B_0(p)]^2 +4B_{\m \n}(p)B^{\m \n}(p) 
	-2 p^2\eta^{\m \n}B_{\m }(p)B_{\n}(p)  =0 \\
	C_T^{(\xi)}|_{O(\l)} &=0
	\end{align}
For the vanishing of the $\xi$-independent part  we used the reduction relation \eqref{Bm reduced} and \eqref{Bmn reduced},
while the $\xi$-dependent part reduces to a scaleless integral.
We can actually extend the vanishing of the Caterpillar topologies to all orders:
	\be
	\begin{gathered}
	\begin{tikzpicture}
         \begin{feynman}
        \vertex (x) at (0, 0);  
        \vertex (z) at (1, 0);
        \vertex (r) at (2, 0);
        \vertex (y) at (3, 0);       
        \diagram {(x) -- [half left] (z) -- [half left] (r)-- [half left] (y) -- [half left] (r)-- [half left] (z) -- [half left] (x)};
        \end{feynman}
      \node[draw, fill=black, scale=0.7, rotate=45] at (x) {};
         \node[draw, fill=black, scale=0.7, rotate=45] at (y) {};
         \node[scale=0.7]at (-0.2,0.3){$\m \n$};
          \node[scale=0.7]at (3.2,0.3){$\r \s$};
         \filldraw [gray] (1.5,0) circle (14pt);
\end{tikzpicture}
	\end{gathered} \rightarrow C_T = 0
	\ee
A short proof of this statement can be given by noticing that
	\begin{align}\label{caterpillar vanishing}
    	\begin{gathered}
	\begin{tikzpicture}[scale = 0.7]
                 \begin{feynman}
                    \vertex (x) at (0, 0);  
                    \vertex (z) at (1, 0);
                    \vertex (r) at (2, 0);
                    \vertex (y) at (3, 0);       
                        \diagram {(x) -- [half left] (z) -- [half left] (r)
                        -- [half left] (y) -- [half left] (r)-- [half left] (z) -- [half left] (x)};
                \end{feynman}
                    \node[draw, fill=black, scale=0.7, rotate=45] at (x) {};
                    \node[draw, fill=black, scale=0.7, rotate=45] at (y) {};
                    	\node[scale=0.7]at (-0.2,0.3){$\m \n$};
           		\node[scale=0.7]at (3.2,0.3){$\r \s$};
                    \filldraw [gray] (1.5,0) circle (14pt);
            \end{tikzpicture}
   	 \end{gathered} &\sim (-i\l_0)^n 
	 \int \frac{\rmd^d k \rmd^d l}{(2\pi)^{2d}}
	 \frac{{\cal V}_T^{(\rm i)}+ \xi {\cal V}_T^{(\xi)}}{k^2 (p+k)^2l^2 (p+l)^2}
	 \int \frac{\rmd^d r_{1}\cdots r_{n-2}}{(2\pi)^{(n-2)d}}f(r_{1},\cdots ,r_{n-2};p^2) \nonumber\\
	 &=C_{T}|_{O(\l_0)}
	 \times (-i\l_0)^{n-1} \int \frac{\rmd^d r_{1}\cdots r_{n-2}}{(2\pi)^{(n-2)d}}
	 f(r_{1},\cdots ,r_{n-2};p^2) =0\, .
	\end{align}
Let us now proceed with the first non-trivial contribution to the EMT's 2-point function that comes at $O(\l^2)$.
This is the first order where the diagrams break free from the vanishing Caterpillar topology, at least in the massless theory.
At this order, three different topologies contribute to the correlation function.
In Figure \ref{fig: three loop topology}, we present the topologies of those diagrams:
	\begin{figure}[H]
     		\centering
    			 \begin{tikzpicture}[scale =0.8]
         			\begin{feynman}
             				\vertex (x) at (0, 0);       
             				\vertex (y) at (3, 0);       
             					\diagram {(x) -- [quarter left] (y) -- [quarter left] (x)};
         			\end{feynman}
           				\draw (2, 0.6) arc [start angle=0, end angle=360,radius=0.5];
      					\node[draw, fill=black, scale=0.7, rotate=45] at (x) {};
         				\node[draw, fill=black, scale=0.7, rotate=45] at (y) {};
				\node[scale=0.7]at (-0.2,0.3){$\m \n$};
           			\node[scale=0.7]at (3.2,0.3){$\r \s$};
     			\end{tikzpicture} \hspace{0.3cm} , \hspace{0.3cm} 
         		\begin{tikzpicture}[scale =0.8]
         			\begin{feynman}
       					\vertex (x) at (0, 0);  
        					\vertex (z) at (1, 0);
        					\vertex (r) at (2, 0);
        					\vertex (y) at (3, 0);       
        						\diagram {(x) -- [half left] (z) -- [half left] (r)-- 
						[half left] (y) -- [half left] (r)-- 
						[half left] (z) -- [half left] (x)};
       		 		\end{feynman}
      					\node[draw, fill=black, scale=0.7, rotate=45] at (x) {};
         				\node[draw, fill=black, scale=0.7, rotate=45] at (y) {};
					\node[scale=0.7]at (-0.2,0.3){$\m \n$};
           				\node[scale=0.7]at (3.2,0.3){$\r \s$};
			\end{tikzpicture} \hspace{0.3cm} , \hspace{0.3cm}
   			\begin{tikzpicture}[scale =0.8]
    				\begin{feynman}
        					\vertex (x) at (0, 0);  
        					\vertex (z) at (1.5, 0.5);
        					\vertex (r) at (1.5, -0.5);
        					\vertex (y) at (3, 0);       
        						\diagram {(x) -- [quarter left] (z) -- 
						[half left] (r)-- [half left] (z) -- [quarter left] (y)-- 
						[quarter left] (r)--[quarter left](x)};
    				\end{feynman} 
      					\node[draw, fill=black, scale=0.7, rotate=45] at (x) {};
         				\node[draw, fill=black, scale=0.7, rotate=45] at (y) {};
					\node[scale=0.7]at (-0.2,0.3){$\m \n$};
           				\node[scale=0.7]at (3.2,0.3){$\r \s$};
			\end{tikzpicture}
     		\caption{Topologies of 3-loop diagrams, contributing at $O(\lambda^2)
     			$ to $\braket{{T}_{\m \n}{T}_{\r \s}}$.  }
     		\label{fig: three loop topology}
	\end{figure}
The topology in the middle in Fig.\ref{fig: three loop topology} is a Caterpillar,
therefore it does not contribute to the 2-point function.
We are left to deal with the other two topologies:
		\be
		\text{"Sun-over-the-hill-I" (SH): }
  		\begin{gathered}
			\begin{tikzpicture}[scale=0.7]
         			\begin{feynman}
             				\vertex (x) at (0, 0);       
             				\vertex (y) at (3, 0);
						\diagram {(x) -- [quarter left] (y) -- [quarter left] (x)};
        				\end{feynman}
           				\draw (2, 0.6) arc [start angle=0, 
						end angle=360, radius=0.5];
      					\node[draw, fill=black, scale=0.7, rotate=45] at (x) {};
         				\node[draw, fill=black, scale=0.7, rotate=45] at (y) {};
					\node[scale=0.7]at (-0.2,0.3){$\m \n$};
           				\node[scale=0.7]at (3.2,0.3){$\r \s$};
     			\end{tikzpicture}
		\end{gathered} = \s_{\rm SH}
	\l_0^2 \int \frac{\rmd^d k}{(2\pi)^d} \frac{V_{\m\n}(-p,k)V_{\r \s} (-p,k)}{k^4 (k-p)^2} S_1(k)
	\ee
	\be
	\text{"Cat-eye" (CE): }
	\begin{gathered}
	\begin{tikzpicture}[scale =0.7]
    				\begin{feynman}
        					\vertex (x) at (0, 0);  
        					\vertex (z) at (1.5, 0.5);
        					\vertex (r) at (1.5, -0.5);
        					\vertex (y) at (3, 0);       
        						\diagram {(x) -- [quarter left] (z) -- 
						[half left] (r)-- [half left] (z) -- [quarter left] (y)-- 
						[quarter left] (r)--[quarter left](x)};
    				\end{feynman} 
      					\node[draw, fill=black, scale=0.7, rotate=45] at (x) {};
         				\node[draw, fill=black, scale=0.7, rotate=45] at (y) {};
					\node[scale=0.7]at (-0.2,0.3){$\m \n$};
           				\node[scale=0.7]at (3.2,0.3){$\r \s$};
			\end{tikzpicture}
			\end{gathered} = \s_{\rm CE}   \l_0^2
			\int \frac{\mathrm{d}^dk \mathrm{d}^d l}{(2\pi)^{2d}} 
			\frac{V_{\m\n}(-p,k)V_{\r \s} (-p,l)}{k^2 (k-p)^2l^2(l-p)^2}
			 B_0\left( k-l\right)
	\ee
where $S_1(k)$ is the 2-loop Sunset integral defined as:	
	\be \label{sun int}
	S_1(k)\equiv \int \frac{\rmd^d l \rmd^d q}{(2\pi)^{2d}} \frac{1}{l^2 q^2 (l+q-k)^2} = b_{1,1}b_{1,2-\frac{d}{2}} (k^2)^{d-3}\, 
	\ee
	and $\s_{\rm SH}$ and $\s_{\rm CE}$ are the corresponding symmetry factors.
	\be \label{3loop symmetry factors}
	\s_{\rm SH} = 2 \, , \, \s_{\rm CE} =3\, .
	\ee
To extract the spin-2 charge, we consider the traces as indicated in \eqref{ct} and use the appropriately 
contracted vertices, see Appendix A. 
The two contributions reduce to the loop integrals
	\begin{align}
	C_{T,{\rm SH}} &=  \s_{\rm SH} \l_0^2 \frac{2}{(d-2)(d-1)} \int \frac{\rmd^dk}{(2\pi)^d} \frac{S_1(k)}{k^4 (k-p)^2}
	 \left( \mathcal{V}_{T}^{(i)}(-p,k,k) + \xi \mathcal{V}_{T}^{(\xi)}(-p,k,k) \right) p^{-d} \\
	 C_{T,{\rm CE}} &= \s_{\rm CE} \l_0^2 \frac{2}{(d-2)(d-1)} \int \frac{\rmd^dk \rmd^d l}{(2\pi)^{2d}} \frac{B_0(k-l)}{k^2 l^2 (k-p)^2 (l-p)^2}
	  \left( \mathcal{V}_{T}^{(i)}(-p,k,l) + \xi \mathcal{V}_{T}^{(\xi)}(-p,k,l) \right)  p^{-d}\, ,
	\end{align}
where as promised, we have divided each contribution into a $\xi$-independent and a $\xi$-dependent part.
Regarding the $\xi$-dependent terms, as we will show in the following, at this order a non-trivial cancellation takes place.
We follow the same steps as those used in the lower-order diagrams, write
	\begin{align}
	C_{T,\text{sun}} &= C_{T,\text{sun}}^{(i)} + \xi C_{T,\text{sun}}^{(\xi)}  \\
	C_{T,\text{cat}} &= C_{T,\text{cat}}^{(i)} + \xi C_{T,\text{cat}}^{(\xi)}\, 
	\end{align}
and examine each term separately.

\subsubsection{$\xi$-independence of $C_T$ to $O(\l^2)$}

We begin with the SH diagram. Expanding the numerator of the integrand and recalling  the renormalization factor $Z_T$ \eqref{Z_T}, we first write
	 \be\label{ctsunxi}
	 \frac{C_{T,{\rm SH}}^{(\xi)}}{Z_T} = -  \s_{\rm SH} \l_0^2 b_{1,2-d/2} \frac{2(d-1)}{(d-2)}\left[ -b_{1,4-d} 
	 +2 b_{1,3-d} \right](p^2)^{d-4} 	
	 \ee
and then exploiting the recursive relation for the $b$-coeffiecient \eqref{recurr b} 
	\be
	b_{1,3-d} = \frac{d-3}{2 d-5}b_{1,4-d} \, ,
	\ee
we obtain the simple form
	\begin{align}\label{ctsunxi reduced}
	 \frac{C_{T,{\rm SH}}^{(\xi)}}{Z_T}
	 =  \s_{\rm SH}\l_0^2 \frac{2(d-1)}{(d-2)}\frac{1}{(2d-5)} b_{1,2-d/2}b_{1,4-d}(p^2)^{d-4} \, .
	\end{align}
We proceed in the same manner with the CE diagrams' $\xi$-dependent part, which is
	\be
	C_{T,{\rm CE}}^{(\xi)} =\s_{\rm CE}\frac{2\l_0^2 p^{-d}}{(d-2)(d+1)}
	\int \frac{\rmd^d k\rmd^d l}{(2\pi)^{2d}} \frac{\mathcal{V}^{(\xi)}_T (-p,k,l)}{k^2 l^2 (k-p)^2 (l-p)^2}B_0(k-l) \, .
	\ee
Expanding the numerator of the integrand, several scaleless integrals emerge, all of which vanish in DR. 
The remaining integrals are primitive $B$-integrals, though some of them involve non-trivial numerators.
For those integrals with numerators different from one, appropriate shifts of the loop momenta must be performed,
followed by the application of the reduction formulae given by \eqref{Bm reduced} and \eqref{Bmn reduced}. 
As a result, we arrive at the following expression:
	\be
	 \frac{C_{T,{\rm CE}}^{(\xi)}}{Z_T} = -\s_{\rm CE}\l_0^2
	\frac{(d-1)}{(d-2)} \frac{4}{3(2d-5)}b_{1,2-\frac{d}{2} }b_{1,4-d}(p^2)^{d -4}\, .
	\ee
Combining the above result with \eqref{ctsunxi reduced}, we obtain:
	\be
	\frac{C_{T,{\rm SH }}^{(\xi)}+C_{T,{\rm CE}}^{(\xi)}}{Z_T} = \left(2\s_{\rm SH}- \s_{\rm CE} \frac{4}{3} \right)b_{1,2-d/2}b_{1,4-d}(p^2)^{d-4}\, .
	\ee
Using the symmetry factors \eqref{3loop symmetry factors} a cancellation between the $\xi$-dependent parts takes place:
		\be
		\frac{C_{T,{\rm SH }}^{(\xi)}+C_{T,{\rm CE}}^{(\xi)}}{Z_T} =0\, .
		\ee
Thus, we have demonstrated that the $O(\l^2)$ contribution to the spin-2 sector charge is independent of the improvement term
in the energy-momentum tensor.
This result is justified to all orders of $\ve$, since the cancellation takes place at the integral level.

\subsubsection{The $O(\l^2)$  spin-2 charge}

The $\xi$-independent part of the SH diagram is given by
	\begin{align}
	C_{T,{\rm SH}}^{(i)}=\s_{\rm SH}\l_0^2 \frac{2 p^{-d}}{(d-2)(d+1)}
	\int \frac{\rmd^d k}{(2\pi)^d} \frac{\mathcal{V}^{(i)}_{T} (-p,k,k)}{k^4 (k-p)^2}S_1(k)\, .
	\end{align}
Using the form  \eqref{calVi} of $\mathcal{V}^{(i)}_{T} (p,k,l)$, for $l=k$, all the terms multiplied by $(k-l)$ vanish automatically.
The remaining loop integrals reduce to simple $B$-integrals. Using the form of $Z_T$ \eqref{Z_T}, and expressing the bare coupling in terms of the renormalized coupling\footnote{The relation between the bare and the renormalized coupling is given by \eqref{bare coupling}.
It is textbook material to show that $Z_\phi =1 + O(\l^2)$, form which implies that $\l_0^2=(\frac{Z_\l}{Z_\phi^2} \l \m^\ve)^2 = (Z_\l \l \m^\ve)^2 +O(\l^4) $}, we obtain:
	\be\label{ctsuni}
	\frac{C_{T,{\rm SH}}^{(i)}}{Z_T}=-  \s_{\rm SH}(Z_\l \l \m^\ve)^2 b_{1,2-\frac{d}{2}}
	\left[\frac{1}{2} b_{1,5-d} -\frac{3}{2} b_{1,4-d} + \frac{7d-16}{8(d-2)}b_{1,3-d} \right]
	\left( p^2 \right)^{d-4}\, .
	\ee
In the $\ve$-expansion this is
	\be\label{ctsunexp}
	\frac{C_{T,{\rm SH}}^{(i)}}{Z_T}=
	-\s_{\rm SH}\frac{\l^2}{(4\pi)^4} \left\{\frac{1}{12 \ve} - \frac{1}{12}\ln \left(\frac{-p^2}{\tilde{\m}^2}\right)
	 + \frac{15}{24}  
	+ \ve
	 \left[\frac{1}{24} \ln^2 \left( \frac{-p^2}{\tilde{\m}^2} \right) -\frac{7}{24}
	  \ln \left(\frac{-p^2}{\tilde{\m}^2} \right)+  \frac{485}{576}-\frac{\pi ^2}{288} \right]
	 \right\}  +O(\ve^2 , \l^3)
	\ee
where the reduced renormalization energy scale is defined in \eqref{tildemu}.

Turning to the CE diagram, its $\xi$-independent part is given by
	\be
	C_{T,{\rm CE}}^{(i)} =\s_{\rm CE} \l_0^2
	\int \frac{\rmd^d k \rmd^d l}{(2\pi)^d} \frac{\mathcal{V}_{T}^{(i)} (-p,k,l) }{k^2 l^2 (k-p)^2 (l-p)^2}B_0(k-l)(p)^{-d} \, .
	\ee
Expanding the numerator, this expression can be expressed as the sum of four terms:
	\begin{align}
	\frac{C_{T,{\rm CE}}^{(i)}}{Z_T}&=-\s_{\rm CE}(Z_\l \l \m^\ve)^2\left[\mathcal{C}^{(1)} +  
	\mathcal{C}^{(2)} +  \mathcal{C}^{(3)}+  \mathcal{C}^{(4)}  \right]\, ,
	\end{align}
with
	\begin{align}
	\label{calC1} \mathcal{C}^{(1)} &=\frac{2  (d-1)}{(d-2)} \left[I\left(-\frac{d}{2} \right) + I\left(1 -\frac{d}{2} \right) p^2 
	 + \frac{(d-2)}{4(d-1)}I\left(2-\frac{d}{2} \right) p^4 \right] p^{-d} \\
\label{calC2}	 \mathcal{C}^{(2)}& = - \frac{8(d-1)}{(d-2)}b_{1,1-\frac{d}{2}}b_{1,3-d}(p^2)^{d-4} \\
\label{calC3}	 \mathcal{C}^{(3)}&= -3 b_{1,2-\frac{d}{2}}b_{1,4-d}(p^2)^{d-4} \\
\label{calC4}	  \mathcal{C}^{(4)}&=  \frac{7d-16}{4(d-2)}b_{1,2-\frac{d}{2}}b_{1,3-d}(p^2)^{d-4}\, .
	\end{align}
The $I$-integrals in \eqref{calC1} are defined as
\begin{align}\label{I-int}
	 I(\a)\equiv \int \frac{\rmd^d k \rmd^d l}{(2\pi)^{2d}} \frac{1}{k^2 l^2 (p+k)^2 (p+l)^2 \left[(k-l)^2 \right]^{\a}} =
	 \int  \frac{\rmd^d k}{(2\pi)^d} \frac{J_{1,1,\a}(-p,k)}{k^2 (k+p)^2} \, ,
	\end{align}
with $\a = -d/2, 1-\tfrac{d}{2},2-\tfrac{d}{2}$ and with the $J$-integral defined as
	\be
	J_{\n_1,\n_2, \n_3}(p_1,p_2) = \int \frac{\rmd^d k}{(2\pi)^d} \frac{1}{k^{2\n_1}(k-p_1)^{2\n_2} (k-p_2)^{2\n_3}}\, .
	\ee
The $I(\a)$ cannot be reduced to more elementary type-$B$ integrals, since $J$ is a master integral.
Out of the three $I$-integrals in \eqref{calC1}, it is necessary to compute explicitly only one of them,
since in \cite{ChetyrkinI} it was proven that they satisfy the following recursion relation:
	\be \label{I-recur}
	I(\a)=- \frac{\a +2 - \frac{d}{2}}{\a +3 -d} I(\a +1)p^2 - 
	\frac{3d-4\a-10}{\a+3-d}b_{1,1+\a} b_{1,3+\a-\frac{d}{2}} (p^2)^{d-4-\a} \,.
	\ee
Applying it to \eqref{calC1}, the coefficient $\mathcal{C}^{(1)}$ gets simplified to
	\be \label{calC1 reduced}
	\mathcal{C}^{(1)}= \frac{(d-6)(d-4)}{6(d-2)(3d-8)}I\left(2- \frac{d}{2} \right)p^{4-d} 
	+\frac{2  (d-1)}{(d-2)}\left[ \frac{10}{3}b_{1,1-\frac{d}{2}}b_{1,3-d} 
	-\frac{1}{3} \frac{5d-14}{4 - 3d/2}b_{1,2-d/2}b_{1,4-d} \right](p^2)^{d -4}\, .
	\ee
The terms in the square brackets in the above expression, as well as those in \eqref{calC2}, \eqref{calC3}, and \eqref{calC4},
are primitive $B$-integrals. Their $\varepsilon$-expansion is straightforward using the coefficients provided in \eqref{B coef}. 
The evaluation of \eqref{calC1} is detailed in Appendix \ref{MBintegrals}.
In the context of $\ve$-expansion, the CE contribution becomes
		\begin{align}\label{ctcatexp}
	\frac{C_{T,{\rm CE}}^{(i)} }{Z_T^{(0)}} =-\s_{\rm CE}\frac{\l^2}{(4\pi)^4}
	&\left\{
	-\frac{1}{18 \ve }+\frac{1}{18} \ln \left( \frac{-p^2}{\tilde{\m}^2} \right)- \frac{10}{27} \right. \nonumber \\
	&\left.+\ve \left[- \frac{1}{36}\ln^2 \left( -\frac{p^2}{\tilde{\m}^2} \right) 
	+ \frac{4}{27} \ln \left( -\frac{p^2}{\tilde{\m}^2} \right)
	- \frac{779}{2592} +\frac{\pi^2}{432}
	 \right] 
	\right\} +O(\ve^2 \l^2,\l^3)\, .
	\end{align}
Thus, the total $O(\l^2)$ contribution is the sum of \eqref{ctsunexp} and \eqref{ctcatexp}:
	\be
	C_{T}|_{O(\l^2)} =\frac{C_{T,{\rm CE}}^{(i)}  + C_{T,{\rm SH}}^{(i)}}{Z_T^{(0)}}
	= -\frac{\l^2}{(4\pi)^4}\left\{\left[\frac{\s_{\rm SH}}{12} - \frac{\s_{\rm CE}}{18} \right] \left[\frac{1}{\ve} 
	+ \ln\left( \frac{-p^2}{\tilde{\m}^2}\right) \right] + \s_{\rm SH}\frac{15}{24} - \s_{\rm CE}\frac{10}{27} \right\} +O(\ve\l^2,\l^3)
	\ee
Recalling the symmetry factors \eqref{3loop symmetry factors}, the divergent parts and the logarithms get cancelled.
Keeping also the $O(\ve)$ terms for later use, we arrive at the following expression:
	\be \label{ct 3loop}
	C_{T}|_{O(\l^2)}=-\frac{5}{36} \frac{\l^2}{(4\pi)^4}  +\frac{5}{36}\frac{\l^2 \ve}{(4\pi)^4} \ln\left( \frac{-p^2}{\tilde{\m}^2}\right) -\frac{169}{216}\frac{\l^2 \ve}{(4\pi)^4} +O(\ve^2\l^2 ,\l^3)
	\ee
The  $O(\l^3)$ terms arise from the renormalization of the coupling constant and will be discussed in the next section.
For $\ve \to 0$ we recover the known result \eqref{FriedanCT}

Let us comment on the result \eqref{ct 3loop}.
First of all, it is trivial to check that the three-loop  charge trivially satisfies the Callan-Symanzik equation
	\be
	\left[ \frac{\partial}{\partial \ln \tilde{\m} }+ \b_\l \frac{\partial}{\partial \l} \right] C_{T}|_{O(\l^2)}= 0 +O(\ve^2\l^2 ,\l^3)
	\ee
Moreover, considering the conformal limit, $(\frac{\l}{(4\pi)^2} = \frac{\ve}{3}+ \cdots)$, of \eqref{ct 3loop} the
 resulting WF central-charge is in complete agreement with the corresponding result  presented in the literature
 \cite{Tassos, HenrikssonCT1, HenrikssonCT, GopakumarCT}
	\be
	C_T^{*} =1 - \frac{5}{324}\ve^2 +O(\ve^3)\, .
	\ee
	
\subsection{$C_T$ at 4-loops}

One of the main goals of this work is to extend the $O(\l^2)$ result at least by one more loop order. 
To motivate this from the purely computational perspective, observe that 
this form satisfies trivially the Callan-Symanzik equation
since the running of $C_T$ is only indirect, coming from the running of $\l$. In order to have a direct running one would need 
logarithmic corrections of the form $\ln \frac{-p^2}{\tilde{\m}^2}$ that arise for the first time at the 4-loop level. 

All the occurring topologies\footnote{The topologies containing tadpoles are omitted, as they vanish identically in DR.} 
are presented in Fig.\ref{fig: four loop topology}.
	\begin{figure}[H]
    \centering
    \begin{tikzpicture}[scale=0.7]
        \begin{feynman}
            \vertex (x) at (0,0);
            \vertex (y) at (3,0);
            \vertex (z) at (1,1);
            \vertex (r) at (2,1);
            \vertex (q) at (1.5,1.5);
            \diagram{(x)--[quarter right](y)};
            \diagram{(x)--[quarter left](z)--[quarter left](q)--[quarter left](z)--(r)--[quarter left](q)--[quarter left](r)--[quarter left](y)};
        \end{feynman}
          \node[draw, fill=black, scale=0.7, rotate=45] at (x) {};
         \node[draw, fill=black, scale=0.7, rotate=45] at (y) {};
    \end{tikzpicture}\hspace{0.1cm} , \hspace{0.1cm}
    \begin{tikzpicture}[scale=0.7]
        \begin{feynman}
            \vertex (x) at (0,0);
            \vertex (z) at (1.5,0);
            \vertex (y) at (3,0);
            \diagram{(x)--[half left](z)--[half left](y)--[half left](z)--[half left](x)};
        \end{feynman}
        \draw (1, 0.6) arc [start angle=0, end angle=360, radius=0.3];
          \node[draw, fill=black, scale=0.7, rotate=45] at (x) {};
         \node[draw, fill=black, scale=0.7, rotate=45] at (y) {};
    \end{tikzpicture} \hspace{0.1cm} , \hspace{0.1cm} 
    \begin{tikzpicture}
    \begin{feynman}[scale=0.7]
    	\vertex (x) at (0,0);
	\vertex (y) at (3,0);
	\diagram{(x) --[quarter left](y) -- [quarter left](x)};
    \end{feynman}
    \node[draw, fill=black, scale=0.7, rotate=45] at (x) {};
         \node[draw, fill=black, scale=0.7, rotate=45] at (y) {};
         \draw[color=black](1,0.7) circle (0.25);
         \draw[color=black](1,0.9) circle (0.15);
    \end{tikzpicture}\hspace{0.1cm} , \hspace{0.1cm}
         \begin{tikzpicture}[scale=0.7]
         \begin{feynman}
        \vertex (x) at (0, 0);  
        \vertex (z) at (1, 0);
        \vertex (r) at (2, 0);
        \vertex (q) at (3,0);
        \vertex (y) at (4, 0);       
        \diagram {(x) -- [half left] (z) -- [half left] (r)-- [half left] (q) -- [half left] (y)-- [half left] (q)  -- [half left] (r)-- [half left] (z) -- [half left] (x)};
        \end{feynman}
      \node[draw, fill=black, scale=0.7, rotate=45] at (x) {};
         \node[draw, fill=black, scale=0.7, rotate=45] at (y) {};
\end{tikzpicture} \hspace{0.1cm} ,\\
\begin{tikzpicture}[scale=0.7]
    \begin{feynman}
        \vertex (x) at (0, 0);  
        \vertex (z) at (1.5, 0.5);
        \vertex (q) at (1.5 ,0);
        \vertex (r) at (1.5, -0.5);
        \vertex (y) at (3, 0);       
        \diagram {(x) -- [quarter left] (z) -- [half left] (q)-- [half left] (r)--[half left] (q)--[half left] (z) -- [quarter left] (y)-- [quarter left] (r)--[quarter left](x)};
    \end{feynman} 
      \node[draw, fill=black, scale=0.7, rotate=45] at (x) {};
         \node[draw, fill=black, scale=0.7, rotate=45] at (y) {};
\end{tikzpicture}  \hspace{0.1cm} , \hspace{0.1cm}
  \begin{tikzpicture}[scale=0.7]
    \begin{feynman}
        \vertex (x) at (0, 0);  
        \vertex (z) at (1.5, 0.5);
        \vertex (r) at (1.5, -0.5);
        \vertex (y) at (2.5, 0);      
        \vertex (k) at(3.5,0 );
        \diagram {(x) -- [quarter left] (z) -- [quarter left] (r)-- [quarter left] (z) --[quarter left] (y)--[quarter left] (r)--[quarter left](x)};
        \diagram{(y)--[half left](k)-- [half left](y)};
    \end{feynman} 
      \node[draw, fill=black, scale=0.7, rotate=45] at (x) {};
         \node[draw, fill=black, scale=0.7, rotate=45] at (k) {};
\end{tikzpicture}\hspace{0.1cm} , \hspace{0.1cm}
\begin{tikzpicture}[scale =0.7]
    \begin{feynman}
        \vertex (x) at (0,0);
        \vertex (y) at (3,0);
        \vertex (z1) at (1,0.5);
        \vertex (z2) at (2,0.5);
        \vertex (r) at(1.5,-0.5);
        \diagram{(x)--[quarter left](z1)--[quarter left](z2)--[quarter left](z1)--(r)--(z2)--[quarter left](y)--[quarter left](r)--[quarter  left](x)};
    \end{feynman}
    \node[draw, fill=black, scale=0.7, rotate=45] at (x) {};
         \node[draw, fill=black, scale=0.7, rotate=45] at (y) {};
\end{tikzpicture}
    \caption{Topologies of 4-loop diagrams, contributing at $O(\l_0^3)$ to $\braket{T_{\m \n} T_{\r \s}}$}
    \label{fig: four loop topology}
\end{figure}
Now, topologies of the form
	\be
	\begin{gathered}
		\begin{tikzpicture}[scale = 0.7]
		\begin{feynman}
		\vertex (x) at (0,0);
		\vertex (y) at (3,0);
		\diagram{(x)--[quarter left](y)--[quarter left](x)};
		\end{feynman}
		 	\node[draw, fill=black, scale=0.7, rotate=45] at (x) {};
         		\node[draw, fill=black, scale=0.7, rotate=45] at (y) {};
         		\node[scale=0.7]at (-0.2,0.3){$\m \n$};
          		\node[scale=0.7]at (3.2,0.3){$\r \s$};
		  \draw[color=black](1.5,1) circle (0.4);
		  \draw[color=black,fill=gray](1.5,1.4) circle (0.2);
		\end{tikzpicture}
	\end{gathered} \sim 
	\int \frac{\rmd^d k}{(2\pi)^d} \frac{V_{\m \n}(p,k)V_{\r \s}(p,k)}{k^4 (p+k)^2} 
	\int \frac{\rmd^dl \rmd^d r_{1, \cdots n}}{(2\pi)^{(n+1)d}} \frac{f(r_1, \cdots,r_n,l)}{l^4}=0
	\ee
in general vanish to all orders, as they reduce to scaleless integrals.	
Throwing away also the Caterpillars, the remaining non-vanishing topologies at order $O(\lambda^3)$ are the following:
    \be
  \text{"Sun-over-the-hill-II" (SI):}  \begin{gathered}
    \begin{tikzpicture}[scale=0.7]
        \begin{feynman}
            \vertex (x) at (0,0);
            \vertex (z) at (1.5,0);
            \vertex (y) at (3,0);
            \diagram{(x)--[half left](z)--[half left](y)--[half left](z)--[half left](x)};
        \end{feynman}
        \draw (1, 0.6) arc [start angle=0, end angle=360, radius=0.3];
          \node[draw, fill=black, scale=0.7, rotate=45] at (x) {};
         \node[draw, fill=black, scale=0.7, rotate=45] at (y) {};
         \node[scale=0.7]at (-0.2,0.3){$\m \n$};
          \node[scale=0.7]at (3.2,0.3){$\r \s$};
    \end{tikzpicture}
    \end{gathered}= i\l_0^3 \s_{\rm SI} \int \frac{\rmd^d k \rmd^d l \rmd^d q \rmd^d r}{(2\pi)^{4d}}
     \frac{V_{\m \n}(-p,k)V_{\r \s}(-p,l) p^{-d}}{k^4 l^2 q^2 r^2(k-p)^2  (l-p)^2 (r+q-k)^2 } 
     \ee
     \be
	\text{"Bobcat-eye"(BC):}\begin{gathered}
	 \begin{tikzpicture}[scale=0.7]
    \begin{feynman}
        \vertex (x) at (0, 0);  
        \vertex (z) at (1, 0.5);
        \vertex (r) at (1, -0.5);
        \vertex (y) at (2, 0);      
        \vertex (k) at(3,0 );
        \diagram {(x) -- [quarter left] (z) -- [quarter left] (r)-- [quarter left] (z) -- [quarter left](y)--[quarter left] (r)--[quarter left](x)};
        \diagram{(y)--[half left](k)-- [half left](y)};
    \end{feynman} 
      \node[draw, fill=black, scale=0.7, rotate=45] at (x) {};
         \node[draw, fill=black, scale=0.7, rotate=45] at (k) {};
         \node[scale=0.7]at (-0.2,0.3){$\m \n$};
          \node[scale=0.7]at (3.2,0.3){$\r \s$};
\end{tikzpicture}
	\end{gathered}=i\l^3 \s_{\rm BC} \int \frac{\rmd^d k \rmd^d l \rmd^d q \rmd^d r}{(2\pi)^{4d}}
	\frac{V_{\m \n}(-p,k)V_{\r\s}(-p,l)  p^{-d}}{k^2 l^2 q^2 r^2 (k-p)^2 (q-p)^2  (l-p)^2 (r-k+q)^2} 
	\ee
     \be
    \text{"Hourglass" (HR):}  \begin{gathered}
		\begin{tikzpicture}[scale=0.7]
   			 \begin{feynman}
        \vertex (x) at (0, 0);  
        \vertex (z) at (1.5, 0.5);
        \vertex (q) at (1.5 ,0);
        \vertex (r) at (1.5, -0.5);
        \vertex (y) at (3, 0);       
        \diagram {(x) -- [quarter left] (z) -- [half left] (q)
        -- [half left] (r)--[half left] (q)--[half left] (z) -- [quarter left] (y)-- 
        [quarter left] (r)--[quarter left](x)};
    \end{feynman} 
      \node[draw, fill=black, scale=0.7, rotate=45] at (x) {};
         \node[draw, fill=black, scale=0.7, rotate=45] at (y) {};
         \node[scale=0.7]at (-0.2,0.3){$\m \n$};
          \node[scale=0.7]at (3.2,0.3){$\r \s$};
	\end{tikzpicture} 
	\end{gathered} =i \l_0^3 \s_{\rm HR}  \int\frac{\mathrm{d}^dk \rmd^d l \rmd^d q \rmd^d r}{(2\pi)^{4d}}
	 \frac{V_{\m \n}(-p,k)V_{\r\s}(-p,l) p^{-d}}{k^2 l^2 q^2 r^2 (k-p)^2 (l-p)^2 (r-k+l)^2 (q-k+l)^2}  
	\ee
	\be
	\text{"Tent" (TE):}	\begin{gathered}
		\begin{tikzpicture}[scale=0.7]
        \begin{feynman}
            \vertex (x) at (0,0);
            \vertex (y) at (3,0);
            \vertex (z) at (1,1);
            \vertex (r) at (2,1);
            \vertex (q) at (1.5,1.5);
            \diagram{(x)--[quarter right](y)};
            \diagram{(x)--[quarter left](z)--[quarter left](q)--[quarter left](z)--(r)--[quarter left](q)--[quarter left](r)--[quarter left](y)};
        \end{feynman}
          \node[draw, fill=black, scale=0.7, rotate=45] at (x) {};
         \node[draw, fill=black, scale=0.7, rotate=45] at (y) {};
          \node[scale=0.7]at (-0.2,0.3){$\m \n$};
          \node[scale=0.7]at (3.2,0.3){$\r \s$};
    \end{tikzpicture}\end{gathered} = i \l_0^3  \s_{\rm TE}
    \int \frac{\rmd^d k \rmd^dl \rmd^d q \rmd^d r}{(2\pi)^{4d}} 
    \frac{V_{\m \n}(-p,k)V_{\r \s}(-p,k)  p^{-d}}{k^4 l^2 q^2 r^2 (k-p)^2  (r-k+l)^2 (q-k+l)^2} 
    \ee 
	\be
	\text{"Cockroach" (CR):}\begin{gathered}
	\begin{tikzpicture}[scale =0.7]
    \begin{feynman}
        \vertex (x) at (0,0);
        \vertex (y) at (3,0);
        \vertex (z1) at (1,0.5);
        \vertex (z2) at (2,0.5);
        \vertex (r) at(1.5,-0.5);
        \diagram{(x)--[quarter left](z1)--[quarter left](z2)--[quarter left](z1)--(r)--(z2)--[quarter left](y)--[quarter left](r)--[quarter  left](x)};
    \end{feynman}
    \node[draw, fill=black, scale=0.7, rotate=45] at (x) {};
         \node[draw, fill=black, scale=0.7, rotate=45] at (y) {};
         \node[scale=0.7]at (-0.2,0.3){$\m \n$};
          \node[scale=0.7]at (3.2,0.3){$\r \s$};
\end{tikzpicture}
	\end{gathered}=i\l^3 \s_{\rm CR} 
	\int \frac{\rmd^d k\rmd^d l \rmd^d q \rmd^d r }{(2\pi)^{4d}}
	\frac{V_{\m \n}(-p,k)V_{\r\s}(-p,l)  p^{-d}}{k^2 l^2 q^2 r^2(k-p)^2  (l-p)^2  (k-r-q)^2 (r-k+l)^2 } 
	\ee
with the symmetry factors given by
	\begin{align}
	\label{4loop sym factors}	\s_{\rm CR }&=2\s_{\rm TE} = 4\s_{ \rm HR} =\s= 8 \\
	\s_{\rm BC}  &= 3 \\
	\s_{\rm SI}&=2
	\end{align}
The reduction of the above four-loop integrals is non-trivial. 
To facilitate the process, we used the \texttt{FIRE6} package
\cite{FIRE}, which automates the reduction of multi-loop integrals to a linear combination of master integrals and a set of primitive $B$-integrals. 
As we will demonstrate in the following, all relevant topologies reduce to the following set of integrals:
	\begin{align}
\label{G1}	G_1 &=B_0(p) \int \frac{\rmd^d k}{(2\pi)^d} \frac{S_1(k)}{(k-p)^2}
	= (b_{1,1})^2 b_{1,2-\frac{d}{2}} b_{1,3-d}(p^2)^{2d-6}\\
\label{G2}	G_2&= \int\frac{\rmd^d k}{(2\pi)^d} [B_0(k)]^2 B_0(k-p) 
	= (b_{1,1})^3b_{4-d,2-\frac{d}{2}}(p^2)^{2d-6}\\
\label{hour master}	 
G_{\rm HR} &= \int \frac{\rmd^d k \rmd^d l \rmd^d q \rmd^d r}{(2\pi)^{4d}} \frac{1}{k^2 l^2 q^2 r^2 (k-p)^2 (l-p)^2 (r-k+l)^2 (q-k+l)^2}  = (b_{1,1})^2 I(4-d)\\
\label{cr master}	G_{\rm CR}&=\int \frac{\rmd^d k \rmd^d l \rmd^d q \rmd^d r}{(2\pi)^{4d}} 
	 \frac{1}{k^2 l^2 q^2 r^2 (k-p)^2 (l-p)^2 (k-r-q)^2 (r-k+l)^2} \, .
	\end{align}
The integrals \eqref{G1} \eqref{G2}  are the primitive ones and their $\ve$-expansion is straightforward 
while \eqref{hour master} and \eqref{cr master} are the corresponding 4-loop master integrals.

Evaluating high loop master integrals is a rather demanding subject on its own.
The 4-loop master integrals \eqref{hour master} \eqref{cr master} have been evaluated in \cite{Chetyrkin4},
using the "Glue and Cut" trick.
However, the results presented by Baikov and Chetyrkin  are not obviously useful for our analysis, as the authors adopt 
a regularization scheme different from the one used here.
More specifically, the authors use the so-called $G$-scheme, which corresponds to setting $b_{1,1} =\frac{2 i}{(4\pi)^2 \ve}$ in $ d = 4- \ve$\footnote{The overall factor of $i$ arises from the analytic continuation from  Euclidean to  Minkowski spacetime,
 while the $\frac{1}{(4\pi)^2}$ comes from the normalization of the loop integral.}
, while neglecting the finite terms.
 To illustrate the impact of the scheme choice, we present the simple example of the 2-loop Caterpillar  integral,
  for $p^2=1$,  that compares the results obtained under the different schemes:
 	\begin{align}
	&\text{G-scheme: } (4\pi)^4(b_{1,1})^2 = -\frac{4}{\ve^2} + O(\ve^2) 
	 \\
	&\text{Our\, scheme: } (4\pi)^4(b_{1,1})^2 = 
	 -\frac{4}{\ve^2} + \frac{4}{\ve}\left[ \ln \left( \frac{-e^\g}{4\pi} \right) -2 \right]
	 + \frac{1}{6} \left[-12\ln^2 \left( \frac{-e^\g}{4\pi} \right) +48\ln \left( \frac{-e^\g}{4\pi} \right) -72 +\pi^2 \right] +O(\ve)
	\end{align}
where the result in the G-scheme is taken from eq. (13) of \cite{Chetyrkin4}. 
\footnote{The analytic continuation to Mikowski spacetime fixes the overall sign.}

In our analysis, we are interested in the precise evaluation of the finite parts of the loop integrals. 
To this end we evaluate the full structure of the integrals, including all finite contributions,
rather than discarding them as is done in some alternative schemes.
Nevertheless, regarding the master integrals, our loop calculation, presented in Appendix \ref{MBintegrals}, reproduces the leading-order divergences found in \cite{Chetyrkin4}, as this contribution is the same in these two schemes.
Moreover, we have repeated the full computation of the spin-2 charge in the G-scheme and obtained the same expression.
We will not present this calculation here but comment that the agreement between the two 
schemes is not obvious and is a strong indication for the scheme-independence of $C_T$. 
\subsubsection{$\xi$-independence of $C_T$ to $O(\lambda^3)$}

At first sight, one may guess that
	\be
	 \begin{gathered}
    \begin{tikzpicture}[scale=0.7]
        \begin{feynman}
            \vertex (x) at (0,0);
            \vertex (z) at (1.5,0);
            \vertex (y) at (3,0);
            \diagram{(x)--[half left](z)--[half left](y)--[half left](z)--[half left](x)};
        \end{feynman}
        \draw (1, 0.6) arc [start angle=0, end angle=360, radius=0.3];
          \node[draw, fill=black, scale=0.7, rotate=45] at (x) {};
         \node[draw, fill=black, scale=0.7, rotate=45] at (y) {};
         \node[scale=0.7]at (-0.2,0.3){$\m \n$};
          \node[scale=0.7]at (3.2,0.3){$\r \s$};
    \end{tikzpicture}
    \end{gathered} 
	+\begin{gathered}
	 \begin{tikzpicture}[scale=0.7]
    \begin{feynman}
        \vertex (x) at (0, 0);  
        \vertex (z) at (1, 0.5);
        \vertex (r) at (1, -0.5);
        \vertex (y) at (2, 0);      
        \vertex (k) at(3,0 );
        \diagram {(x) -- [quarter left] (z) -- [quarter left] (r)-- [quarter left] (z) --[quarter left] (y)--[quarter left] (r)--[quarter left](x)};
        \diagram{(y)--[half left](k)-- [half left](y)};
    \end{feynman} 
      \node[draw, fill=black, scale=0.7, rotate=45] at (x) {};
         \node[draw, fill=black, scale=0.7, rotate=45] at (k) {};
         \node[scale=0.7]at (-0.2,0.3){$\m \n$};
          \node[scale=0.7]at (3.2,0.3){$\r \s$};
\end{tikzpicture}
	\end{gathered} 
	 \sim i\l_0 \left[\begin{gathered}
			\begin{tikzpicture}[scale=0.7]
         			\begin{feynman}
             				\vertex (x) at (0, 0);       
             				\vertex (y) at (3, 0);
						\diagram {(x) -- [quarter left] (y) -- [quarter left] (x)};
        				\end{feynman}
           				\draw (2, 0.6) arc [start angle=0, 
						end angle=360, radius=0.5];
      					\node[draw, fill=black, scale=0.7, rotate=45] at (x) {};
         				\node[draw, fill=black, scale=0.7, rotate=45] at (y) {};
					\node[scale=0.7]at (-0.2,0.3){$\m \n$};
           				\node[scale=0.7]at (3.2,0.3){$\r \s$};
     			\end{tikzpicture}
		\end{gathered}
		+
		\begin{gathered}
	\begin{tikzpicture}[scale =0.7]
    				\begin{feynman}
        					\vertex (x) at (0, 0);  
        					\vertex (z) at (1.5, 0.5);
        					\vertex (r) at (1.5, -0.5);
        					\vertex (y) at (3, 0);       
        						\diagram {(x) -- [quarter left] (z) -- 
						[half left] (r)-- [half left] (z) -- [quarter left] (y)-- 
						[quarter left] (r)--[quarter left](x)};
    				\end{feynman} 
      					\node[draw, fill=black, scale=0.7, rotate=45] at (x) {};
         				\node[draw, fill=black, scale=0.7, rotate=45] at (y) {};
					\node[scale=0.7]at (-0.2,0.3){$\m \n$};
           				\node[scale=0.7]at (3.2,0.3){$\r \s$};
			\end{tikzpicture}
			\end{gathered}   \right]B_0(p)\, .
	\ee
Although this guess is not totally precise, it leads to the suspicion that the sum of these two diagrams 
may be $\xi$-independent, as we have seen that the $\xi$-dependent contributions from the lower-order diagrams cancel out. 
In the following, we confirm this guess via an explicit loop calculation.

Starting from the SI diagram, its $\xi$-dependent contribution is given by
	\be
	C_{T,{\rm SI}}^{(\xi)} = i \l_0^3 \s_{\rm SI}\frac{2 p^{-d}}{(d-2)(d+1)}
	\int \frac{\rmd^d k \rmd^d l \rmd^d q \rmd^d r}{(2\pi)^{4d} }
	 \frac{\mathcal{V}^{(\xi)}_T(-p,k,l)}{k^4l^2 q^2 r^2 (k-p)^2  (l-p)^2 (r+q-k)^2}\, ,
	\ee
which, after reduction reduces to	
	\be
	C_{T,{\rm SI}}^{(\xi)} =- i \l_0^3 \s_{\rm SI}\frac{2 p^{-d}}{4(d-2)(d+1)} \frac{1}{2(3-d)}G_1\, .
	\ee
The BC topology on the other hand is given by
	\be
	C_{T,{\rm BC}}^{(\xi)} = 
	i\l^3 \s_{\rm BC} \frac{2p^{-d}}{(d-2)(d+1)} \int \frac{\rmd^d k\rmd^d l \rmd^d q \rmd^d r}{(2\pi)^{4d}}
	\frac{\mathcal{V}_T^{(\xi)} (-p,k,l)  }{k^2 l^2 q^2 r^2(k-p)^2 (q-p)^2 (l-p)^2(r+q-k)^2}  \, 
	\ee
and reduces to
	\be
	C_{T,{\rm BC}}^{(\xi)} = 
	i\l^3 \s_{\rm BC} \frac{2p^{-d}}{4(d-2)(d+1)} \frac{1}{3(3-d)}G_1\, .
	\ee
Substituting the symmetry factors, we get the claimed cancellation:
	\be
	C_{T, {\rm SI}}^{(\xi)}+ C_{T, {\rm BC}}^{(\xi)}=  0.
	\ee 
We expect that the $\xi$-dependent parts of the remaining three topologies will cancel each other out.
Indeed, the $\xi$-dependent parts of the remaining diagrams evaluate to
	\begin{align}
	C_{T,{\rm CR}}^{(\xi)}
	&=i\l_0^3 \s_{\rm CR} \frac{2p^{-d}}{4(d-2)(d+1)}  \frac{(15-2 d) d-26}{d (6 d-35)+50} G_{2} \\
	C_{T,{\rm HR}}^{(\xi)}
	&= i \l_0^3  \s_{\rm HR} \frac{2 p^{-d}}{4(d-2)(d+1)}2\frac{d (2 d-13)+22}{d (6 d-35)+50}G_{2}\\
	C_{T,{\rm TE}}^{(\xi)}
	&= i \l_0^3  \s_{\rm TE} \frac{2 p^{-d}}{4(d-2)(d+1)} \frac{(d-6)}{3 d-10} G_{2}\, .
	\end{align}
Taking into account the symmetry factors \eqref{4loop sym factors}, we arrive at the expected cancellation:
	\be
	C_{T,{\rm HR}}^{(\xi)}+C_{T,{\rm TE}}^{(\xi)} +C_{T,{\rm CR}}^{(\xi)} = 0  \,.
	\ee
 \subsubsection{The $O(\l^3)$ spin-2 charge}
 
The 4-loop contribution to the spin-2 charge is determined by the $\xi$-independent 
parts of the loop diagrams and the corrections arising from the lower-order contribution due to the renormalization of the coupling constant.
The contributions arising from the 4-loop diagrams are:
	\begin{align}
	C_{T,{\rm SI}}^{\rm (i)}&=
	i\l_0^3 \s_{\rm SI} \frac{2p^{-d}}{4(d-2)(d+1)}\frac{(d-2)}{8 (d-3) (d-1)}G_1 \\
	C_{T,{\rm BC}}^{\rm (i)}&=
	-i\l_0^3 \s_{\rm BC} \frac{2p^{-d}}{4 (d-2)(d+1)}\frac{(d-2)}{12 (d-3) (d-1)}G_1\\
	C_{T,{\rm CR}}^{\rm(i)} &= i\l_0^3 \s_{\rm CR} \frac{2p^{-d}}{(d-2)(d+1)}
	\left[ \frac{(d-5) (d-4) (d-2) p^4}{8 (d-3) (d-1) (2 d-5)} G_{\rm CR}
	+\frac{(d-2) (2 d-11) (3 d-8) }{8 (d-3)^2 (d-1)}G_{1}  \right.
	\nonumber \\
	&\left.
	+\frac{(d (d (d (2 (1309-78 d) d-16823)+52181)-78554)+46080) }{16 (5-2 d)^2 (d-3) (d-1) (3 d-10)}
	G_2
	\right]\\
	C_{T,{\rm HR}}^{\rm (i)} &= i\l_0^3 \s_{\rm HR} \frac{2p^{-d}}{4(d-2)(d+1)} \left[
	\frac{(d-5) (d-4) (d-2)  p^4}{8 (d-3) (d-1) (2 d-5)}G_{\rm HR} \right. \nonumber \\
	 &\left.+\frac{\left(d \left(d \left(d \left(-40 d^2+622 d-3737\right)+10883\right)-15398\right)+8480\right)}
	{8 (5-2 d)^2 (d-3) (d-1) (3 d-10)} G_{2}\right]\\
	C_{T,{\rm TE}}^{\rm (i)} &= i\l_0^3 \s_{\rm TE} \frac{2p^{-d}}{4(d-2)(d+1)}
	\left[\frac{(d ((178-17 d) d-472)+320) }{48 (d-4) (d-1) (3 d-10)}  G_2\right]\, .
	\end{align}
A welcome feature of the calculation of these integrals is the fact that the BC and SI terms cancel each other.
It is therefore remarkable that out of the seven topologies initially appearing in Figure~\ref{fig: four loop topology}, 
only three contribute to the determination of the spin-2 charge.
We parametrize the spin-2 charge in the following convenient way:
	\be
	C_T|_{O(\l^3)} = \frac{C_{T,O(\l^3)}^{(0)}}{Z_T^{(0)}}= i\s \l_0^3\frac{1}{Z_T^{0}} 
	\left[ \bar{G}_{\rm HR} +
	\bar{G}_{\rm CR} +\bar{G}_{1} +\bar{G}_{2} \right] +\frac{\l^3}{(4\pi)^6} \bar{G}_\l\, ,
	\ee
where the TE integral turns out to be primitive, and
	\begin{align}
\label{barGHR}	\bar{G}_{\rm HR}&=
	\frac{2p^{-d}}{4(d-2)(d+1)} \frac{1}{4} \frac{(d-5) (d-4) (d-2)  p^4}{8 (d-3) (d-1) (2 d-5)}G_{\rm HR} \\
\label{barGCR}	\bar{G}_{\rm CR}
	&=\frac{2p^{-d}}{4(d-2)(d+1)}\frac{(d-5) (d-4) (d-2) p^4}{8 (d-3) (d-1) (2 d-5)}G_{\rm CR}  \\
\label{barG1}	\bar{G}_{1}&=
	\frac{2p^{-d}}{4(d-2)(d+1)}\frac{(d-2) (2 d-11) (3 d-8) }{8 (d-3)^2 (d-1)}G_{1} \\
\label{barG2}	\bar{G}_{2}&=
	-\frac{2p^{-d}}{4(d-2)(d+1)}\frac{(d-6) (d-2) (d (d (281 d-2813)+9340)-10264) }{48 (d-4) (d-3) (d-1) (2 d-5) (3 d-10)}	G_2 \, ,
	\end{align}
with $\bar{G}_\l$ arising from the running of the coupling constant. 
To obtain its explicit form, we expand the charge of the
previous order to $O(\l^3)$, using the counter-term $Z_\lambda$.
In this way, we obtain
	\be
	C_{T}|_{O(\l^2)} =-\frac{\l^2}{(4\pi)^4} \frac{5}{36}
	 +\frac{5}{36}\frac{\l^2 \ve}{(4\pi)^4} \ln\left( \frac{-p^2}{\tilde{\m}^2}\right)
	  -\frac{169}{216}\frac{\l^2 \ve}{(4\pi)^4} 
	  +\frac{\l^3}{(4\pi)^6} 
	  \left[ -\frac{5}{6 \ve} + \frac{5}{6} \ln \left( -\frac{p^2}{\tilde{\m}^2} \right) 
	  - \frac{169}{36}  \right] +O(\ve^2,\l^4)\, ,
	\ee
from which it follows that
	\be \label{Gl}
	\bar{G}_\l = -\frac{5}{6 \ve} + \frac{5}{6} \ln \left( -\frac{p^2}{\tilde{\m}^2} \right) \, 
	  - \frac{169}{36}\, .
	\ee
In the $\ve$-expansion the contributions from the diagrams are
	\begin{align}
\label{barGHRexp}	
	i \s \l_0^3 \frac{\bar{G}_{ \rm HR}}{Z_T^{(0)}} 
	&=  \frac{\s \l^3}{(4\pi)^6} 
	\left\{
	- \frac{1}{36 \ve^2} 
	+ \frac{1}{ \ve} \left[  \frac{1}{24} \ln \left( \frac{-p^2}{\tilde{\m}^2} \right)-\frac{11}{108}\right] 
	- \frac{1}{32} \ln^2 \left( \frac{-p^2}{\tilde{\m}^2} \right)  
	+\frac{11}{72} \ln \left( \frac{-p^2}{\tilde{\m}^2} \right)
	-\frac{19}{81}+\frac{\pi ^2}{576}
	\right\} \\
	\label{barGCRexp}
	i \s \l_0^3 \frac{\bar{G}_{ \rm CR}}{Z_T^{(0)}} &=
	\frac{\s \l^3}{(4\pi)^6 }
	\left\{- \frac{1}{18\ve^2}  
	+ \frac{1}{ \ve} \left[ \frac{1}{12} \ln \left( \frac{-p^2}{\tilde{\m}^2} \right) - \frac{25}{108} 
	\right] 
		-\frac{1}{16}\ln^2 \left( \frac{-p^2}{\tilde{\m}^2} \right)
	 +  \frac{25}{72}  \ln \left( \frac{-p^2}{\tilde{\m}^2} \right)
	- \frac{397 }{648} +\frac{\pi^2}{288}
	\right\} \\
	\label{barG1exp}
	i\s\l_0^3 \frac{\bar{G}_{1}}{Z_T^{(0)}}&=\frac{\s\l^3}{(4\pi)^6}
	 \left[ \frac{1}{6 \ve}-\frac{ 1}{4} \ln\left( \frac{-p^2}{\tilde{\m}^2} \right) + \frac{29}{36}\right] \\
	\label{barG2exp}
	 i\s\l_0^3 \frac{\bar{G}_2}{Z_{T}^{(0)}}&=
	 \frac{\s \l^3}{ (4\pi)^6}	
	 \left\{ \frac{1}{12 \ve ^2} 
	 - \frac{1}{\ve}\left[ \frac{1}{8} \ln \left( \frac{-p^2}{\tilde{\m}^2} \right) -\frac{13}{48} \right]
	 + \frac{3}{32}\ln^2 \left( \frac{-p^2}{\tilde{\m}^2} \right)
	 -\frac{13}{32} \ln \left( \frac{-p^2}{\tilde{\m}^2} \right) + \frac{517}{576}-\frac{\pi ^2}{192} 
	 \right\}\, .
	\end{align}
All the necessary details for the derivation of \eqref{barGHRexp} and \eqref{barGCRexp} are presented in Appendix \ref{MBintegrals}.

Summing all the contributions, the $\frac{1}{\ve^2}$ and $\frac{1}{\ve}$ terms, the overlapping divergences, the 
terms proportional to $\pi^2$ and the double logarithms all get cancelled.
So we obtain:
	\begin{align} \label{ct order 3}
	 C_{T,O(\l^3) } =\frac { C_{T,O(\l^3) }^{(0)} }{Z_T^{(0)}} 
	 &= \frac{\l^3}{(4\pi)^6}  \left[- \frac{5}{12} \ln\left( \frac{-p^2}{\tilde{\m}^2} \right) 
	+\frac{155}{72}
	 \right] +O(\ve \l^3 , \l^4)\, .
	\end{align}
As promised, the renormalization factor $Z_T^{(0)}$, determined by the $\mathcal{O}(1)$ diagram, was sufficient to
absorb the occurring divergences.
In some sense, this outcome is expected, since the presence of a non-trivial divergence at the interacting level would imply the
emergence of a non-vanishing anomalous dimension, $\frac{\rmd \ln Z_T}{\rmd \ln \tilde{\mu}} \neq 0$,
which is forbidden by the conservation of the energy-momentum tensor.
Another interesting feature of this result is that the running of the central charge is governed by logarithmic corrections that do not introduce an anomalous dimension to the correlation function.
This is in contrast with the standard QFT picture where logarithmic corrections typically signal the presence of an anomalous dimension.

The total spin-two charge up to order $\l^3$ is given by:
	\be \label{Ct total}
	C_T= 1 - \frac{5}{36}\frac{\l^2}{(4\pi)^4} 
	- \frac{7}{36}\frac{\l^3}{(4\pi)^6} 
	+ \frac{169}{216} \frac{\l}{(4\pi)^4}
	\left(-\ve \l  + \frac{3\l^2}{(4\pi)^2} \right) 
	-\frac{5}{36} \frac{\l}{(4\pi)^4} \left(-\ve \l + \frac{3\l^2}{(4\pi)^2}\right)
	 \ln \left(- \frac{p^2}{\tilde{\m}^2}\right) \, ,
	 \ee
which can be written in the more convenient form	 
	\be\label{CTfinal}
	C_T=1 - \frac{5}{36}\frac{\l^2}{(4\pi)^4} 
	- \frac{7}{36}\frac{\l^3}{(4\pi)^6}  + \frac{\l\b_\l}{(4\pi)^4} \frac{5}{36} \left[ \frac{169}{30} 
	- \ln \left(- \frac{p^2}{\tilde{\m}^2}\right)  \right] +O(\ve^2 \l^2,\ve \l^3 , \l^4)\, .
	\ee
This is the main result of this work.
	
It is easy to show that the above result satisfies the Callan-Symanzik equation:
	\begin{align}
	\tilde{\m} \frac{\partial}{\partial \tilde{\m}} C_T& =  \frac{5}{18} \frac{\l}{(4\pi)^2} \left[- \ve \l +3 \frac{\l^2}{(4\pi)^2} \right]  +O(\ve^2 \l^2,\ve \l^3 , \l^4) \simeq \frac{5}{18} \frac{\l}{(4\pi)^4} \b_\l C_T +O(\ve^2 \l^2,\ve \l^3 , \l^4) \\
	\b_\l \frac{\partial}{\partial \l} C_T &= -\frac{5}{18} \frac{\l}{(4\pi)^2} \left[- \ve \l +3 \frac{\l^2}{(4\pi)^2} \right]  +O(\ve^2 \l^2,\ve \l^3 , \l^4) \simeq -\frac{5}{18} \frac{\l}{(4\pi)^4} \b_\l C_T +O(\ve^2 \l^2,\ve \l^3 , \l^4)
	\end{align}
from which it follows that  the leading order, perturbative eigenvalue introduced in \eqref{ct eigenvalue eq} is given by
	\be
	e_T = -\frac{5}{18} \frac{\l \b_\l}{(4\pi)^4} + +O(\ve^2 \l^2,\ve \l^3 , \l^4)\, .
	\ee
Let us make some additional remarks on the final result.  
We observe that the QFT spin-2 charge naturally separates into two distinct sectors: the conformal sector and the RG-flow sector.  
The former encodes the information about the central charge at fixed points, while the latter captures the scale dependence of the correlator.  
The presence of the logarithmic correction is a manifestation of the direct running of the charge, whereas the $\beta$-function introduces an indirect dependence through the running of the coupling.

It is now evident that expression \eqref{Ct total} possesses a smooth CFT limit, as the RG-flow sector vanishes identically when the system reaches the WF fixed point.  
Specifically, for $\frac{\lambda}{(4\pi)^2} = \frac{\lambda^*}{(4\pi)^2} = \frac{\varepsilon}{3} + \frac{17}{81}\varepsilon^2$, we recover the $\mathcal{O}(\varepsilon^3)$ central charge of the Wilson–Fisher CFT, as presented in \cite{HenrikssonCT, GopakumarCT}, 
	\be
	C_T^{*}= 1 - \frac{5}{324} \ve^2 - \frac{233}{8748}\ve^3 +O(\ve^4) \, .
	\ee

\section{Applications}
\subsection{Normalization of the  spin-$0$ charge}

The normalization of the spin-zero charge affects physical quantities.
In~\cite{spin0} (see \eqref{theta theta}), the 2-point function of the trace operator was determined up to an overall normalization constant. Therefore, it is important to constrain it as much as possible. 
The value of $C_T$ has been normalized to its value at the Gaussian fixed point,
which was set to 1. Likewise, $C_\Theta$ can be normalized to this value.

To set the stage, we examine the structure of the energy-momentum tensor in more detail. 
We adopt a different perspective from the one presented in \cite{spin0}, as we define the EMT in generic $d$-dimensions rather than $d=4$.
This allows us to define the trace  of EMT, upon applying the equations of motion, as
	\be \label{trace in d}
	\Theta^{(0)}= \frac{d-4}{4} \phi \Box \phi\, .
	\ee
The above classical expression can be promoted to a bare operator and subjected to the standard renormalization procedure:
classical $\to$ bare $\to$ renormalized.
This was not the case in \cite{spin0}, since for $d=4$ the trace vanishes identically, as a result the first step of the standard renormalization chain was not justified.
This prompted us to construct the trace operator as a linear combination of the operators with mass dimensions equal to four, multiplied by an overall $\b_\l^2$, to ensure the vanishing of the correlator at the fixed points.
Turning back to the $d$-dimensional formalism, the expression \eqref{trace in d} renders the trace operator explicitly dependent on the space-time dimensionality, which, in turn, influences the $ \ve$-expansion of the contributions to the bare correlation function.
The two-point function of the trace is given by
	\begin{align} \label{Thetatheta oper}
	\braket{\Theta^{(0)}\Theta^{(0)} } 
	&= \frac{(d-4)^2}{16} \braket{(\phi_0 \Box \phi_0) (\phi_0 \Box \phi_0)}\, .
	\end{align}
The evaluation of the correlation function in the above expression is conducted by applying a perturbative series in the bare coupling constant.
Since we are interested in the renormalized version of  \eqref{Thetatheta oper}, it is convenient to express the bare correlation function in terms of the renormalized $\l$,
which is realized with the use of \eqref{bare coupling}.
As we will show, the renormalization factor $Z_\l$ renders the two-point function of the trace proportional to $\b_\l^2$, 
where the terms linear in $\ve$ arise from the loop calculation.
Apart from the explicit dependence on $d$, another important aspect to consider is the appearance of contact terms. These terms arise from the d'Alabertian of the $\phi \Box \phi$ operator.
In the case of the two-point function, the contact terms are one-point functions multiplied by $\d$-functions.
Fortunately, in the context of the massless $\l  \phi^4$-theory, the one-point functions are 
evaluated by scaleless integrals and as such vanish identically.

There are three equivalent ways to extract the $\braket{\Theta\Theta}$ correlator. 
The first is to evaluate the path integral by introducing a source coupled to the $\phi \Box \phi$ operator. 
The second is to act with the boxes on the diagrams of the four-point function $\braket{\phi \phi \phi \phi}$ 
and then consider appropriate limits.
The third is to exploit the equation of motion and evaluate the loop diagrams occurring in the $\braket{\phi_0^4 \phi_0^4}$ correlator.
We follow the second route, evaluating $\braket{(\phi_0 \Box \phi_0)(\phi_0\Box \phi_0)}$ by taking appropriate limits.

As a first step, we act with the derivatives directly on the Feynman diagrams of the 4-point function $\braket{\phi \phi  \phi \phi}$.
Then, by using the identity $\Box_x \D_{x,y} = \delta(x - y)$, where $\D_{x,y} = \braket{\phi_0(x) \phi_0(y)}$ , we identify the resulting contact terms. Finally, we take the relevant limit and retain only the contributions that do not involve any $ \delta$-functions.
As shown in \cite{spin0}, the  first non-vanishing contribution to $\braket{(\phi_0 \Box \phi_0) (\phi_0 \Box \phi_0)}$ is of order $\l_0^2$ 
and is given by the Watermelon integral
\begin{align}
	\lim_{x _1 \to x}\lim_{y_1 \to y} \Box_{y_1}\Box_{x_1}\left[
	\begin{gathered}
	\begin{tikzpicture}
				\begin{feynman}[scale = 1.5]
				\vertex (x) at (0,0.3){\tiny$x$};
				\vertex (y) at (0,0) {\tiny$y$};
				\vertex (x1) at (1,0.3){\tiny$x_1$};
				\vertex (y1) at (1,0){\tiny$y_1$};
				\diagram {(x)--(x1)};
				\diagram{(y)--(y1)};
 			\end{feynman}
			\draw (0.7,0.45) circle (0.2 cm);
			\end{tikzpicture}
	\end{gathered}+
	\begin{gathered}
	\begin{tikzpicture}
		\begin{feynman}[scale=1.5]
		\vertex (x) at (0,0.3){\tiny$x$};
		\vertex (y) at (0,0) {\tiny$y$};
		\vertex (x1) at (1,0.3){\tiny$x_1$};
		\vertex (y1) at (1,0){\tiny$y_1$};
		\vertex (r) at (0.3,0.15);
		\vertex (z) at (0.7,0.15);
		\diagram{(x)--(r)--(y)};
		\diagram{(r)--[half left](z)--[half left](r)};
		\diagram{(x1)--(z)--(y1)};
		\end{feynman}
	\end{tikzpicture}
	\end{gathered}
	 \right]
	\sim  \l_0^2 (\D_{x,y})^4 \, .
	\end{align}
The above result corresponds to the Watermelon topology on the left in Fig. \ref{watermelon}.
Taking into account the symmetry factors
\footnote{In \cite{spin0}, we considered only one of the three possible ways to close the external legs in order to determine the bare correlator. As a result, an overall multiplicative factor of three is missing.}
and performing a Fourier transformation, the correlator reduces to:
	\begin{align}
	\braket{(\phi_0 \Box \phi_0) (\phi_0 \Box \phi_0)}=3 \cdot \frac{2}{3} \l_0^2  W(p)  \, ,
	\end{align}
where
	\be
	W(p)= \int \frac{\rmd^d k}{(2\pi)^d} \frac{S_1(k)}{(k+p)^2} =  b_{1,1}b_{1,2-\frac{d}{2}}b_{1,3-d}(p^2)^{\frac{3d}{2} -4 }\, .
	\ee
	\begin{figure}[H]
		\begin{center}
			\begin{tikzpicture}
				\begin{feynman}
					\vertex (x) at (0,0);
					\vertex(y) at (2,0);
					\diagram{(x)--[quarter left](y)--[quarter left](x)--[half left](y)--[half left](x)};
				\end{feynman}
					\node[draw, fill=black, scale=0.7] at (x) {};
					\node[draw, fill=black, scale=0.7] at (y) {};
			\end{tikzpicture} 
			\hspace{1cm}
			\begin{tikzpicture}
				\begin{feynman}
				\vertex (x) at (0,0);
				\vertex(y) at (2,0);
				\vertex(z) at(1,0);
				\diagram{(x)--[half left](y)--[quarter right](z)--[quarter right](x)--[quarter right](z)--[quarter right](y)--[half left](x)};
				\end{feynman}
				\node[draw, fill=black, scale=0.7] at (x) {};
				\node[draw, fill=black, scale=0.7] at (y) {};
			\end{tikzpicture}
			
		\caption{The Watermelon (on the left) and the Ninja-turtle (on the right) topologies of $\braket{\phi_0^4 \phi_0^4}$.}
		\label{watermelon}
		\end{center}
	\end{figure}
In an analogous manner, one can show that the $O(\l_0^3)$ topologies reduce to the Ninja-turtle diagram, on the right in Fig.\ref{watermelon}:
	\be
	\lim_{x_1 \to x}\lim_{y_1 \to y}\Box_{x_1}\Box_{y_1}\left[
	\begin{gathered}
	\begin{tikzpicture}
				\begin{feynman}
				\vertex (x) at (0,0.1);
				\vertex (y) at (2,0.1);
				\vertex (z) at (0,-0.1);
				\vertex(z1) at (2,-0.1);
				\vertex (r1) at (0.4,0.1);
				\vertex (r2) at (1.6,0.1);
				\vertex (r) at (1,0.6);
				\diagram{(x)--(r1)--[quarter left](r) -- [quarter left](r1)--(r2)--[quarter right](r)--[quarter right](r2)--(y)};
				\diagram{(z)--(z1)};
 			\end{feynman}
			\end{tikzpicture} 
			\end{gathered}+
			\begin{gathered}
			\begin{tikzpicture}
				\begin{feynman}
				\vertex (x) at (0,0);
				\vertex (y) at (2,0);
				\vertex (z) at (0,-0.5);
				\vertex(z1) at (2,-0.5);
				\vertex (r1) at (0.5,-0.25);
				\vertex (r2) at (1,-0.25);
				\vertex (r) at (1.5,-0.25);
				\diagram{(x)--(r1)--[half left](r2) -- [half left](r)--[half left](r2)--[half left](r1)--(z)};
				\diagram{(y)--(r)--(z1)};
 			\end{feynman}
			\end{tikzpicture} 
			\end{gathered}+
			\begin{gathered}
			\begin{tikzpicture}
				\begin{feynman}[scale = 1.5]
				\vertex (x) at (0.7,0);
				\vertex (y) at (1.3,0);
				\vertex (z) at (0.8,-0.2);
				\vertex(z1) at (1.2,-0.2);
				\vertex (r1) at (1,-0.4);
				\vertex (r2) at (0.7,-0.5);
				\vertex (r) at (1.3,-0.5);
				\diagram{(x)--(z)--[quarter left](z1)--[quarter left](z)};
				\diagram{(y)--(z1)--(r1)--(r2)--(r1)--(r)--(r1)--(z)};
 			\end{feynman}
			\end{tikzpicture}
	\end{gathered} \right]  \sim i \l_0^3 (\D_{x,y})^2 \int \rmd^d z (\D_{x,z})^2(\D_{y,z} )^2\, .
		\ee
In momentum space, the total bare two-point function of the $\phi_0 \Box \phi_0$ operator is therefore given by
	\be\label{k3k3}
		\braket{(\phi_0 \Box \phi_0)(\phi_0 \Box \phi_0)} = 
		3 \left[\frac{2}{3}\l_0^2 W(p) + \frac{8}{3}i \l_0^3 Q(p) \right]\, ,
	\ee
where $Q(p)$ is the loop integral corresponding to the Ninja-turtle topology
	\be
	Q(p)= (b_{1,1})^2 b_{1,4-d}b_{1,5-\frac{3d}{2}} (p^2)^{2d-6}\, .
	\ee
An interesting observation is that one could have arrived precisely at \eqref{k3k3}
by evaluating $\frac{\l_0^2}{36} \braket{\phi_0^4 \phi_0^4}$, 
which is a manifestation of the validity of the equation of motion at the operatorial level. 
Thus, the two-point function of the trace, expressed in terms of the renormalized coupling constant is given by:
	\begin{align} \label{theta0theta01}
	\braket{\Theta^{(0)} \Theta^{(0)}} 
		&=\frac{(d-4)^2}{16} 2 \left[   (Z_\l)^2 \l^2 \m^{2\ve} W(p) + 4i (Z_\l \l \m^\ve)^3 Q(p)\right]\, ,
	\end{align}
where $Z_\l$ is the renormalization factor of the coupling constant $\l$:
	\be
	Z_\l = 1  + \d_\l = 1 -i \frac{3}{2}\l \m^\ve B_0(p)|_{p^2 = -\tilde{\m}^2}  = 1 + \frac{3\l}{(4\pi)^2 \ve}+ \cdots
	\ee
Substituting $Z_\l$ and expanding in powers of $\ve$, the term in the squared brackets in \eqref{theta0theta01} becomes
	\be
		\frac{2}{3}  \left[   (Z_\l)^2 \l^2 \m^{2\ve} W(p) + 4i (Z_\l \l \m^\ve)^3 Q(p)\right] =
		-\frac{2}{3(4\pi)^6} \frac{1}{\ve} \frac{i}{18} \left[\l^2 - 6 \frac{\l^3}{(4\pi)^2 \ve} 
		+O(\ve \l^2 ,\l^3) \right] \sim -\frac{2}{3(4\pi)^6} \frac{1}{\ve^3} \frac{i}{18} \b_\l^2\, .
	\ee
The $\frac{1}{\ve^3}$ term in the above expression combined with the $(d-4)^2$ leaves an overall $\frac{1}{\ve}$ term.
Expanding the full bare 2-point function we arrive at the following expression:
	\be
	\braket{\Theta^{(0)} \Theta^{(0)}} =-\frac{i}{144(4\pi)^6}\frac{1}{\ve}\left[\ve^3 \l^2 -6\ve \frac{\l^3}{(4\pi)^2} + O(\ve^3 \l^2 ,\ve^2 \l^3) \right] \sim
	-\frac{i}{144(4\pi)^6}\frac{1}{\ve} \left[ \b_\l^2 + + O(\ve^3 \l^2 ,\ve^2 \l^3)  \right]p^4\, .
	\ee
As shown in the previous section, the total EMT is (re)normalized by a $Z_T^{(0)}$ which is determined by the $O(1)$ diagrams of the $\braket{T_{\m \n} T_{\r \s}}$:
	\be
	Z_{T}^{(0)} = - \frac{b_{1,1}}{4(d^2-1)}\, .
	\ee
Adapting this normalization we conclude that
	\be
	\braket{\Theta \Theta} = \frac{5}{24}\frac{1}{(4\pi)^4}
	 \left[ \ve^2 \l^2  -6\ve \frac{\l^3}{(4\pi)^2} + O(\ve^3 \l^2 ,\ve^2 \l^3)\right] p^4\, 
	\ee
or more conveniently
	\be
	\braket{\Theta \Theta} \equiv \frac{1}{Z_T^{(0)}} \braket{\Theta^{(0)} \Theta^{(0)}} = 	 \frac{5}{24(4\pi)^4}\b_{\l}^2p^4 +O(\l^4)\, .
	\ee
Compairing this with \eqref{theta theta}, the normalization constant $c$ gets fixed to
	\be
	c^2= \frac{5}{24(4\pi)^4}\, .
	\ee
Taking the limit $\ve \to 0$ of the  renormalized expression we get
	\be
	\braket{\Theta \Theta}_{\ve \to 0} = 0 + O(\l^4)\, ,
	\ee
where the $O(\l^4)$ contribution comes from the purely 4-dimensional part of $\b_\l^2$.
This result is consistent with the one presented in \cite{spin0}. 
The advantage of working with a trace in $d $-dimensions is that the $\b_\l^2$ term emerges naturally from an explicit loop calculation. However, the drawback is that reproducing the logarithmic corrections in \cite{spin0} requires a higher loop renormalization.

\subsection{The RG flow of $C_T$} 

A monotonicity theorem for the RG flow in generic renormalizable field theories is well established in two dimensions, known as the $c$-theorem~\cite{Zamolodchikov}, {where the $c$-function decreases  monotonically and becomes stationary at the fixed points}.
 This theorem is constructed from the correlation functions of the EMT, exploiting the relatively simple tensorial structure in two dimensions.
In dimensions $d > 2$, an analogous proof of a $c$-theorem is lacking. The primary obstacle is the increased number of independent tensor structures in the two-point function $\langle T_{\mu \nu} T_{\rho \sigma} \rangle$~\cite{Cardy}. For $d = 4$, a monotonicity theorem is instead provided by the $a$-theorem~\cite{a-theorem}, which adopts a distinct perspective from the two-dimensional case, focusing on the trace anomaly coefficient $a$ rather than the spin-2 charge $C_T$.

In~\cite{spectralTT}, it was stated that the $O(\l^2)$ spin-two charge $C_T$ is incompatible with a $c$-function,
since it is not stationary at the WF fixed point.
Let us re-evaluate this conclusion in light of the 4-loop result \eqref{CTfinal}.
We define the parameter $x$ as 
	\be
	x\equiv -\frac{p^2}{\tilde{\m}^2}\, ,
	\ee
in terms of which the extremal values of $C_T$ are at:
\begin{align}
\frac{ \partial C_T}{\partial \l} = 0 \Rightarrow 
	\begin{cases}
	\l_{\rm max} &= 0 \,, \quad \text{global maximum} \\
	\l_{\rm min} &= \frac{28 \ve + 60}{1395 - 270 \ln (x)} + \frac{2 \ve }{9} \,, \quad \text{global minimum}
	\end{cases}
\end{align}
Clearly whether $\l_{\rm min}$ coincides with $\l^*$ depends on the value of $x$.
In particular $\l_\min =\l^*$  for $x= e^{\frac{128}{85 \ve +45}-\frac{2}{\ve }+\frac{31}{6}}\vert_{\ve=1}\equiv x^*$.
We can construct RG flow curves, for example by keeping $x$ fixed, which is a particular way to flow, among others.
	\begin{figure}[H]
		\centering
		\begin{tikzpicture}[scale =1]
			\begin{axis}[
	        		axis lines = left,
     			xlabel = {$\l$},
	        		ylabel = {$C_T$},
        			xmin = 0, xmax = 0.8,
       			ymin =0 ,ymax = 1.2,
        			domain = 0:0.6,
   			samples = 500,
			thick,
			legend pos = outer north east,
			xtick=\empty,
           		 ytick=\empty
			]
   \addplot[gray,dotted, very thick]{1+0.201442*x^2 - 3.0016*x^3 };\addlegendentry{ \, $\l_{\min} > \l^*$}
    \addplot[gray, very thick]{1- 4.80457*x^2 +5.3384*x^3};\addlegendentry{ \, $\l_{\min} =\l^*$}
    \addplot[gray,dashed, very thick]{1-9.81057*x^2 +13.6784*x^3};\addlegendentry{ \, $\l_{\min} < \l^*$}
          \node at (axis cs:0.6,0.426) {\Large $\star$};
          \node at (axis cs: 0.01,1) {\Large $\bullet$};
               \end{axis}
\end{tikzpicture}
\caption{Qualitive plot of the running of $C_T$ for $d=4-\ve$. { Each curve in this plot is  generated by keeping the parameter $x$ fixed. The dotted curve corresponds to $x>x^*$, the solid curve  to $x= x^*$ and the dashed curve to $x<x^*$.}}
\label{RGcurves}
\end{figure}
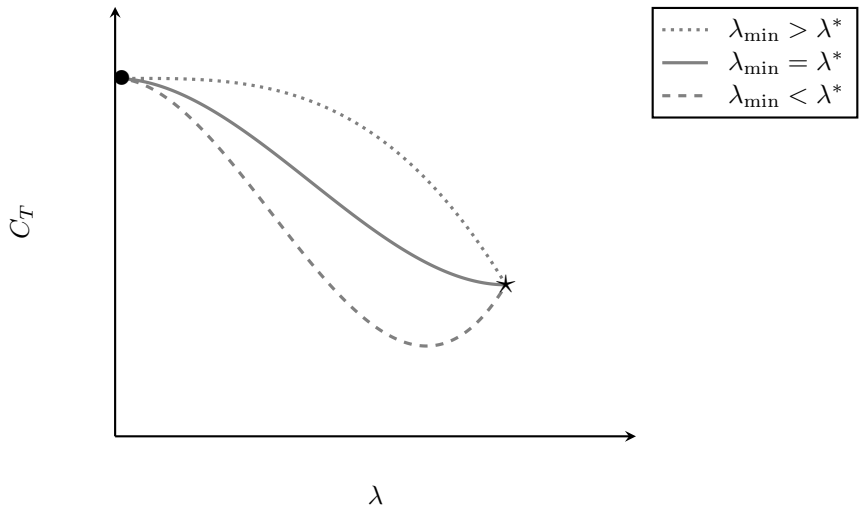
We then distinguish three kinematic regimes:
\begin{itemize}
    \item For $x = x^*$, $C_T$ is monotonic throughout the RG flow and is also stationary at the Wilson–Fisher fixed point.
    
    \item For $x < x^*$, the charge satisfies a weak version of the $c$-theorem, namely $C_T^{\rm (UV)} > C_T^{\rm (IR)}$, 
    but it is neither monotonic throughout the RG flow nor stationary at the fixed points.
    
    \item For $x > x^*$, the charge still satisfies a weak version of the $c$-theorem and is monotonic along the flow; however, 
    it is not stationary at the fixed points. 
\end{itemize}
Even though the 3-loop result of \cite{spectralTT} does not have a kinematic dependence, its
RG flow falls in the third category, the dotted curve in Fig. \ref{RGcurves}.

An alternative way to draw the RG flow is to keep instead of $x$, the external momentum fixed, in which case $C_T$
becomes a function of $\tilde \mu$. We have checked numerically that the three types of flows in 
Fig. \ref{RGcurves} persist.

\subsection{Holographic cosmology}
Another  application comes from holographic cosmology.
Before discussing the specifics of the application, 
we will outline the key cosmological observables against which we will test our results.
Two important observed quantities are the power spectra of the scalar and tensor fluctuations of the CMB. 
Adapting standard cosmological parametrization, these power spectra are given by:
	\begin{align} \label{cosm scalar}
	\D_S^2(p) &= \left. \D_S^2 \right|_{p^2 =-\tilde{\m}^2}\left( \frac{p}{\tilde{\m}} \right)^{n_s(p)-1} \\
	\label{cosm tensor}
	\D_T^2(p) &= \left. \D_T^2 \right|_{p^2 =-\tilde{\m}^2}\left( \frac{p}{\tilde{\m}} \right)^{n_T(p)}\, ,
	\end{align}
where $p$ is the observed wave number (or equivalently frequency) and $\tilde{\m}$ is a chosen pivot frequency.
As suggested by the chosen notation, the pivot frequency will be identified with the DR renormalization energy scale, and 
the wave number with the external momentum of the correlator.
The two quantities $n_S$ and $n_T$ that characterize the power spectra are the 
scalar and tensor CMB fluctuation spectral indices.

The scalar spectral index $n_S$ has been directly measured and given by \cite{BICEP}:
	\be
	n_S^{({\rm obs})} \simeq 0.964\, .
	\ee
The tensor spectral index is not  directly measured yet, but its value is constrained 
\cite{ntconstr}:
	\be
	-1.37 \lesssim n_T^{({\rm obs})} \lesssim 0.42\, .
	\ee 
A negative valued $n_T$ typically indicates a  slow-roll type of inflationary scenario, while a non-negative value possibly some kind of alternative inflationary scenario.
A third observable is the ratio of the tensor over the scalar fluctuations, with an upper bound constraint \cite{BICEP}:
	\be\label{defr}
	r = \frac{\D_T^2(p)}{\D_S^2(p)}\, , \hskip .5cm r^{(\rm obs)} \lesssim  0.036\, .
	\ee
Focusing from now on the holographic point of view, let us suppose that the dual QFT to the inflationary epoch is a real scalar with a $\l\phi^4$ interaction.
The holographic system by its nature then requires the coefficients $C_T$ and $C_\Theta$ to be evaluated near the WF fixed point. 
This is also supported by the almost scale-invariant CMB, observed by the experiments.

The "distance" $d_\l = \l -\l^*$ from the fixed point where the physical system is supposed to be sitting is,
in principle, a free parameter.
It can be however fixed from a precisely measured quantity, such as the scalar index $n_S$.
Then a number of other cosmological parameters will be predicted, given a pair of chosen eigenvalues $e_\Theta$ and $e_T$.
Alternatively, there could be dual information about this distance coming from the bulk, where the small breaking of scale invariance
may have a thermal origin \cite{IrgesThermal}. For a bulk with a scalar in a dS background, it will be related to the intrinsic temperature, proportional to the dS curvature.
We will return to these questions in more detail in the near future.
Here we will just present some simple consequences of this idea, some of which have already appeared in \cite{Ising}.

The basis for the holographic interpretation of cosmological parameters was proposed by \cite{Maldacena1}.
The connection of these parameters with the anomalous dimension of composite operators in the dual QFT (a scalar $\phi^4$ theory like here) was made in \cite{Larsen}.
More details of this connection for a general QFT were given in \cite{HolographicUniverse, HolographyforCosmology}, whose analysis can be summarized by the following relation
between the CMB scalar and tensor power spectra and the QFT charges $C_\Theta$ 
and $C_T$:
	\begin{align} \label{QFT scalar}
	\D_S^2(p) &= -\frac{1}{16\pi^2}\frac{1}{C_\Theta (p)} \\
	\label{QFT tensor}
	\D_T^2(p) &= -\frac{2}{\pi^2}\frac{1}{C_T (p)}\, .
	\end{align}
	
We will therefore expand below the three cosmological observables around the fixed point.
Before doing so, we can make one general comment. 
From the definition \eqref{defr} and the holographic formulae \eqref{QFT scalar} and \eqref{QFT tensor} the ratio $r$ can be written as
	\begin{align}
	r= \frac{ 1}{8} \frac{C_\Theta (p)}{C_T(p)}\, ,
	\end{align}
from which we can infer that since $C_\Theta\sim \b_\l^2$ and $C_T = 1 +\cdots$, the ratio will be small enough to be compatible with observations sufficiently near the interacting fixed point
where $\beta_\l\to 0$, for generic values of the charges.
We can now expand the above expression for small values of the deviation $d_\l $ from the fixed point:
	\begin{align} \label{r pert}
	r= \frac{1}{8}\left\{ \left. \frac{C_\Theta}{C_T}\right\vert_{\l=\l^*}
	+  d_\l \left[\frac{\partial}{\partial \l} \frac{C_\Theta}{C_T}\right]_{\l =\l^*}
	+ \frac{d_\l^2}{2}  \left[\frac{\partial^2}{\partial \l^2} \frac{C_\Theta}{C_T}\right]_{\l =\l^*} +
	 \cdots \right\} 
	\end{align}
The first term is zero at the fixed point.
The linear term in $d_\l$ also vanishes since the derivate of the ratio is zero at the fixed point. 
Thus, the parameter $r$ is determined by the second derivative.
Using \eqref{Theta eigenvalue eq} and \eqref{ct eigenvalue eq},  \eqref{r pert}  reduces to
	\be
	r= \frac{1}{8} \frac{d_\l^2}{2}  \left\{\frac{C_\Theta}{C_T}
	\left[\frac{e_\Theta}{\b_\l^2} - \frac{e_\Theta' \b_\l'}{\b_\l^2} \right] \right\}_{\l=\l^*} +O(d_\l^3)
	\ee
with the prime denoting a $\lambda$-derivative.
Recalling that 
	\be
	C_\Theta =\frac{  \braket{ \Theta \Theta} }{(d-1)^2}= \frac{5}{24(4\pi)^4} \frac{\b_{\l}^2}{(d-1)^2} +O(\ve^3 \l^2,\ve^2 \l^3,\l^4)\, ,
	\ee
this becomes
	\be \label{r pert contr}
	r = \frac{1}{8} \frac{d_\l^2}{2} \frac{5}{24(4\pi)^4 (3-\ve)^2} \frac{1}{C_T^*} \left[ e_\Theta^* -(e_\Theta')^* (\b_\l')^* \right]\, .
	\ee
Expanding in powers of $\ve$ and then setting $\ve \to 1$ we obtain:
	\be
	r \simeq  8 \times 10^{-9} d_\l^2\, ,
	\ee
which is easily compatible with current observational constraints.

The situation is less clear for the other two indices.
One can see that it is actually the eigenvalues $e_\Theta$ and $e_T$ that determine them:
	\begin{align}
	n_S&=1-e_\Theta  = 1- 
	\left( e_\Theta |_{\l^*} 
	+ d_{\l} \left. \frac{\partial e_\Theta}{\partial \l} \right|_{\l^*} + \cdots \right)\\
	\label{nt=-et} n_T&=-e_T=-\left( e_T |_{\l^*} 
	+ d_{\l} \left. \frac{\partial e_T}{\partial \l} \right|_{\l^*} + \cdots \right)\, .
	\end{align}
Let us test our perturbative formulae on these expressions as well. 
From the leading order expression of $e_\Theta$ \eqref{Theta eigenvalue} we obtain:
	\be \label{ns pert contr}
	n_S = 1 - 2\left(\G_{\phi^4}^* 
	- \left. \frac{\partial \G_{\phi^4}} {\partial \l} \right|_{\l^*} d_\l + \cdots \right) 
	\ee
It is easy to see that this can not lead to acceptable values, meaning that
the perturbative $e_\Theta$ does not reproduce the cosmological data for $n_S$.

Regarding the spectral index of the tensor fluctuations,
since $e_T \sim \b_\l$ the leading order contribution turns out to be linear in $d_\l$:
	\be  \label{nt pert contr}
	n_T= \frac{5}{18 (4\pi)^4}\l^* \left. \frac{ \partial \b_\l}{\partial \l} \right|_{\l^*}d_\l +O(d_\l^2) \simeq 5\times 10^{-3} d_\l \, ,
	\ee
which lies within the  allowed values, while its negative sign most likely has implications 
for the nature of the dual inflationary scenario.

Some comments on the formulae \eqref{r pert contr},  \eqref{ns pert contr} and \eqref{nt pert contr}.
All three observables are essentially determined by the eigenvalues $e_\Theta$ and $e_T$
(with the latter proportional to the $\b_\lambda$).
The $\b$-function for $d=4-\ve$ has positive slope in the vicinity of the WF fixed point (see Fig.\ref{beta function 4-ve}), so  $n_T$ is negative. As metioned the pertubative $e_\Theta$ fails to reproduce the observed value for $n_S$.
A possible way around this is to recall an earlier statement we made, that general values of $e_\Theta$ and $e_T$ 
define through equations \eqref{Theta eigenvalue eq} and \eqref{ct eigenvalue eq} general RG-flows towards the WF fixed point.
By general we mean here "not necessarily perturbative". For example in \cite{Ising} it is proposed that the RG-flow characterized by  
$e_\Theta = 2\gamma_\phi$ explains well the value of $n_S$, near the WF point, where it becomes the critical exponent $\eta\simeq 0.036$. 
Supposing an RG flow characterized by $e_\theta=\eta\simeq 0.036$ and $(e_\Theta')^* \simeq 0$, we get
	\begin{align}
	n_S&\simeq 1- 0.036=0.964  \\
	r &\simeq  4 \times 10^{-9}
	\end{align}
while $ n_T$ retains its value given by \eqref{nt pert contr}, as it is determined solely by the $\beta$-function.

\section{Conclusions}
 From a 4-loop calculation, we extracted the $\mathcal{O}(\lambda^3)$ contribution to the spin-2 charge, $C_T$, and demonstrated that it  decomposes into two sectors: a conformal and an RG-sector. 
 The latter is proportional to the $\beta$-function therefore cannot be accessed via purely CFT methods.
 As such, a QFT framework is needed to extract information about this RG-sector.
Furthermore, we verified that the spin-2 charge satisfies an eigenvalue equation analogous to that of the spin-0 charge, with a distinct eigenvalue $e_T$ proportional to the $\beta$-function. Also, $C_T$ exhibits a smooth CFT limit: 
it becomes regulating scale independent as the system reaches a fixed point, thereby restoring scale invariance.
In addition, we showed that the $\mathcal{O}(\lambda^3)$ corrections introduce an explicit dependence of $C_T$ on the external momenta, affecting its behavior along the RG flow. 
 As a possible physical application, we applied our results to the holographic cosmology senario and illustrated that the three main 
 observables ($n_S$, $r$, and $n_T$) can be expressed in terms of the eigenvalues $e_\Theta$ and $e_T$.
While the perturbative value of $e_\Theta$ does not reproduce the observed value pf $n_S$, we showed that an alternative choice for this eigenvalue brings all three observables within the allowed window from observations.
A self-contained analysis in the bulk is required to determine whether this choice can be justified.

\begin{center}{\bf\large Acknowledgements}\end{center}
N.I. and L.K. acknowledge Claudio Corianò for  discussions.  
L.K. would also like to thank Claudio Corianò for his warm hospitality during a visit to Lecce.  
L.K. is supported by a scholarship from the Research Committee of the National Technical University of Athens.

\newpage
\begin{appendix}
\numberwithin{equation}{section}
\begin{center}{\bf\Huge Appendices}\end{center}
\section{Notation}
\subsection{Table of symbols}
\begin{table}[H]
    \centering
    \begin{tabular}{|c|c|}
\hline 
${\rm \bf Symbol}$ & ${\rm \bf Meaning}$  \\
\hline \hline   
$\blacklozenge$ & ${T_{\m \n}^{(0)} \, \,\, \text{insertion}}$ \\ \hline
${\blacksquare}$  & $\phi_0^4 \, \,\, \text{insertion} $ \\ \hline
\scalebox{1.5}{$\bullet$}  & $\rm Gaussian \, fixed \, point$ \\ \hline
\scalebox{1.5}{$\star$}  & $\rm Wilson-Fischer \, fixed \, point$ \\ \hline

\end{tabular}
    \caption{Summary of the various symbols used in the text and in Figures.}
    \label{tab:my_label}
\end{table}
\subsection{The action}\label{quantumaction}
\begin{figure}[H]
	\centering
		\begin{tikzpicture}[scale =0.6]
			\begin{axis}[
	        		 axis x line =center,
           		 axis y line = left ,
        			xmin = 0, xmax = 1.2,
       			ymin =-0.27 ,ymax = 0.1,
        			domain = 0:1.1,
   			samples = 500,
			thick,
			legend pos = outer north east,
			xtick=\empty,
           		 ytick=\empty,
			 xlabel = {$\l$},
	        		ylabel = {$\b_\l$},
			xlabel style={at={(axis description cs:1.05, 0.7)}},
        			ylabel style={at={(axis description cs: 0.2,1.05)}, rotate=-90}
			]
   \addplot[gray, thick]{-x+ x^2 };
              \node at (axis cs:1,0) {\Large $\star$};
          \node at (axis cs: 0.02,0) {\Large $\bullet$};
               \end{axis}
	\end{tikzpicture}
	\caption{\small Qualitive plot of the $\b$-function for $d=4-\ve$. 
	The slope of the $\b$-function in thw vicinity of the WF fixed point is positive.}
	\label{beta function 4-ve}
	\end{figure}
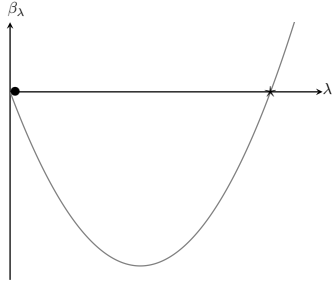

As we are interested in the renormalized version of the 2-point function of the EMT,
it is convenient to express the bare correlation function in terms of the renormalized $\l$.
To achieve this, we introduce the renormalization factors of the coupling constant and the primary field,
$Z_\l$ and $Z_\phi$, respectively, to write down the Lagrangean  in terms of renormalized quantities:
	\be
	S_{0}=\int \rmd^d x \left[ \frac{1}{2} (\partial \phi_0)^2 - \frac{\l_0}{4!}\phi_0^4 \right] =
	\int \rmd^d x \left[ \frac{Z_\phi}{2} (\partial \phi)^2 - Z_\l\frac{\l \m^{4-d}}{4!}\phi^4\right] \, .
	\ee
The relation between the bare and renormalized coupling constant for $d=4-\ve$ is
	\be\label{bare coupling}
	\l_0 = \frac{Z_\l}{Z_\phi^2} \l \m^{\ve}\, .
	\ee
The evaluation of the renormalization factors is reviewed in the Appendix of \cite{spin0}.
The key point of the analysis is the appropriate renormalization condition of the correlation functions.
To get consistent results, we should properly define the renormalization scale.
The choice of the renormalization condition stems from the renormalization of the coupling constant:
	\be
	\braket{\phi \phi \phi \phi}_{\rm amp}= -i \l -i\l \d_\l -  
	i \frac{3}{2} \l^2\left[ \frac{2}{\ve} - \ln\left(- \frac{p^2 e^\g}{\m^2 e^2 (4\pi)} \right)\right] + O(\ve)
	\ee
The  above expression implies that   it is convenient to apply the renormalization condition at the reduced renormalization energy scale 
	\be
	\braket{\phi \phi \phi \phi}_{\rm amp}= -i \l \, ,  \,\text{at } \,  s=t=u =-\tilde{\m}^2 \, , \ve \to 0
	\ee
where
		\be\label{tildemu}
	\tilde{\m}^2 = \frac{e^2 4\pi}{e^\g} \m^2 \, 
	\ee
and $s,t,u$ are the 3 channels of the $\phi \phi \to \phi \phi$ process. 	
Applying the renormalization above, we obtain the counterterm of the coupling constant:
	\be\label{Zlambda}
	Z_\l = 1 -i\frac{3}{2}\l \m^\ve B_{0}(p)|_{p^2 =- \tilde{\m}^2}=1 -i\frac{3}{2}\l \m^\ve b_{1,1}(-\tilde{\m}^2)^{-\ve/2}\, ,
	\ee
where $B_0(p)$ is the standard one-loop integral defined in \eqref{spin0 gen int} and $b_{1,1}$ is given by \eqref{B coef}.
Choosing the reduced renormalization scale does not affect the RG equations since
	\be
	\frac{\rmd}{\rmd \ln \m} = \frac{\rmd}{\rmd \ln \tilde{\m} } \, .
	\ee
As the bare $\l_0$ is independent of the renormalization scale,
	\be
	\frac{\rmd \l_0}{\rmd \ln \m} =0 \, ,
	\ee
it follows that the $\b$-function for $d=4-\ve$ is given by
	\be
	\b_\l \equiv \frac{\rmd \l}{ \rmd \ln \m} = -\ve \l + \frac{3\l^2}{(4\pi)^2} -\frac{17}{3} \frac{\l^4}{(4\pi)^4}
	+ \cdots
	\ee
whose behaviour is shown in Fig. \ref{beta function 4-ve}.

We will also need the operator anomalous dimensions
\be
\Gamma_{\phi^2} = \frac{\lambda}{(4\pi)^2}+ \cdots\, , \hskip 1cm \Gamma_{\phi^4} = \frac{6\lambda}{(4\pi)^2}+ \cdots
\ee
\subsection{Vertex of the energy-momentum tensor }
We define the action in a generic curved background, where the scalar field is conformally coupled with the metric, as
	\be
	S_{\rm curved}= \int \rmd^d x\sqrt{|g|} \left[\frac{1}{2} g^{\m \n}\partial_\m \phi \partial_\n \phi - \frac{\l}{4!} \phi^4 + \frac{1}{2}\xi R\phi^2 \right]\, ,
	\ee
where $\xi=\frac{(d-2)}{4(d-1)}$.
The equation of motion is 
	\be
	\frac{\d S_{\rm curved}}{\d\phi} =0 \Rightarrow  \nabla_\a \nabla^\a \phi + \frac{\l}{6}\phi^3 + \xi R \phi =0\, .
	\ee
In the flat space limit we get
	\be
	\Box \phi + \frac{\l}{6}\phi^3 =0\, .
	\ee
Employing the equations of motion and the  $F$ and$F_{\m \n}$ identities, the improved EMT reduces to:
	\be
	T_{\m \n} =\left(\frac{1}{2}-\xi\right) \partial_\m \partial_\n \phi^2 
	+ \eta_{\m \n} \left( \xi - \frac{1}{4} \right) \Box \phi^2 - \phi \partial_\m \partial_\n \phi
	+ \frac{\eta_{\m \n}}{4} \phi \Box \phi  \, .
	\ee
The $T_{\m\n} \to \phi \phi$ vertex is defined as:

	\begin{align} \label{Tmn vertex}
		&\begin{gathered}
			\begin{tikzpicture}
			\begin{feynman}
				\vertex (x) at (0,0);
				\vertex (y) at (1,0.5);
				\vertex (z) at (1,-0.5);
				\diagram{(z)--(x)--(y)};
			\end{feynman}
			 	\node[draw, fill=black, scale=0.7, rotate=45] at (x) {};
				\node[scale=0.7]at (0,0.3){$\m \n$};
			\end{tikzpicture}
		\end{gathered}
		\equiv  (2\pi)^d \d\left(\sum_{j=1}^{3} p_j\right) V_{\m \n}(p_1,p_2) 
		 =\int \rmd^dx_{1,2,3}e^{i\sum_{j=1}^{3} p_j \cdot x_j} \frac{1}{2}
		\frac{\d^2 T_{\m \n}(x_1)}{\d \phi(x_2) \d \phi( x_3)} \nonumber \\
		&=(2\pi)^d \d\left(\sum_{j=1}^{3} p_j\right) \frac{1}{2}
		\left[(p_1+p_2)_{\m}(p_1+p_2)_{\n} +p_{2\m}p_{2\n}
		 - \frac{1}{4}\eta_{\m \n}(p_2^2+(p_1+p_2)^2)\right. \nonumber \\
		&\hspace{4 cm} \left.-2 \left(\frac{1}{2}-\xi \right)p_{1\m}p_{1\n} 
		 -2 \left(\xi - \frac{1}{4} \right)\eta_{\m \n}p_1^2
		 \right]\, .
	\end{align}
The above vertex definition is equivalent to the one obtained via the limit method.
We provide the derivation. Start from
\begin{align} 
		& (2\pi)^d \d\left(\sum_{j=1}^{3} p_j\right) V_{\m \n}(p_1,p_2)\frac{i}{p_2^2}\frac{i}{p_3^2} \equiv \frac{1}{2}
		\int \rmd^d x\rmd^d y \rmd^d z e^{ip_1x}e^{ip_2y}e^{ip_3z} 
		\left[ \begin{gathered}
			\begin{tikzpicture}
			\begin{feynman}
				\vertex (x) at (0,0) node[left] {\tiny{$x$}};
				\vertex (y) at (1,0.5) {\tiny{$y$}};
				\vertex (z) at (1,-0.5){\tiny{$z$}};
				\diagram{(z)--(x)--(y)};
			\end{feynman}
			 	\node[draw, fill=black, scale=0.4, rotate=45] at (x) {};
				\node[scale=0.7]at (0,0.3){$\m \n$};
			\end{tikzpicture}
		\end{gathered}\right]
		 \end{align}
and evaluate the vertex in configuration space:
	\begin{align}
		&\begin{gathered}
			\begin{tikzpicture}
			\begin{feynman}
				\vertex (x) at (0,0) node[left] {\tiny{$x$}};
				\vertex (y) at (1,0.5) {\tiny{$y$}};
				\vertex (z) at (1,-0.5){\tiny{$z$}};
				\diagram{(z)--(x)--(y)};
			\end{feynman}
			 	\node[draw, fill=black, scale=0.4, rotate=45] at (x) {};
				\node[scale=0.7]at (0,0.3){$\m \n$};
			\end{tikzpicture}
		\end{gathered}= \nonumber \\
		&  \left\{ \lim_{x_1 \to x} \left[ -\partial_{\m}^{(x)} \partial_\m^{(x)} + \frac{\eta_{\m \n}}{4}\Box_x \right] 
		+\left[\left(\frac{1}{2}-\xi\right) \partial_\m \partial_\n
		+ \eta_{\m \n} \left( \xi - \frac{1}{4} \right) \Box_x \right] \lim_{x_1 \to x}
		\right\}
		\left[\begin{gathered}
			\begin{tikzpicture}
			\begin{feynman}
				\vertex (x) at (0,0.3) {\tiny{$x$}};
				\vertex (y) at (1,0.3) {\tiny{$y$}};
				\vertex (z) at (1,-0.3){\tiny{$z$}};
				\vertex (x1) at (0,-0.3){\tiny{$x_1$}};
				\diagram{(z)--(x1)};
				\diagram{(y)--(x)};
			\end{feynman}
			\end{tikzpicture}
		\end{gathered}\right]  +{(y\leftrightarrow z)}=  \nonumber \\
	&-\D_{x,z}\partial_{\m }\partial_\n\D_{x,y} -\frac{i \d(x-y)}{4}\D_{x,z}
	 +\left(\frac{1}{2}-\xi\right)\partial_{\m \n}\D_{x,y}\D_{x,z} 
	 + \eta_{\m \n} \left( \xi - \frac{1}{4} \right) \Box_x \D_{x,y}\D_{x,z}+ (y\leftrightarrow z)
	\end{align}
Applying a Fourier transformation, it follows that
	\begin{align}
	V_{\m \n}(p_1,p_2)= \frac{1}{2}
		\left[(p_1+p_2)_{\m}(p_1+p_2)_{\n} +p_{2\m}p_{2\n}
		 - \frac{1}{4}\eta_{\m \n}(p_2^2+(p_1+p_2)^2)\
		 -2 \left(\frac{1}{2}-\xi \right)p_{1\m}p_{1\n} 
		 -2 \left(\xi - \frac{1}{4} \right)\eta_{\m \n}p_1^2
		 \right]\, ,
	\end{align}
which is in agreement with \eqref{Tmn vertex}. Τhe trace of the vertex is given by:
	\be
	\eta^{\m \n}V_{\m \n}(p,k) =
	\frac{1}{2}\left[ -\frac{d-4}{4}(k^2 + (k+p)^2) - 2(d-1)\left( \xi - \frac{(d-2)}{4(d-1)} \right)p^2 \right]\, .
	\ee	
Another useful expression for the evaluation of the spin-2 charge is 
	\be
	V_{\m \n}(p,k)V^{\m \n} (p,l)- \frac{1}{d-1}\eta^{\m \n}\eta^{\r \s}V_{\m \n}(p,k)V_{\r \s} (p,l) =
	\mathcal{V}_T^{(i)}(p,k,l) + \xi \mathcal{V}_T^{(\xi)}(p,k,l)\, ,
	 \ee
where
	\begin{align}
\label{calVi}	\mathcal{V}_{T}^{\rm (i)}(p,k,l) &= \frac{1}{4} \left[(k-l)^4 + p^2(k-l)^2 + \frac{(d-2)}{4(d-1)}p^4 
	-(k-l)^2 \left( k^2 + (k+p)^2 +( k \leftrightarrow l )\right) \right. \nonumber \\
	 &-3 \frac{(d-2)}{8(d-1)}p^2\left(k^2 + (k+p)^2  + ( k \leftrightarrow l )  \right)  
	 +\frac{7d-16}{16(d-1)} \left(k^2 (l+p)^2 + ( k \leftrightarrow l )   \right) \nonumber \\
	 &\left. +\frac{7d-16}{16(d-1)} \left(k^2l^2 +(k+p)^2(l+p)^2 \right) +k^2 (k+p)^2 + l^2 (l+p)^2 \right]\\
\label{calVxi}	\mathcal{V}_T^{(\xi)}(p,k,l)&=\frac{1}{4}
	 \left[(k+p)^4+k^4 -2k^2(k+p)^2-\frac{p^2}{2}(k^2 +(k+p)^2) +(k \leftrightarrow l) \right]\, .
	\end{align}
\section{Some useful loop integrals }\label{IntegralsB}

The 1-loop integrals that enter the computation at $O(\l^2)$ are:
\begin{align}
B(\a, \b ; p) &= \int\frac{\mathrm{d}^dk}{(2\pi)^d} \frac{1}{\left[k^2 \right]^{\a} \left[(p-k)^2 \right]^{\b}}\\
B_{\m }(\a , \beta ; p)&= \int \frac{\mathrm{d}^dk}{(2\pi)^d} \frac{k_\m }{\left[k^2 \right]^{\a} \left[(p-k)^2 \right]^{\b}}\\
B_{\m \n  }(\a , \beta ; p)&= \int \frac{\mathrm{d}^dk}{(2\pi)^d} \frac{k_\m k_\n }{\left[k^2 \right]^{\a} \left[(p-k)^2 \right]^{\b}}
\end{align}

\subsection{Rank-0  \texorpdfstring{$B$}{B}-integral}
We begin with the spin-0 loop integral, which can be evaluated straightforwardly. After introducing a Feynman parameter we can evaluate this general 1-loop integral:
\be\label{spin0 gen int}
B(\a,\b;p)=  \frac{i^{1-d}}{(4\pi)^{d/2}}\frac{\G \left(\frac{d}{2}-\a \right) \G \left(\frac{d}{2}-\b \right) \G \left(\a + \b -\frac{d}{2} \right)}{\G(\a) \G(\b) \G(d-\a-\b)} (p^2)^{\frac{d}{2}-\a -\b}\, , \, \a,\b>0.
\ee
We define the very important coefficient $b_{\a, \b}$ as:
	\be \label{B coef}
		b_{\a, \b}=\frac{i^{1-d}}{(4\pi)^{d/2}}
		\frac{\G \left(\frac{d}{2}-\a \right) \G \left(\frac{d}{2}-\b \right)
		 \G \left(\a + \b -\frac{d}{2} \right)}{\G(\a) \G(\b) \G(d-\a-\b)} \, .
	\ee
A useful identity for the $b$ coefficients is:
	\be \label{recurr b}
	b_{\a, \b +1 } =
	 \frac{(\a +\b +1-d)(\a+\b-d/2)}{\b(\b+1-d/2)}b_{\a,\b}\, .
	\ee
So the general 1-loop integral is given by
	\be
		B(\a,\b;p)= b_{\a, \b}(p^2)^{\frac{d}{2}-\a -\b}\, .
	\ee
The simplest 1-loop scalar integral is $B(1,1;p^2)$:
\be\label{B0}
B_0(p)\equiv B(1,1;p) \, ,
\ee
which is equivalent to the $B_0$-integral in the Passarivo-Veltman language \cite{PaVe}.

After some algebra, we can  reduce  $B(\a,\b;p^2)$ to $B_0(p^2)$:
\be\bal
B(\a,\b;p)&=(-1)^{-\a -\b} \frac{\G(d-2)}{\left[\G \left(\frac{d}{2} \right) \right]^2 \G \left( 2-\frac{d}{2} \right)} \frac{\G\left( \frac{d}{2}-\a \right) \G\left( \frac{d}{2}-\b \right)\G\left( \a+\b -\frac{d}{2} \right)}{\G(\a) \G(\b) \G(d-\a-\b)} (-p^2)^{2-\a-\b}B_0(p)\\
&= (-1)^{-\a-\b} \frac{\left(d-\a-\b \right)_{\a+\b-2} \left(2-\frac{d}{2} \right)_{\a+\b-2} }{\G(\a) \G(\b)\left(\frac{d}{2}-\a \right)_{\a} \left(\frac{d}{2}-\b \right)_{\b} } (-p^2)^{2-\a -\b} B_0(p)\, ,
\eal\ee
where $(x)_n$ is the Pochhammer symbol.

With the use of \eqref{spin0 gen int}, we exctract the following recurrence relations:
\begin{align}
B(\a+1 ,\b ;p) &= \frac{1}{p^2} \frac{\left(\frac{d}{2}-\a-\b\right)(d-\a-\b-1)}{\a \left(1+\a-\frac{d}{2} \right)} B(\a,\b;p) , \, \, \a,\b >0\, ,\\
\label{rec rel b}B(\a,\b+1;p)&=\frac{1}{p^2} \frac{\left(\frac{d}{2}-\a-\b\right)(d-\a-\b-1)}{\b \left(1+\b-\frac{d}{2} \right)} B(\a,\b;p) , \, \, \a,\b >0.
\end{align}
The simplest 2-loop integral is the sunset integral:
\be \label{S1 to B0}
S_1(p) = \int \frac{\mathrm{d}^dk \mathrm{d}^d l} {(2\pi)^{2d}}\frac{1}{k^2 l^2 (l+k-p)^2} = \int \frac{\mathrm{d}^d k}{(2\pi)^d} \frac{B_0(k-p)}{k^2}\, , 
\ee
which can be expressed as 
\be \label{S1}
S_1(p) =b_{1,1} b_{1,2-\frac{d}{2}} (p^2)^{d-3}\, .
\ee
In the $\ve$-expansion the sunset integral is given by \footnote{In the Appendix A  of \cite{spin0} there is a typo in the $\ve$ expansion of the sunset integral, giving an $\frac{1}{\ve}$ term instead a $\frac{1}{2\ve}$.}: 
	\be
	S_1(p)= \frac{p^2}{(4\pi)^4} \left[ \frac{1}{2\ve}  - \frac{1}{2} \ln(p^2) + \cdots\right]
	\ee

\subsection{Rank-1  \texorpdfstring{$B$}{B}-integral}
\be\label{Bm reduced}
B_{\m}(\a,\b;p)=\frac{p_\m}{2}\frac{d-2\a}{d-\a -\b}B(\a,\b;p)\, .
\ee
\subsection{Rank-2  \texorpdfstring{$B$}{B}-integral}
\be\label{Bmn reduced}\bal
B_{\m \n}(\a, \b ;p)=&\left[p_\m p_\n(d-2\a +2) + \eta_{\m \n}p^2\frac{\b  - \frac{d}{2}}{\frac{d}{2}-\a-\b+1}\right]\frac{(d-2\a)}{4(d-\a-\b+1)(d-\a-\b)} B(\a,\b;p)\, .
\eal\ee

\subsection{Integrals of type $J$}

Τhe 1-loop integral with $3$ propagators
	\be \label{Jn1n2n3}
		J_{\n_1, \n_2 ,\n_3}( p_1,p_2)=
		 \int \frac{\mathrm{d}^d k}{(2\pi)^{d}}
		 \frac{1}{k^{2\n_1} (k-p_1)^{2 \n_2} (k-p_2)^{2 \n_3}}\, ,
	\ee
is of central importance. It has been extensively studied in \cite{Boosint}. An interesting derivation, 
using the constraints of conformal symmetry on the 3-point function of scalar operators, is presented in \cite{Claudioint}.
In our analysis however, we instead use the Mellin-Barnes representation
	\begin{align}\label{J MB repr}
	J_{\n_1, \n_2 ,\n_3}(p_1,p_2)
	=\frac{i^{1-d}}{(4\pi)^{d/2}} \frac{(p_3^2)^{d/2-\n_{tot}}}{\G(\n_1) \G(\n_2)\G(\n_3) \G(d-\n_{tot})}
	\iint_{MB} \frac{\rmd s \rmd t}{(2\pi i)^2} 
	\mathcal{M}(s,t,\n_1 \n_2 ,\n_3)\left( \frac{p_1^2}{p_3^2}\right)^s \left( \frac{p_2^2}{p_3^2} \right)^t\, ,
	\end{align}
with
	\be \label{J int MB}
	\mathcal{M}(s,t,\n_1 \n_2 ,\n_3) = \G(-s) \G(-t) 
	\G(\n_1 +\n_2+\n_3+s+ t -\tfrac{d}{2} )
	\G(\tfrac{d}{2}-\n_1 -\n_2-s)
	\G(\frac{d}{2}-\n_1 -\n_3-t ) \G(\n_1 + s +t) \, .
	\ee
As a product of $\G$-functions, the above expression has multiple poles located along the real axis.
Specifically, the poles occur at the points where the arguments of the $\G$-functions become negative integers. We classify these poles into two categories: increasing poles and decreasing poles.
For generic values of the indices $\n_i$, the poles are organized as follows:	
	\begin{align}
	\text{increasing poles: }
	& s= n \,, \, n \in \mathbb{N} \\
	&t= n \,, \, n \in \mathbb{N} \\
	& s = \frac{d}{2}-\n_1 -\n_2 +\n \,, \, n \in \mathbb{N} \\
	 &t = \frac{d}{2}-\n_1 -\n_3 +\n \,, \, n \in \mathbb{N} \\
	\text{decreasing poles: } 
	&s+t = -\n_1 - n \,, \, n \in \mathbb{N} \\
	&s+t =\frac{d}{2} -\n_1 -\n_2 -\n_3 - n \,, \, n \in \mathbb{N}
 	\end{align}
Moreover, the last three terms in \eqref{J int MB} arise from interchanging the integrals over $s$ and
$t$ with the integral over the Feynman parameters. This interchange is justified provided that the Feynman parameter integral remains finite. Consequently, the MB representation in \eqref{J MB repr} is well-defined when the following set of inequalities is satisfied:
	\begin{align} \label{set of ineq J}
	s&< \frac{d}{2} -\n_1 -\n_2 \\
	t&< \frac{d}{2}-\n_1 -\n_3 \\
	s+t &> - \n_1
	\end{align}
The integral on $s$ and $t$ is a contour integral, defined such that  it separates 
the increasing from the decreasing poles of \eqref{J int MB}.
In particular, the MB integrals have the following form:	
	\be
	\iint_{MB} = \int_{s_0 -i \infty}^{s_0+i \infty} \int_{t_0 -i \infty}^{t_0+i \infty}\, ,
	\ee
where the values of $s_0$ and $t_0$ are determined by the set of inequalites \eqref{set of ineq J}. 
\section{Evaluation of loop integrals with the Mellin-Barnes method}\label{MBintegrals}

\subsection{Evaluation of ${\cal C}^{(1)}$ in eq. \eqref{calC1 reduced}}

The final step in obtaining the Cat-eye contribution is the evaluation of the $I\left(2 - \frac{d}{2}\right)$ integral.
The Mellin--Barnes (MB) representation \eqref{J int MB} of the $J $-integral implies that
	\be
	I(2-d/2) = \frac{i^{1-d}}{(4\pi)^{d/2} \G \left(2-\frac{d}{2}\right) \G \left( \frac{3d}{2}-4 \right)}
	\int \frac{1}{k^2 \left[(k+p)^2 \right]^{5-d} } 
	\iint_{MB} \mathcal{M}(s,t) \left( \frac{p^2}{(k+p)^2}\right)^s \left( \frac{k^2}{(k+p)^2}\right)^t\, ,
	\ee
with
	\be
	\mathcal{M}(s,t)=\G (-s) \G (-t) \G \left(\frac{d}{2}-s-2\right) \G (d-t-3) \G (s+t+1) \G (-d+s+t+4)\, .
	\ee
	The integrals in $s$ and $t$ are integrals over contours on the complex plane that are parallel to the imaginary axis.
The contours of integration are defined so that they separate the increasing poles of $\mathcal{M}(s,t)$ 
from its decreasing poles. 

\begin{figure}[h]
		\begin{center}
			\begin{tikzpicture}[scale=1.5]
				\def\d{3.8}
				\draw[gray ,fill =gray!60] (\d-4,0) -- (\d/2-2,0) -- (\d/2 -2,-\d/2+2+\d-4) -- cycle;
				\draw[black] (\d/2-2,-1.5)--(\d/2-2,1.4)node[above ]{\tiny$s=\frac{d}{2}-2 $} ;
				\draw[black] (\d/2-1,-1.5)--(\d/2-1,1.2)node[anchor=south ]{\tiny$s=\frac{d}{2}-1 $} ;
				\draw[black] (0,-1.5)--(0,1.2)node[above right ]{\tiny$s=0$} ;
								
				\draw[black] (-1.5,\d-3)--(1.5,\d-3)node[anchor=west]{\tiny$t=d-3$} ;
				\draw[black] (-1.5,0)--(1.5,0)node[anchor=west]{\tiny$t=0$} ;
				\draw[black] (-1.5,1)--(1.5,1)node[anchor=west]{\tiny$t=1$} ;
				\draw[dashed] (-1.5,0.5)--(1.5,0.5)node[anchor=west]{\tiny$t_0'$} ;
				\draw[black] plot[domain= -1.5:0.6 , samples= 2](\x,-1-\x)node[below right]{\tiny$s+t=- 1 $} ;
				
				\draw[black] plot[domain= -1.5:1 , samples= 2](\x,\d-4-\x)node[below right]{\tiny$s+t=d-4 $} ;
				\draw[black,thick, dotted] (3*\d/2-6,1.2)--(3*\d/2-6,-1.5)node[below]{\tiny$s=\frac{3d}{2}-6 $} ;
				\draw[black,thick, dotted] (2,1-\d/2)--(-1.5,1-\d/2) node[left]{\tiny$t=1-\frac{d}{2}$} ;
				\draw[black, thick, dotted] plot[domain= 2:-0.5 , samples= 2](\x,3*\d/2-5-\x)node[above left]{\tiny$s+t=\frac{3d}{2}-5 $} ;
				
				\filldraw[black] (\d/2-2-0.03,-0.03) circle(0.4pt)node[right] {\tiny$(s_0,t_0)$};
				\filldraw[black](\d/2-2-0.03,0.5) circle(0.4pt);				
				\filldraw[black](0.1,0.5) circle(0.4pt)node[above right] {\tiny$(s'_0,t'_0)$};
				\draw[ thick,->](\d/2-2-0.03,-0.03)--(\d/2-2-0.03,0.5);
				\draw[ thick,->](\d/2-2-0.03,0.5)--(0.1,0.5);
		\end{tikzpicture}
		\caption{\small The MB representation is well defined for $(s_0,t_0)=(\frac{d}{2}-2-\frac{\ve}{10},- \frac{\ve}{10})$. We define the point $(s_0',t_0')=(\kappa,\frac{1}{2})$ with $0< \kappa < 1$.
		The commutation of the integrals is allowed in the region defined by the dotted triangle. } 
		\label{I(2-d/2) shift}
	\end{center}
	\end{figure}
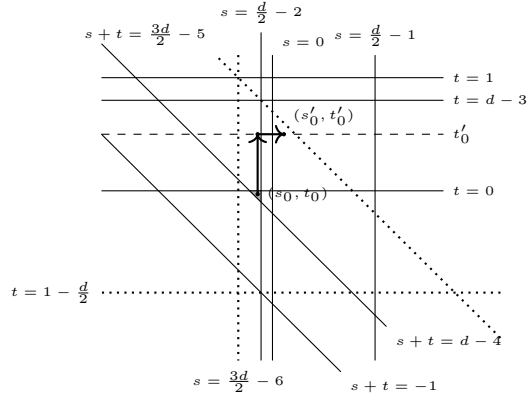

In Fig.\ref{I(2-d/2) shift},  we plot these poles on the $\text{Re}\{s\}-\text{Re}\{t\}$ plane, 
with each sector of poles corresponding to a solid line.
The vertical and horizontal lines are the increasing poles and the tilted lines are the decreasing poles.
As already discussed, the MB integral is well defined  when the set of inequalities \eqref{set of ineq J} is satisfied
	\begin{align}
	s_0&<\frac{d}{2}-2\\
	t_0&< d-3\\
	s_0+t_0&> d-4 \, ,
	\end{align}
while the poles are automatically seperated when
	\begin{align}
	s_0&<\frac{d}{2}-2\\
	t_0&< 0\\
	s_0+t_0&> d-4 \, .
	\end{align}
Pictorially, the solution to these inequalities is the region inside the filled triangle in Fig.\ref{I(2-d/2) shift}.
For $d=4-\ve$ such points are
	\be \label{s0,to cat}
	(s_0,t_0)=\left( \tfrac{d}{2}-2-\tfrac{\ve}{r}, -\tfrac{\ve}{r}\right) \, , \,  r>4 \, .
	\ee
Therefore the MB- contour for the case at hand is defined as follows:
	\be
	\iint_{MB} = \int _{s_0 - i \infty}^{s_0+i \infty} \int _{t_0 - i \infty}^{t_0+i \infty} \frac{\rmd s \rmd t}{(2\pi i)^2} \, .
	\ee
Then the $I$-integral becomes
	\begin{align} \label{I(2-d/2)MB}
	I(2-d/2) = \frac{i^{1-d}}{(4\pi)^{d/2} \G \left(2-\frac{d}{2}\right) \G \left( \frac{3d}{2}-4 \right)}
	  \int \frac{\rmd^d k}{(2\pi)^d}
	\int _{s_0 - i \infty}^{s_0+i \infty}\int _{t_0 - i \infty}^{t_0+i \infty} \frac{\rmd s \rmd t}{(2\pi i)^2} \frac{\mathcal{M}(s,t) (p^2)^s }{(k^2)^{1-t} [(k+p)^2]^{5-d+s+t}}\, .
	\end{align}
	The next step we would like to take is to perform the loop integral, which requires to interchange
the loop integral over $k$ with the $s,t$ contour integrals. 
To interchange however the $k$ with the contour integrals, the loop integral must be finite.
Commuting the integrals, there occurs an integral of type $B$ and in particular the $B(1-t,5-d+s+t ;p)$.
To check the finiteness of the result we use the generic form of the $B$-integrals \eqref{B coef}, 
for $\a=1-t$ and $\b=5-d+s+t $ and we demand the numerator to be non-singular. 
This results in the following set of inequalities:
 	\begin{align}
	s+t< \frac{3d}{2}-5 \nonumber\\
	s> \frac{3d}{2}-6\nonumber \\
	t>1-\frac{d}{2} \nonumber\, ,
	\end{align}
which is solved in the regime of points enclosed by the dotted triangle in Fig.\ref{I(2-d/2) shift}.
The chosen point  $(s_0, t_0)$ lies already in this area, ensuring that the interchange of the integrals is allowed.
Τhus, \eqref{I(2-d/2)MB} reduces to the following simple form:
	\be \label{I-commute}
	I(2-d/2) = \frac{i^{1-d} (p^2)^{\frac{3d}{2} -6}}{(4\pi)^{d/2} \G \left(2-\frac{d}{2}\right) \G \left( \frac{3d}{2}-4 \right)}
	\int _{s_0 - i \infty}^{s_0+i \infty}\int _{t_0 - i \infty}^{t_0+i \infty} 
	\frac{\rmd s \rmd t}{(2\pi i)^2} \tilde{\mathcal{M}}(s,t) \, ,
	\ee
where
	\be
	\tilde{\mathcal{M}}(s,t)= \mathcal{M}(s,t) b_{1-t,5-d+s+t}\, .
	\ee
The remaining step is to evaluate the double complex integral. However, the $\ve$-expansion 
of \eqref{I-commute} reveals one more point that must be handled carefully.
 In the limit $\ve \to 0$, the initially chosen values $s_0$ and $t_0$ approach the poles at $s = 0$ and $t = 0$, respectively.
 Therefore, it becomes necessary to shift the contours from the point  $(s_0, t_0)$ to a new point $(s_0', t_0')$,
 such that the integration contours avoid these poles in the $\ve \to 0$ limit.

\begin{figure}[h]
		\begin{center}
			\begin{tikzpicture}[scale=0.8]
				\draw[black,  thick , ->](0,-2.2)--(0,3)node[anchor=south]{Im} ;
				\draw[black,  thick , ->](-2,0)--(3.5,0)node[anchor=west]{Re} ;
				\draw[black, thick , ->] (-1.2,-2)--(-1.2,2)node[anchor=east]{$c_0$} ;
				\draw[black, thick ] (-1.2,2)--(-1.2,2.7);
				\draw[dashed,black,  thick ,->] (-1.2,2.7)--(-0.7,2.7)node[anchor=south]{$r^+$} ;
				\draw[dashed,black,  thick ] (-0.7,2.7)--(0.5, 2.7);
				\draw[dashed,black,  thick, -> ](0.5, -2)--(-0.7,-2)node[anchor=north]{$r^-$} ;
				\draw[dashed,black,  thick ](-0.7,-2)--(-1.2 ,-2);
				\draw[black,  thick , ->] (0.5,-2)--(0.5,2)node[anchor=west]{$c_0'$};
				\draw[black, thick ] (0.5,2)--(0.5,2.7);
				\filldraw[ fill = white] (-1,0) circle (2pt) ;
				\filldraw[ fill = white] (1,0) circle (2pt) ;
				\filldraw[ fill = white] (3,0) circle (2pt) ;
				\filldraw[ fill = white] (0,0) circle (2pt) ;
				\filldraw[ fill = white] (2,0) circle (2 pt) ;
			\end{tikzpicture}
			\caption{\small Qualitative picture of the contour shift $c_0\to c_0'$. The resulting closed contour is $C=c_0+r^+-c_0'+r^-$ and 
			the value of the integral along it is $\int_C=-2\pi i \, {\rm Res} \{\rm poles\}$. This means that $\int_{c_0} =
			\int_{c_0'} -   \Res {\rm poles}$.}
			\label{contour shift}
				\end{center}				 
	\end{figure}
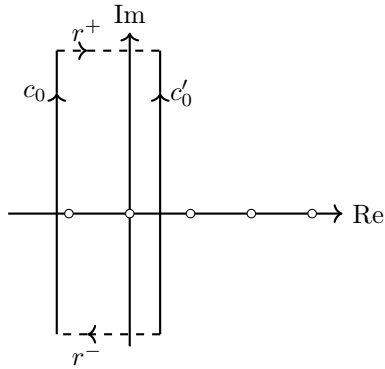

This contour deformation is performed in two steps: \( (s_0, t_0) \to (s_0, t_0') \to (s_0', t_0') \), as illustrated in Fig.~\ref{I(2-d/2) shift}. The integral evaluated at \( (s_0', t_0') \) differs from the one originally defined at \( (s_0, t_0) \) by a sum of residues collected during the deformation.
To extract these residues, we proceed as follows: we define a closed contour $C$ formed by the original and the shifted open contours at $t_0$ and $t_0' $, respectively, and connect them via horizontal segments $r^\pm$ 
defined along $t = \mathbb{R} \pm i\infty$.
This closed contour construction is depicted in Fig.~\ref{contour shift}.
Since $\tilde{\mathcal{M}}$ is a product of Gamma functions, the identity
	\be
	\G(x \pm i\infty) = 0 \,, \, x \in \mathbb{R} \, ,
	\ee
implies that the contributions from the horizontal segments $r^\pm$ are equal to zero. 
As a consequence, the difference between the integrals along the original and shifted contours equals
the sum of the residues at the intermediate poles enclosed by the deformation.
Thus we obtain:
	\begin{align}
	\iint_{s_0,t_0}\tilde{\mathcal{M}}(s,t) &=
	 \iint_{s_0',t_0'}\tilde{\mathcal{M}}(s,t)  -  \int_{t_0'}\Res{\tilde{\mathcal{M}} (s,t)}|_{s=0}  - \int_{t_0'}\Res{\tilde{\mathcal{M}} (s,t)}|_{s=\frac{d}{2}-2} \nonumber \\
	 & -\int_{s_0'}\Res{\tilde{\mathcal{M}}(s,t) }|_{t=0}
	 +\Res{\tilde{\mathcal{M}}(s,t) }|_{s=0,t=0}+\Res{\tilde{\mathcal{M}} (s,t)}|_{s=\frac{d}{2}-2,t=0}
	\end{align}
where we employed the following condensed notation
	\be
	\iint_{s_0,t_0} \equiv 
	\int_{s_0 - i\infty}^{s_0 +i\infty} \int_{t_0 - i\infty}^{t_0 +i\infty} \frac{\rmd s \rmd t}{(2\pi i)^2}\, .
	\ee
After evaluating the residues, multiplying with the prefactor in front of the $I$ integral 
in \eqref{calC1 reduced} and expanding in powers of $\ve$, we obtain:
	\begin{align}
	\frac{(d-6)(d-4)}{6(d-2)(3d-8)}I\left(2- \frac{d}{2} \right)p^{4-d} =
	\frac{1}{(4\pi)^4} \left[- \frac{1}{18 \ve} 
	+ \frac{1}{18} \ln \left( \frac{-p^2e^\g}{4\pi}\right) -\frac{18}{72}- \frac{1}{72}\int_{t_0'} \G(1-t) \G(t)\G(1+t)\G(-t) \right] + O(\ve)
	\end{align}
The $\frac{1}{\ve}$ term above is determined by the double residues, while the finite and the $O(\ve)$ terms are determined by the double residues and the single residues at $s= \frac{d}{2}$ and $s= 0$. The $\tilde{\cal M}$ and the residue at $t=0$, contribute to $O(\ve^2)$.
To evaluate the integral on $t$ we use the following two identities of the $\G$-function:
	\begin{align}
	\G(1-t)\G(t)&=-\G(-t)\G(1+t) \\
	\G(1-t)\G(t)&= \frac{\pi}{\sin(\pi t)} \, , \, t \not\in \mathbb{Z}
	\end{align}
and the limit:
	\be
	\lim_{t \to t_0' \pm i \infty} \cot(\pi t) = \mp i\, .
	\ee
Since $t_0' \in( 0,1)$, the contour of integration does not meet any pole. So we can evaluate the integral on $t$:
	\begin{align}\label{MBcateye2}
	\int_{t_0'-i \infty}^{t_0' + i \infty}  \frac{\rmd t}{2\pi i} \G (1-t) \G (-t) \G (t) \G (t+1) &= 
	- \int_{t_0'-i \infty}^{t_0' + i \infty} \frac{\rmd t}{2\pi i}  \frac{\pi^2}{\sin^2 (\pi t)} \, , \, t=t_0' + iy \nonumber \\
	&=-\int_{-\infty}^{+\infty} \frac{\rmd y}{2\pi } \frac{\pi^2}{\sin^2 (\pi (t_0'+i y))}=-1
	 	\end{align}
We then get
	\be \label{I-exp}
	\frac{(d-6)(d-4)}{6(d-2)(3d-8)}I\left(2- \frac{d}{2} \right)p^{4-d} =	\frac{1}{(4\pi)^4} \left[- \frac{1}{18 \ve} 
	+ \frac{1}{18} \ln \left( \frac{-p^2e^\g}{(4\pi)}\right) -\frac{17}{72} \right] +O(\ve)\, .
	\ee
Keeping the $O(\ve)$ terms and adding the contributions from the primitive integrals in \eqref{calC1 reduced}, we arrive at the following expression:
	\begin{align}
	\mathcal{C}^{(1)}&=\frac{1}{(4\pi)^4} \left\{- \frac{1}{18 \ve} 
	+ \frac{1}{18} \ln \left( \frac{-p^2e^\g}{(4\pi)}\right) -\frac{125}{216} \right. \nonumber \\
	&\left. +\frac{\ve}{2592} 
	 \left[ -72 \ln^2 \left( \frac{-p^2 e^\g}{4\pi} \right) + 1500  \ln \left( \frac{-p^2 e^\g}{4\pi} \right)+108A^{(0)}+6 \pi ^2 -6287
	 \right]\right\}\, ,
	\end{align}
where
	\be \label{A0 int}
	A^{(0)} = -\pi^2 \int_{t_0'-i\infty}^{t_0'+i\infty} \frac{\rmd t}{2\pi i} \csc^2 (\pi t) H_{t-1} 
	 \, , \, t_0' \in (0,1)  \, ,
	\ee
where $H_t$ is the harmonic number
	\be
	H_{t} = \psi^{(0)}(1+t) + \g\, ,
	\ee
with $\psi^{(0)}(t)$ the zeroth order Polygamma function
	\be
	\psi^{(0)}(t) = \frac{\rmd}{\rmd t} \ln (\G (t)) \, .
	\ee
The $A^{(0)}$ integral turns out to be equal to one:
	\be
	A^{(0)}=1\,\, .
	\ee
Therefore, the ${\cal C}^{(1)}$ finally becomes:
	\begin{align}
	\mathcal{C}^{(1)}&=\frac{1}{(4\pi)^4} \left\{- \frac{1}{18 \ve} 
	+ \frac{1}{18} \ln \left( \frac{-p^2e^\g}{(4\pi)}\right) -\frac{125}{216} \right. \nonumber \\
	&\left. +\frac{\ve}{2592} 
	 \left[ -72 \ln^2 \left( \frac{-p^2 e^\g}{4\pi} \right) + 1500  \ln \left( \frac{-p^2 e^\g}{4\pi} \right)-6179+6 \pi ^2 
	 \right]\right\}\, .
	\end{align}
\subsection{The master-Hourglass integral}

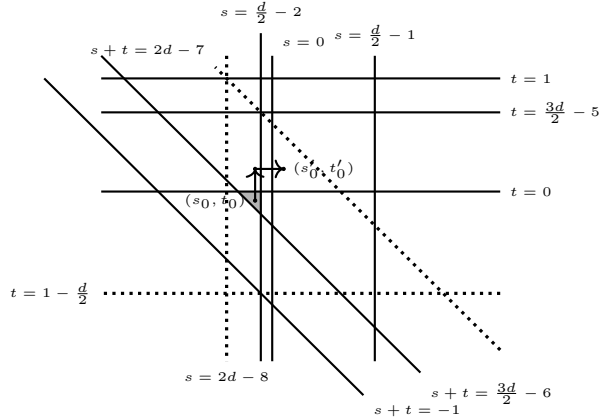
\begin{figure}[H]
		\begin{center}
			\begin{tikzpicture}[scale=1.5]
				\def\d{3.8}
				\draw[gray ,fill =gray!60] (3*\d/2-6,0) -- (\d/2-2,0) -- (\d/2 -2,-\d/2+2+3*\d/2-6) -- cycle;
				
				\draw[black, thick] (\d/2-2,-1.5)--(\d/2-2,1.4)node[above ]{\tiny$s=\frac{d}{2}-2 $} ;
				\draw[black, thick] (\d/2-1,-1.5)--(\d/2-1,1.2)node[anchor=south ]{\tiny$s=\frac{d}{2}-1 $} ;
				\draw[black, thick] (0,-1.5)--(0,1.2)node[above right ]{\tiny$s=0$} ;

				
				\draw[black, thick] (-1.5,3*\d/2-5)--(2,3*\d/2-5)node[anchor=west]{\tiny$t=\frac{3d}{2}-5$} ;
				\draw[black, thick] (-1.5,0)--(2,0)node[anchor=west]{\tiny$t=0$} ;
				\draw[black, thick] (-1.5,1)--(2,1)node[anchor=west]{\tiny$t=1$} ;

				\draw[black, thick] plot[domain= -2:0.8 , samples= 2](\x,-1-\x)node[below right]{\tiny$s+t=- 1 $} ;
				\draw[black, thick] plot[domain= -1.5:1.3 , samples= 2](\x,3*\d/2-6-\x)node[below right]{\tiny$s+t=\frac{3d}{2}- 6 $} ;
				\draw[black,very thick, dotted] (2*\d-8,1.2)--(2*\d-8,-1.5)node[below]{\tiny$s=2d-8 $} ;
				\draw[black, very thick, dotted] (2,1-\d/2)--(-1.5,1-\d/2) node[left]{\tiny$t=1-\frac{d}{2}$} ;
				\draw[black,very thick, dotted] plot[domain= 2:-0.5 , samples= 2](\x,2*\d-7-\x)node[above left]{\tiny$s+t=2d-7 $} ;
				
				\filldraw[black] (\d/2-2-0.05,-0.08) circle(0.4pt)node[left] {\tiny$(s_0,t_0)$};
				\filldraw[black](\d/2-2-0.05,0.2) circle(0.4pt);
				\filldraw[black](0.1,0.2) circle(0.4pt)node[ right] {\tiny$(s'_0,t'_0)$};
				\draw[ thick,->](\d/2-2-0.05,-0.08)--(\d/2-2-0.05,0.2);
				\draw[ thick,->](\d/2-2-0.05,0.2)--(0.1,0.2);
		\end{tikzpicture}
		\caption{\small  ${ \rm Re}\{s\}-{\rm Re}\{t\}$ plane fot the $I(4-d)$ integral.} 
		\label{I(4-d) shift}
	\end{center}
	\end{figure}

Here we evaluate the 4-loop master-Hourglass integral associated with the topology in \eqref{hour master},
	\be\label{GHR no exp}
	G_{\rm HR} = (b_{1,1})^2  I(4-d) \, .
	\ee
As already noted, the MB representation provides an efficient framework for this computation.
The evaluation procedure mirrors that followed for $I(2-\frac{d}{2})$, and the corresponding contour deformation steps are summarized in Fig.\ref{I(4-d) shift}. The resulting expression is
	\be
	G_{\rm HR} = \frac{ i^{1-d} (b_{1,1})^2 }{(4\pi)^{d/2}\G(4-d)\G(2d-6)} \int\frac{\rmd^d k}{(2\pi)^d} 
	\int \frac{\rmd s \rmd t}{(2\pi i)^2} \frac{ {\cal M}(s,t,1,1,4-d) (p^2)^s }{[k^2]^{1-t} [(k-p)^2 ]^{ 7-\frac{3d}{2} +s+t } }\, ,
	\ee 
where ${\cal M}$ is defined in \eqref{J int MB}:
	\be
	{\cal M}(s,t,1,1,4-d) =\G (-s) \G (-t) \G \left(\frac{d}{2}-s-2\right) \G \left(\frac{3 d}{2}-t-5\right) \G (s+t+1) \G \left(-\frac{3 d}{2}+s+t+6\right)\, .
	\ee
The integrals on $s$ and $t$ are along contours that run along straight lines parallel to the imaginary axis and
	\be
	\int_{MB} \frac{\rmd s \rmd t}{(2\pi i)^2} = \int_{s_0 -i \infty}^{s_0 +i \infty} \int_{t_0 -i \infty}^{t_0 +i \infty}  \,\frac{\rmd s \rmd t}{(2\pi i)^2}\, .
	\ee
The values of $s_0$ and $t_0$ lie within the filled triangular region in Fig.\ref{I(4-d) shift}.
A convenient choice of initial points $(s_0,t_0)$ is 
	\be
	(s_0,t_0) = \left(\frac{d}{2}-2 - \frac{\ve}{r} , \frac{\ve}{r} \right) \, , r>2\, .
	\ee
In analogy with the procedure followed for the CE topology, the next step involves commuting the loop integral over \( k \) with the contour integrals over \( s \) and \( t \). This interchange of integration order is justified only if the corresponding loop integral is finite. Upon commuting the integrals, the loop integral yields a factor of 
$B\left(1-t, \frac{3d}{2} - 7 + s + t; p\right),$ which is finite under the following set of inequalities:
	\begin{align}
		t &> 1 - \frac{d}{2}, \\
		s &> 2d - 8, \\
		s + t &< 2d - 7.
	\end{align}
The region defined by these constraints is illustrated as the dotted triangle in Fig.~\ref{I(4-d) shift}. The initially chosen values $(s_0, t_0)$ lie within this triangle, thus validating the interchange of integrations.
As a result, the $ I(4-d) $ integral takes the form:
	\be \label{}
	I(4-d) = \frac{i^{1-d} (p^2)^{2d-8}}{(4\pi)^{d/2} \G(4-d) \G(2d - 6)} 
	\int_{s_0 - i\infty}^{s_0 + i\infty} \int_{t_0 - i\infty}^{t_0 + i\infty} \frac{\mathrm{d}s\, \mathrm{d}t}{(2\pi i)^2}
	\tilde{\mathcal{N}}(s,t),
	\ee
with the integrand given by
	\be
	\tilde{\mathcal{N}}(s,t) = b_{1-t,\, 7 - \frac{3d}{2} + s + t} \, \mathcal{M}(1,1,4-d,s,t).
	\ee

To have a well-defined $\ve \to 0$ limit, we should properly shift the contour of integration, as shown in Fig.\ref{I(4-d) shift}. 
The contour deformations are performed as indicated by Fig.\ref{contour shift}.
	\begin{align}
	\iint_{s_0,t_0}\tilde{\mathcal{N}}(s,t) &=
	 \iint_{s_0',t_0'}\tilde{\mathcal{N}}(s,t)  -  \int_{t_0'}\Res{\tilde{\mathcal{N}} (s,t)}|_{s=0}  - \int_{t_0'}\Res{\tilde{\mathcal{N}} (s,t)}|_{s=\frac{d}{2}-2} \nonumber \\
	 & -\int_{s_0'}\Res{\tilde{\mathcal{N}}(s,t) }|_{t=0}
	 +\Res{\tilde{\mathcal{N}}(s,t) }|_{s=0,t=0}+\Res{\tilde{\mathcal{N}} (s,t)}|_{s=\frac{d}{2}-2,t=0}
	\end{align}
Expanding in powers of $\ve$, the master-Hourglass becomes
\begin{align} \label{GHR1}
	G_{\rm HR} = \frac{1}{(4\pi)^8}
	& \left\{
	\frac{8}{3\ve^4} + \frac{4}{3 \ve^3} \left[10 - 
	\ln \left( \frac{-p^2 e^\g}{4\pi} \right) \right] \right. \nonumber \\
	&\left.
	-\frac{2}{3\ve^2} \left[ - 8\ln^2 \left( \frac{-p^2 e^\g}{4\pi} \right) +40 
	\ln \left( \frac{-p^2 e^\g}{4\pi} \right) -58 + \frac{\pi^2}{3}  \right] +O\left( \frac{1}{\ve} \right)
	\right\}\, ,
	\end{align}
with the constant part of $\frac{1}{\ve^2}$ term determined by integrals analogous to the \eqref{A0 int}.
The $\frac{1}{\ve^4}$ term in \eqref{GHR1} is in agreement with the one presented by Chetyrkin in \cite{Chetyrkin4}, as it should. 

To obtain \eqref{barGHRexp}, one multiplies the expression in \eqref{GHR no exp} by the prefactors given in \eqref{barGHR}, divides by \eqref{Z_T}, and then performs the $\ve$-expansion.

\subsection{The master-Cockroach integral}

Here we evaluate the 4-loop master integral associated with the Cockroach topology in \eqref{cr master}:
	\be \label{cr to j}
	G_{\rm CR}= 
	 \int \frac{B_0(k-r) J_{1,1,1}(p,k-r)}{k^2 r^2 (k-p)^2} = \int B_0(r)[ J_{1,1,1}(p,-r)]^2
	  = b_{1,1}\int \frac{[ J_{1,1,1}(p,-r)]^2 }{(r^2)^{ 2-\frac{d}{2} } }\, .
	\ee
As for the Hourglass topology, we employ the framework of the Mellin–Barnes representation. 
However, evaluating this master integral is significantly more intricate than the previous cases.
The primary complication arises from the presence of two $J$-integrals in the integrand,
which necessitates a more subtle treatment of the complex contour integrals involved in the MB representation
	\be\label{GCR no exp}
	G_{\rm CR}= b_{1,1}\left( \frac{i^{1-d}}{(4\pi)^{d/2} \G(d-3)} \right)^2 
	\int \frac{\rmd^d r}{(2\pi)^d}  
	\int_{ u_0, v_0} \int_{s_0, t_0}
	\frac{{\cal M}(s,t) {\cal M} (u,v) (p^2)^{s+u}}
	{ \left[ r^2 \right]^{ 2-\frac{d}{2} -t-v} \left[(r+p)^2 \right]^{ 6- d +s+t+u+v} }\, ,
	\ee
where 
	\be
	{\cal M}(s,t)\equiv {\cal M}(s,t,1,1,1) \, ,
	\ee
with the function ${\cal M}$ defined in \eqref{J int MB}.

As always, the first step is to commute the loop integral with the complex integrals.
This interchange of the order of the integrations results in a primitive $B$-integral, which should be finite.
The occuring primitive integral is 
	\be
	B(2-\frac{d}{2}-t-v, 6-d+s+t+u+v;p)\, ,
	\ee
which is finite when the following set of inequalities is satisfied:
	\begin{align}
\label{cr ineq 7}	t+v>2-d \\
\label{cr ineq 8}	s+t+u+v< \frac{3d}{2} -6 \\
\label{cr ineq 9}	s+u> 2d-8\, .
	\end{align}
The poles of the primitive integral depend both on $(s,t)$ and $(u,v)$. 
To assess the region in which the $B$-integral is finite, we adopt the following approach:
we fix $(u,v)$  at their initial values $(u_0,v_0)$, as determined by the MB representation, and then examine whether a contour deformation is necessary on the $(s,t)$-plane.

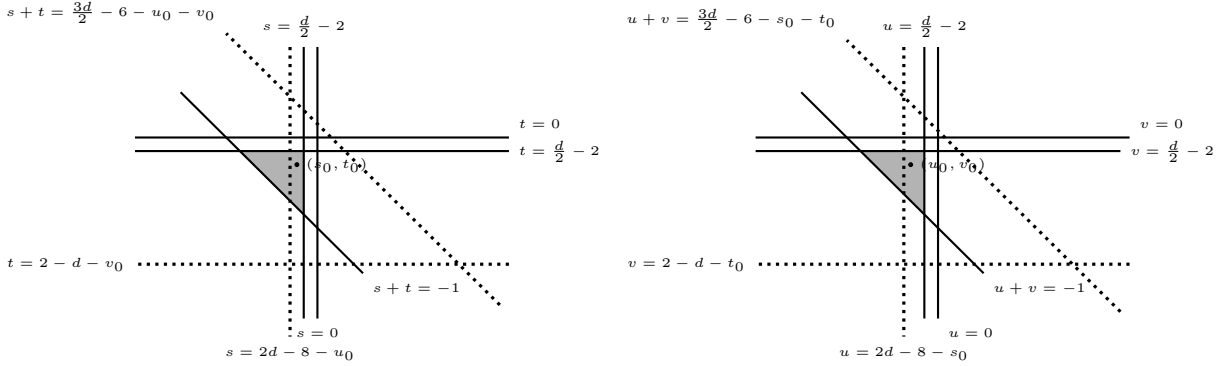
\begin{figure}[H]
		\begin{center}
			\begin{tikzpicture}[scale=1.2]
				\def\d{3.7}
				\draw[gray ,fill =gray!60] (\d/2-2,\d/2-2) -- (\d/2-2,-1-\d/2+2) -- (-1-\d/2+2,\d/2-2) -- cycle;
				\draw[black, thick] (\d/2-2,-2)--(\d/2-2,1)node[above]{\tiny$s=\frac{d}{2}-2 $} ;
				\draw[black, thick] (0,1)--(0,-2)node[below]{\tiny$s=0 $} ;
				\draw[black, thick] (-2,\d/2-2)--(2.1,\d/2-2)node[right]{\tiny$t=\frac{d}{2}-2$} ;
				\draw[black, thick] (-2,0)--(2.1,0)node[above right]{\tiny$t=0$} ;
				\draw[black, thick] plot[domain= -1.5:0.5 , samples= 2](\x,-1-\x)node[below right]{\tiny$s+t=- 1 $} ;
				\draw[black, very thick , dotted](2*\d-8-\d+4,1)-- (2*\d-8-\d+4,-2.2) node[below]{\tiny $s=2d-8-u_0$};
				\draw[black, very thick , dotted](2.1,2-\d-\d+4)--(-2,2-\d-\d+4) node[left]{\tiny $t=2-d-v_0$};
				\draw[black,very thick, dotted] plot[domain= 2:-1 , samples= 2](\x,3*\d/2-6-2*\d+8-\x)node[above left]{\tiny$s+t=\frac{3d}{2}-6 -u_0 -v_0 $} ;
				\filldraw[black] ( 3*\d/4-3,\d-4) circle(0.6 pt) node[right]{\tiny $ (s_0,t_0)$};
		\end{tikzpicture}
		\begin{tikzpicture}[scale=1.2]
				\def\d{3.7}
				\draw[gray ,fill =gray!60] (\d/2-2,\d/2-2) -- (\d/2-2,-1-\d/2+2) -- (-1-\d/2+2,\d/2-2) -- cycle;
				\draw[black, thick] (\d/2-2,-2)--(\d/2-2,1)node[above]{\tiny$u=\frac{d}{2}-2 $} ;
				\draw[black, thick] (0,1)-- (0,-2) node[below right]{\tiny$u=0$} ;

				\draw[black, thick] (-2,\d/2-2)--(2,\d/2-2)node[right]{\tiny$v=\frac{d}{2} -2$} ;
				\draw[black, thick] (-2,0)--(2.1,0)node[above right]{\tiny$v=0$} ;
				\draw[black, thick] plot[domain= -1.5:0.5 , samples= 2]
				(\x,-1-\x)node[below right]{\tiny$u+v=- 1 $} ;
				\draw[black, very thick , dotted]
				(2*\d-8-3*\d/4+3,1)-- (2*\d-8-3*\d/4+3,-2.2) node[below]{\tiny $u=2d-8-s_0$};
				\draw[black, very thick , dotted](2.1,2-\d-\d+4)--(-2,2-\d-\d+4) node[left]{\tiny $v=2-d-t_0$};
				\draw[black,very thick, dotted] 
				plot[domain= 2:-1 , samples= 2]
				(\x,3*\d/2-6-3*\d/4+3-\d+4-\x)node[above left]{\tiny$u+v=\frac{3d}{2}-6 -s_0 -t_0 $} ;
				\filldraw[black] (\d-4,\d-4) circle(0.6pt) node [ right]{\tiny $ (u_0,v_0)$};
				\end{tikzpicture}
	\caption{\small To generate the plots above, we chose $s_0= -\frac{3\ve}{4}$  and $t_0=u_0 =v_0 =-\ve$ } 
	\label{cr shifts}
	\end{center}
	\end{figure}

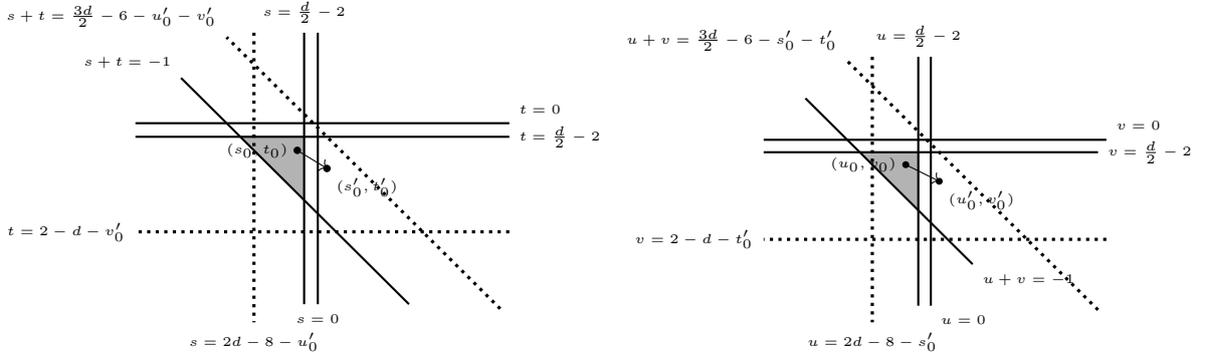
\begin{figure}[h]
		\begin{center}
			\begin{tikzpicture}[scale=1.2]
				\def\d{3.7}
				\draw[gray ,fill =gray!60] (\d/2-2,\d/2-2) -- (\d/2-2,-1-\d/2+2) -- (-1-\d/2+2,\d/2-2) -- cycle;
				\draw[black, thick] (\d/2-2,-2)--(\d/2-2,1)node[above]{\tiny$s=\frac{d}{2}-2 $} ;
				\draw[black, thick] (0,1)--(0,-2)node[below]{\tiny$s=0 $} ;
				\draw[black, thick] (-2,\d/2-2)--(2.1,\d/2-2)node[right]{\tiny$t=\frac{d}{2}-2$} ;
				\draw[black, thick] (-2,0)--(2.1,0)node[above right]{\tiny$t=0$} ;
				\draw[black, thick] plot[domain= 1:-1.5 , samples= 2](\x,-1-\x)node[above left]{\tiny$s+t=- 1 $} ;
				\draw[black, very thick , dotted](2*\d-8-0.1,1)-- (2*\d-8-0.1,-2.2) 
				node[below]{\tiny $s=2d-8-u_0'$};
				\draw[black, very thick , dotted](2.1,2-\d+0.5)--(-2,2-\d+0.5) node[left]{\tiny $t=2-d-v_0'$};
				\draw[black,very thick, dotted] plot[domain= 2:-1 , samples= 2](\x,3*\d/2-6-0.1+0.5-\x)node[above left]{\tiny$s+t=\frac{3d}{2}-6 -u_0' -v_0' $} ;
				\filldraw[black] ( 3*\d/4-3,\d-4) circle(1 pt) node[left]{\tiny $ (s_0,t_0)$};
				\draw[black, ->]  ( 3*\d/4-3,\d-4)-- (0.1,-0.5);
				\filldraw[black] (0.1,-0.5) circle(1pt) node[below right]{\tiny $ (s_0',t_0')$};
		\end{tikzpicture}
		\begin{tikzpicture}[scale=1.1]
				\def\d{3.7}
				\draw[gray ,fill =gray!60] (\d/2-2,\d/2-2) -- (\d/2-2,-1-\d/2+2) -- (-1-\d/2+2,\d/2-2) -- cycle;
				\draw[black, thick] (\d/2-2,-2)--(\d/2-2,1)node[above]{\tiny$u=\frac{d}{2}-2 $} ;
				\draw[black, thick] (0,1)-- (0,-2) node[below right]{\tiny$u=0$} ;

				\draw[black, thick] (-2,\d/2-2)--(2,\d/2-2)node[right]{\tiny$v=\frac{d}{2} -2$} ;
				\draw[black, thick] (-2,0)--(2.1,0)node[above right]{\tiny$v=0$} ;
				\draw[black, thick] plot[domain= -1.5:0.5 , samples= 2](\x,-1-\x)node[below right]{\tiny$u+v=- 1 $} ;
				\draw[black, very thick , dotted](2*\d-8-0.1,1)-- (2*\d-8-0.1,-2.2) node[below]{\tiny $u=2d-8-s_0'$};
				\draw[black, very thick , dotted](2.1,2-\d+0.5)--(-2,2-\d+0.5) node[left]{\tiny $v=2-d-t_0'$};
				\draw[black,very thick, dotted] plot[domain= 2:-1 , samples= 2](\x,3*\d/2-6-0.1+0.5-\x)node[above left]{\tiny$u+v=\frac{3d}{2}-6 -s_0' -t_0' $} ;
				\filldraw[black] (\d-4,\d-4) circle(1pt) node[left]{\tiny $ (u_0,v_0)$};
				\draw[black, ->](\d-4,\d-4) --(0.1,-0.5);
				\filldraw[black] (0.1,-0.5) circle(1pt) node[below right]{\tiny $ (u_0',v_0')$};
				\end{tikzpicture}
					\caption{\small To generate the plots above, we chose $s_0'= u_0'  =0.1$ and $t_0'=v_0'=-0.5$ } 
	\label{cr shifts2}
	\end{center}
	\end{figure}

Choosing $(u_0, v_0) = (-\ve, -\ve)$, which ensures a well-defined Mellin-Barnes representation in the $u$-$v$ plane,  
the set of inequalities that guarantee the finiteness of the primitive integral, combined with those that justify the validity  
of the Mellin representation in the $s$-$t$ plane, reduce to:
	\begin{align}
	-\ve< s_0 < - \frac{\ve}{2} \\
	-2+2\ve <t_0 < \frac{-\ve}{2} \\
	-1< s_0 +t_0 <\frac{\ve}{2}.
	\end{align}
The solution to the above corresponds to a stripe that merges in the $s\text{-}t$ plane,
bounded by the lines $ s = 2d - 8 - u_0 $, $ s = \frac{d}{2} - 2 $, $t= \frac{d}{2}-2$ and $s+t = -1$.
Within this region, the commutation of the integrals is valid without requiring any contour deformation.
By selecting $(s_0,t_0)= (-\frac{3\ve}{2}, -\ve)$, an analogous stripe is formed in the $u$-$v$, within which the fixed values $(u_0,v_0)$ reside.
This ensures that all relevant sets of inequalities are simultaneously satisfied.
These conditions are summarized in the plots shown on the left in Fig.~\ref{cr shifts}.
Accordingly, the master-Cockroach integral takes the form:
	\be
	G_{\rm CR} =b_{1,1}\left( \frac{i^{1-d}}{(4\pi)^{d/2} \G(d-3)} \right)^2 (p^2)^{2d-8}  
	\int_{t_0-i \infty}^{t_0+ i \infty} \int_{v_0-i \infty}^{v_0+ i \infty} 
	\int_{s_0-i \infty}^{s_0+ i \infty} \int_{u_0-i \infty}^{u_0+ i \infty}
	\frac{\rmd t \rmd v \rmd s \rmd u}{(2\pi i)^4} \tilde{ \cal M} (s,t,u,v)\, ,
	\ee
where
	\be
	\tilde{ \cal M} (s,t,u,v)= b_{2-\frac{d}{2}-t-v, 6-d+s+t+u+v}{\cal M}(s,t) {\cal M} (u,v)\, .
	\ee
The final step is to evaluate the complex contour integrals. When setting $\varepsilon = 0$, the integration contours intersect with poles, as illustrated by the plots on the right in Fig.~\ref{cr shifts}. 
To address this, the contours must be deformed prior to taking the limit $\varepsilon \to 0$. 
A subtlety that distinguishes this case from the Hourglass topology is that the triangle defining the region where the interchange of integrations is justified is itself affected by the contour shifts. 
Therefore, the deformations in both planes must be applied simultaneously. 
These shifts are illustrated in Fig.~\ref{cr shifts2}.
Applying the residue theorem, the multiple complex integral becomes:
	\begin{align}
	\int_{t_0 v_0 s_0 u_0} \tilde{{\cal M}}=\int_{t_0' v_0'}& \left\{ \int_{s_0' t_0'} \tilde{ {\cal M} }
	-\int_{s_0'} \Res{ \tilde{{\cal M } } }|_{u=0} - \int_{s_0'} \Res{ \tilde{{\cal M } } }|_{u=\frac{d}{2}-2 }  
	-\int_{u_0'} \Res{ \tilde{{\cal M } } }|_{s=0} -\int_{t_0'} \Res{ \tilde{{\cal M } } }|_{s= \frac{d}{2}-2 } 
	\right. \nonumber\\
	&\left. + \Res{ \tilde{{\cal M } } }|_{s=0, u=0} + \Res{ \tilde{{\cal M } } }|_{s=0, u=\frac{d}{2}-2 }
	+ \Res{ \tilde{{\cal M } } }|_{s=\frac{d}{2} -2, u=0} + \Res{ \tilde{{\cal M } } }|_{s=\frac{d}{2}-2, u=\frac{d}{2}-2 }
	\right\}\, .
	\end{align}
In the context of $\ve$- expansion we obtain:
	\be
	G_{\rm CR} = \frac{1}{(4\pi)^8} \left\{ 
	\frac{4}{3 \ve^4}  
	- \frac{2}{3 \ve^3} \left[-11 + 4 \ln \left( \frac{-p^2 e^\g}{4\pi} \right)    \right] 
	+ \frac{1}{\ve^2} \left[ \frac{8}{3}\ln^2 \left( \frac{-p^2 e^\g}{4\pi} \right)
	- \frac{44}{3}\ln \left( \frac{-p^2 e^\g}{4\pi} \right)  +\frac{71}{3} - \frac{\pi^2}{9} \right]
	 \right\}\, ,
	\ee
where all the terms are determined by integrals over the complex variables $t$ and $v$, with the integrands consisting of powers of 
$\Gamma$-functions and harmonic numbers. The $\frac{1}{\ve^4}$ term is in agreement with the result presented in \cite{Chetyrkin4}, as it should.

To obtain \eqref{barGCRexp}, one multiplies the expression in \eqref{GCR no exp} by the prefactors given in \eqref{barGCR}, divides by \eqref{Z_T}, and then performs the $\ve$-expansion.

\end{appendix}

\newpage
\bibliographystyle{JHEP}
\bibliography{refmain}
\end{document}